\documentclass{bioinfo}
\copyrightyear{2016}
\pubyear{2016}

\usepackage{algorithmic,algorithm}

\usepackage[square]{natbib}

\definecolor{ddarkbrown}{rgb}{0.5,0.2,0.05} \definecolor{bbluegray}{rgb}{0.05,0,0.5}

\newcommand{\BEAS}{\begin{eqnarray*}}
\newcommand{\EEAS}{\end{eqnarray*}}
\newcommand{\BEA}{\begin{eqnarray}}
\newcommand{\EEA}{\end{eqnarray}}
\newcommand{\BEQ}{\begin{equation}}
\newcommand{\EEQ}{\end{equation}}
\newcommand{\BIT}{\begin{itemize}}
\newcommand{\EIT}{\end{itemize}}
\newcommand{\BNUM}{\begin{enumerate}}
\newcommand{\ENUM}{\end{enumerate}}

\newcommand{\BA}{\begin{array}}
\newcommand{\EA}{\end{array}}

\newcommand{\ones}{\mathbf 1}

\newcommand{\diag}{\mathop{\bf diag}}

\newcommand{\QED}{~~\rule[-1pt]{6pt}{6pt}}

\usepackage{amsmath}

\newcommand{\beginsupplement}{%
        \setcounter{table}{0}
        \renewcommand{\thetable}{S\arabic{table}}%
        \setcounter{figure}{0}
        \renewcommand{\thefigure}{S\arabic{figure}}%
        \setcounter{page}{1}
     }
\usepackage{subcaption}
\usepackage{multirow}
\usepackage{placeins}

\usepackage{tabulary}
\newcolumntype{K}[1]{>{\centering\arraybackslash}p{#1}}

\begin{document}
\bibliographystyle{apalike}
\firstpage{1}

\title[A spectral algorithm for fast \textit{de novo} layout of uncorrected long nanopore reads]{A spectral algorithm for fast \textit{de novo} layout of uncorrected long nanopore reads}

\author[Recanati, Br{\"u}ls, d'Aspremont]{Antoine Recanati\,$^{\text{\sfb 1}}$, Thomas Br{\"u}ls\,$^{\text{\sfb 2}, \text{\sfb 3}, \text{\sfb 4}}$ and Alexandre d'Aspremont\,$^{\text{\sfb 1}}$}
\address{$^{\text{\sf 1}}$CNRS \& D.I., UMR 8548, \'Ecole Normale Sup\'erieure, Paris, France. \\
$^{\text{\sf 2}}$Commissariat \`a l'Energie Atomique et aux Energies Alternatives, Direction de la Recherche Fondamentale, Genoscope, Evry, France.\\
$^{\text{\sf 3}}$UMR 8030, Centre National de la Recherche Scientifique, Evry, France.\\
$^{\text{\sf 4}}$Universit\'e Paris-Saclay, Evry, France.}

\maketitle

\begin{abstract}
\section{Motivation:} New long read sequencers promise to transform sequencing and genome assembly by producing reads tens of kilobases long. However, their high error rate significantly complicates assembly and requires expensive correction steps to layout the reads using standard assembly engines.
\section{Results:} We present an original and efficient spectral algorithm to layout the uncorrected nanopore reads, and its seamless integration into a straightforward overlap/layout/consensus (OLC) assembly scheme. The method is shown to assemble Oxford Nanopore reads from several bacterial genomes into good quality ($\sim$99\% identity to the reference) genome-sized contigs, while yielding more fragmented assemblies from the eukaryotic microbe \textit{Sacharomyces cerevisiae}.

\section{Availability and implementation:}

https://github.com/antrec/spectrassembler

\section{Contact:}
antoine.recanati@inria.fr
\end{abstract}

\section{Introduction}
\textit{De novo} whole genome sequencing seeks to reconstruct an entire genome from randomly sampled sub-fragments whose order and orientation within the genome are unknown. The genome is
oversampled so that all parts are covered multiple times with high probability.

High-throughput sequencing technologies such as Illumina substantially reduce sequencing cost at the expense of read length, which is typically a few hundred base pairs long (bp) at best. Yet, \textit{de novo} assembly is challenged by short reads, as genomes contain repeated sequences resulting in layout degeneracies when read length is shorter or of the same order than repeat length \citep{Pop04}.

Recent long read sequencing technologies such as PacBio's SMRT and Oxford Nanopore Technology (ONT) have spurred a renaissance in \textit{de novo} assembly as they produce reads over 10kbp long \citep{KorenOneChr}. However, their high error rate ($\sim$15\%) makes the task of assembly difficult, requiring complex and computationally intensive pipelines.

Most approaches for long read assembly address this problem by correcting the reads prior to performing the assembly,
while a few others integrate the correction with the overlap detection phase, as in the latest version of the Canu pipeline \citep{Koren:Canu} (former Celera Assembler \citep{MyersCA}).

\emph{Hybrid techniques} combine short and long read technologies: the accurate short reads are mapped onto the long reads, enabling a consensus sequence to be derived for each long read and thus providing low-error long reads (see for example \citet{NaS2015}). This method was shown to successfully assemble prokaryotic and eukaryotic genomes with PacBio \citep{KorenHybrid} and ONT \citep{goodwin2015oxford} data. \emph{Hierarchical assembly} follows the same mapping and consensus principle but resorts to long read data only, the rationale being that the consensus sequence derived from all erroneous long reads matching a given position of the genome should be accurate provided there is sufficient coverage and sequencing errors are reasonably randomly distributed: for a given base position on the genome, if 8 out of 50 reads are wrong, the majority vote still yields the correct base. Hierarchical methods map long reads against each other and derive, for each read, a consensus sequence based on all the reads that overlap it. Such an approach was implemented in HGAP \citep{ChinHGAP} to assemble PacBio SMRT data, and more recently by \citet{Loman}, to achieve complete \textit{de novo} assembly of \textit{Escherichia coli} with ONT data exclusively.

Recently, \citet{Li:Miniasm} showed that it is possible to efficiently perform \textit{de novo} assembly of noisy long reads in only two steps, without any dedicated correction procedure: all-vs-all raw read mapping (with minimap) and assembly (with miniasm). The miniasm assembler is inspired by the Celera Assembler and produces unitigs through the construction of an assembly graph. Its main limitation is that it produces a draft whose error rate is of the same order as the raw reads.

Here, we present a new method for computing the layout of raw nanopore reads, resulting in a simple and computationally efficient protocol for assembly. It takes as input the all-vs-all overlap information (\textit{e.g.} from minimap, MHAP \citep{BerlinMHAP} or DALIGNER \citep{MyersDALIGN}) and outputs a layout of the reads (\textit{i.e.} their position and orientation in the genome).
Like miniasm, we compute an assembly from the all-vs-all raw read mapping, but achieve improved quality through a coverage-based consensus generation process, as in nanocorrect \citep{Loman}, although reads are not corrected individually in our case.

The method relies on a simple spectral algorithm akin to Google's PageRank \citep{page1999pagerank} with deep theoretical underpinnings, described in \S\ref{subsec:layout}.
It has successfully been applied to consecutive-ones problems arising in physical mapping of genomes \citep{Atkins96}, ancestral genome reconstructions \citep{ANGES}, or the locus ordering problem \citep{cheema2010thread}, but to our knowledge has not been applied to \textit{de novo} assembly problems.
In \S\ref{subsec:assembly}, we describe an assembler based on this layout method, to which we add a consensus generation step based on POA \citep{LeePOA}, a multi-sequence alignment engine. Finally, we evaluate this pipeline on prokaryotic and eukaryotic genomes in \S\ref{sec:results}, and discuss possible improvements and limitations in \S\ref{sec:discuss}.

\begin{methods}
\section{Methods}

\subsection{Layout computation}\label{subsec:layout}
We lay out the reads in two steps. We first sort them by position, i.e., find a permutation $\pi$ such that read $\pi(1)$ will be positioned before read $\pi(2)$ on the genome. Then, we iteratively assign an exact position (i.e., leftmost basepair coordinate on the genome) to each read by using the previous read's position and the overlap information.

The key step is the first one, which we cast as a seriation problem, i.e. we seek to reconstruct a linear order between $n$ elements using unsorted, pairwise similarity information \citep{Atkins,Fogel}. Here the $n$ elements are the reads, and the similarity information comes from the overlapper (\textit{e.g.} from minimap).

The seriation problem is formulated as follows. Given a pairwise similarity matrix $A_{ij}$, and assuming the data has a serial structure, \textit{i.e.} that there exists an order $\pi$ such that $A_{\pi(i) \pi(j)}$ decreases with $|i - j|$,  seriation seeks to recover this ordering $\pi$ (see Figure~\ref{fig:spectral}
for an illustration). If such an order $\pi$ exists, it minimizes the 2-SUM score,
\BEQ\label{eq:2-sum}
\text{2-SUM} (\pi) = \sum_{i,j=1}^{n} A_{ij} \left(\pi(i) - \pi(j) \right)^2,
\EEQ
and the seriation problem can be solved as a minimization over the set of permutation vectors \citep{Fogel}. In other words, the permutation $\pi$ should be such that if $A_{ij}$ is high (meaning that $i$ and $j$ have a high similarity), then $\left(\pi(i) - \pi(j) \right)^2$ should be low, meaning that the positions $\pi(i)$ and $\pi(j)$ should be close to each other. Conversely, if $A_{ij}=0$, the positions of $i$ and $j$ in the new order may be far away without affecting the score.

When using seriation to solve genome assembly problems, the similarity $A_{ij}$ measures the overlap between reads $i$ and $j$. In an ideal setting with constant read length and no repeated regions, two overlapping reads should have nearby positions on the genome. We therefore expect the order found by seriation to roughly match the sorting of the positions of the reads.
\begin{figure}[ht]
	\centering
	\begin{tabular}{cc}

		\includegraphics[width=0.22\textwidth]{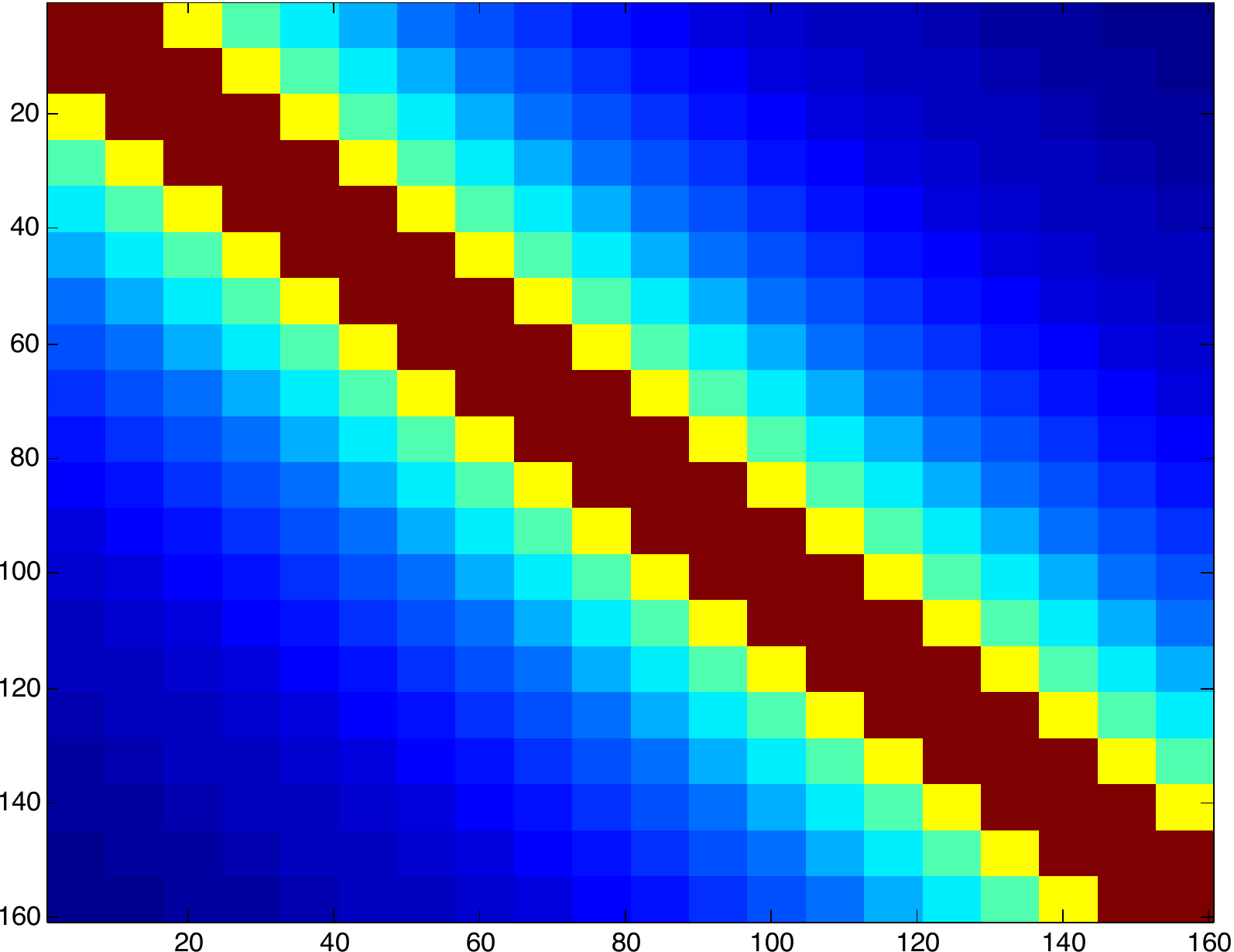}
    &
		\includegraphics[width=0.22\textwidth]{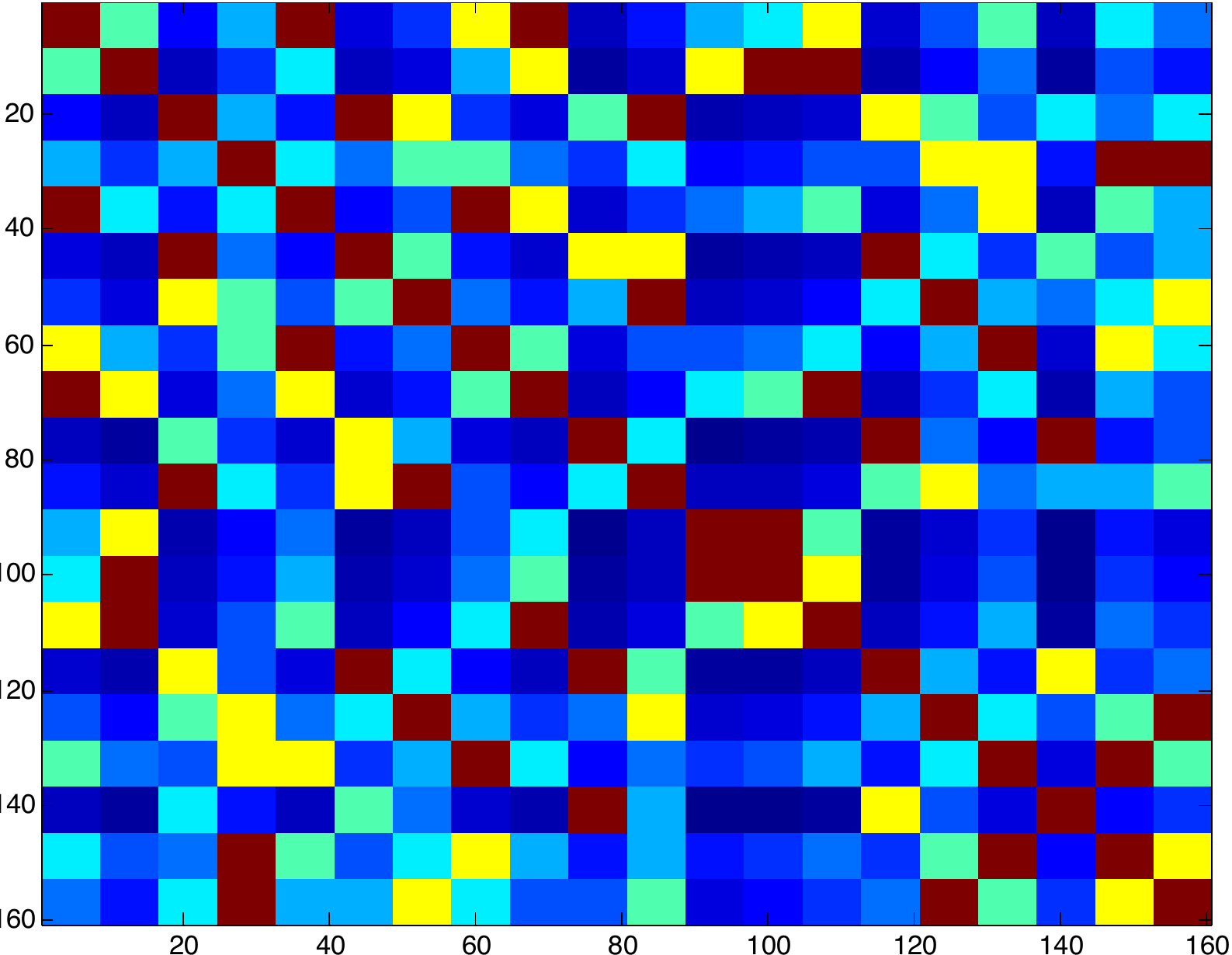}
	\end{tabular}
	\caption{A similarity matrix reordered with the spectral algorithm. The original matrix (left) has values that decrease when moving away from the diagonal (red : high value, blue : low value). It is randomly permuted (right), and the spectral algorithm will find back the original ordering.}
	\label{fig:spectral}
\end{figure}

The problem of finding a permutation over $n$ elements is combinatorial. Still, provided the original data has a serial structure, an exact solution to seriation exists in the noiseless case \citep{Atkins} using spectral clustering, and there exist several convex relaxations allowing explicit constraints on the solution \citep{Fogel}.

The exact solution is directly related to the well-known spectral clustering algorithm. Indeed, for any vector $\mathbf{x}$, the objective in~\eqref{eq:2-sum} reads
\[
\sum_{i,j=1}^{n} A_{ij} \left(x_i - x_j \right)^2 = x^T L_A x, \quad L_A=\diag(A\ones)-A
\]
where $L_A$ is the Laplacian matrix of $A$. This means that the 2-SUM problem amounts to
\[
\min_{\pi}\pi^T L_A \pi
\]
where $\pi$ is a permutation vector. Roughly speaking, the spectral clustering approach to seriation relaxes the constraint ``$\pi$ is a permutation vector'' into ``$\pi$ is a vector of $\mathbb{R}^n$ orthogonal to the constant vector $\ones=(1,...,1)^T$'' with fixed norm. The problem then becomes
\[
\min_{\{\ones^T\pi=0,\,\|\pi\|_2=1\}}\pi^T L_A \pi
\]
This relaxed problem is an eigenvector problem. Finding the minimum over normalized vectors $x$ yields the eigenvector associated to the smallest eigenvalue of $L_A$, but the smallest eigenvalue, $0$, is associated with the eigenvector $\ones$, from which we cannot recover any permutation. However, if we restrict $x$ to be orthogonal to $\ones$, the solution is the second smallest eigenvector, called the Fiedler vector. A permutation is recovered from this eigenvector by sorting its coefficients: given $\mathbf{x} = (x_1, x_2,  ..., x_n)$, the algorithm outputs a permutation $\pi$ such that $x_{\pi(1)} \leq x_{\pi(2)} \leq ... \leq x_{\pi(n)}$. This procedure is summarized as Algorithm \ref{alg:spectral}.

In fact, \citep{Atkins} showed that under the assumption that $A$ has a serial structure, Algorithm \ref{alg:spectral} solves the seriation problem exactly, i.e. recovers the order $\pi$ such that $A_{\pi(i) \pi(j)}$ decreases with $|i - j|$. This means that we solve the read ordering problem by simply solving an extremal eigenvalue problem, which has low complexity (comparable to Principal Component Analysis (PCA)) and is efficient in practice (see Supplementary Figure~\ref{suppfig:runtimeReordering} and Table~\ref{supptab:runtime}).

\begin{algorithm}[ht]
\footnotesize
\caption{Spectral ordering}\label{alg:spectral}
\begin{algorithmic} [1]
\REQUIRE Connected similarity matrix $A \in \mathbb{R}^{n \times n}$
\STATE Compute Laplacian $L_A=\diag(A\ones)-A$
\STATE Compute second smallest eigenvector of $L_A$, $\mathbf{x^*}$
\STATE Sort the values of $\mathbf{x^*}$
\ENSURE Permutation $\pi : \mathbf{x^*}_{\pi(1)} \leq \mathbf{x^*}_{\pi(2)} \leq ... \leq \mathbf{x^*}_{\pi(n)}$
\end{algorithmic}
\end{algorithm}

Once the reads are reordered, we can sequentially compute their exact positions (basepair coordinate of their left end on the genome) and orientation. We assign position 0 and strand ``+'' to the first read, and use the overlap information (position of the overlap on each read and mutual orientation) to compute the second read's position and orientation, etc. More specifically, when computing the position and orientation of read $i$, we use the information from reads $i-1,...,i-c$ to average the result, where $c$ roughly equals the coverage, as this makes the layout more robust to misplaced reads.
Note that overlappers relying on hashing, such as minimap and MHAP, do not generate alignments but still locate the overlaps on the reads, making this positioning step possible.
Thanks to this ``polishing'' phase, we would still recover the layout if two neighboring reads were permuted due to consecutive entries of the sorted Fiedler vector being equal up to the eigenvector computation precision, for example.

\subsection{Consensus generation}\label{subsec:assembly}
We built a simple assembler using this layout idea and tested its accuracy. It is partly inspired by the nanocorrect pipeline of \citet{Loman} in which reads are corrected using multiple alignments of all overlapping reads.
These multiple alignments are performed with a Partial Order Aligner (POA) \citep{LeePOA} multiple-sequence alignment engine.
It computes a consensus sequence from the alignment of multiple sequences using a dynamic programming approach that is efficient when the sequences are similar (which is the case if we trim the sequences to align their overlapping parts).
Specifically, we used SPOA, a Single Instruction Multiple Data implementation of POA developed in \citet{racon2016}.

The key point is that we do not need to perform multiple alignment using all reads, since we already have a layout. Instead, we can generate a consensus sequence for, say, the first 3000 bp of the genome by aligning the parts of the reads that are included in this window with SPOA, and repeat this step for the reads included in the window comprising the next 3000 bp of the genome, etc. In practice, we take consecutive windows that overlap and then merge them to avoid errors at the edges, as shown in Figure~\ref{fig:consensus}. The top of the figure displays the layout of the reads broken down into three consecutive overlapping windows, with one consensus sequence generated per window with SPOA. The final assembly is obtained by iteratively merging the window $k$+1 to the consensus formed by the windows $1,\ldots,k$.

The computational complexity for aligning $N$ sequences of length $L$ with POA, with an average divergence between sequences $\epsilon$, is roughly $O(m N L^2)$, with $m\simeq(1+2 \epsilon)$. With $10\%$ of errors, $m$ is close to 1. If each window of size  $L_{w}$ contains about $C$ sequences, the complexity of building the consensus in a window is $O( m C L_{w}^2)$. We compute $L_{g}/L_{w}$ consensus windows, with $L_g$ the length of the genome (or contig), so the overall complexity of the consensus generation is $O( m C L_{g} L_{w})$.
We therefore chose in practice a window size relatively small, but large enough to prevent mis-assemblies due to noise in the layout, $L_{w}=3$kbp.

\begin{figure}[t]
\centering
\includegraphics[width=.45\textwidth]{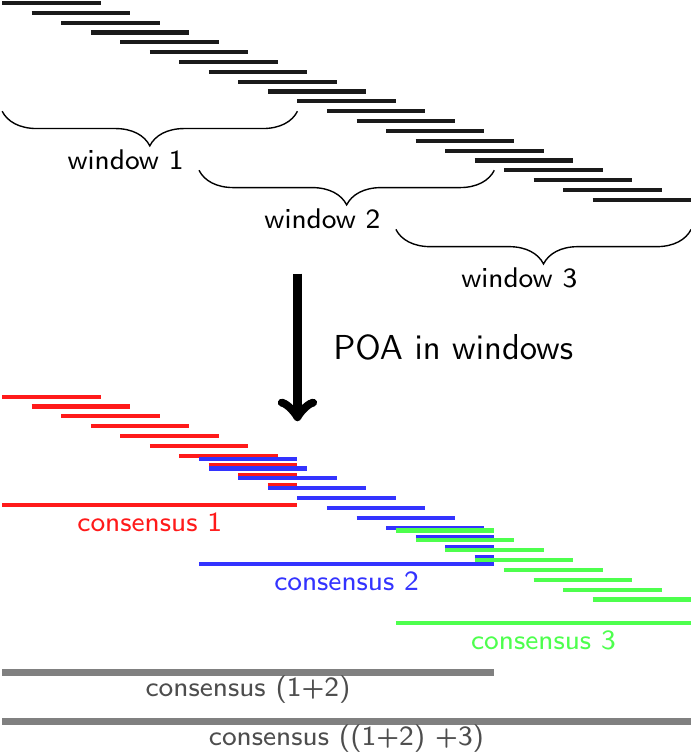}
\center
\caption{Consensus generation. Given the layout, the genome is sliced into overlapping windows, and a consensus is computed in each window. The final consensus is then obtained by merging the consensus windows.}\label{fig:consensus}
\end{figure}

\subsection{Overlap-based similarity and repeats handling}\label{subsec:outliers}

In practice, we build the similarity matrix $A$ as follows.
Given an overlap found between the i-th and j-th reads, we set $A_{ij}$ equal to the overlap score (or number of matches, given in tenth column of minimap or fourth column of MHAP output file).
Such matrices are sparse: a read overlaps with only a few others (the number of neighbors of a read in the overlap graph roughly equals the coverage).
There is no sparsity requirement for the algorithm to work, however sparsity lowers RAM usage since we store the $n \times n$ similarity matrix with about $n \times C$ non-zero values, with $C$ the coverage.
In such cases, the ordered similarity matrix is band diagonal.

\begin{figure}[tbh]
	\centering
	\begin{tabular}{cc}
		\includegraphics[width=0.23\textwidth]{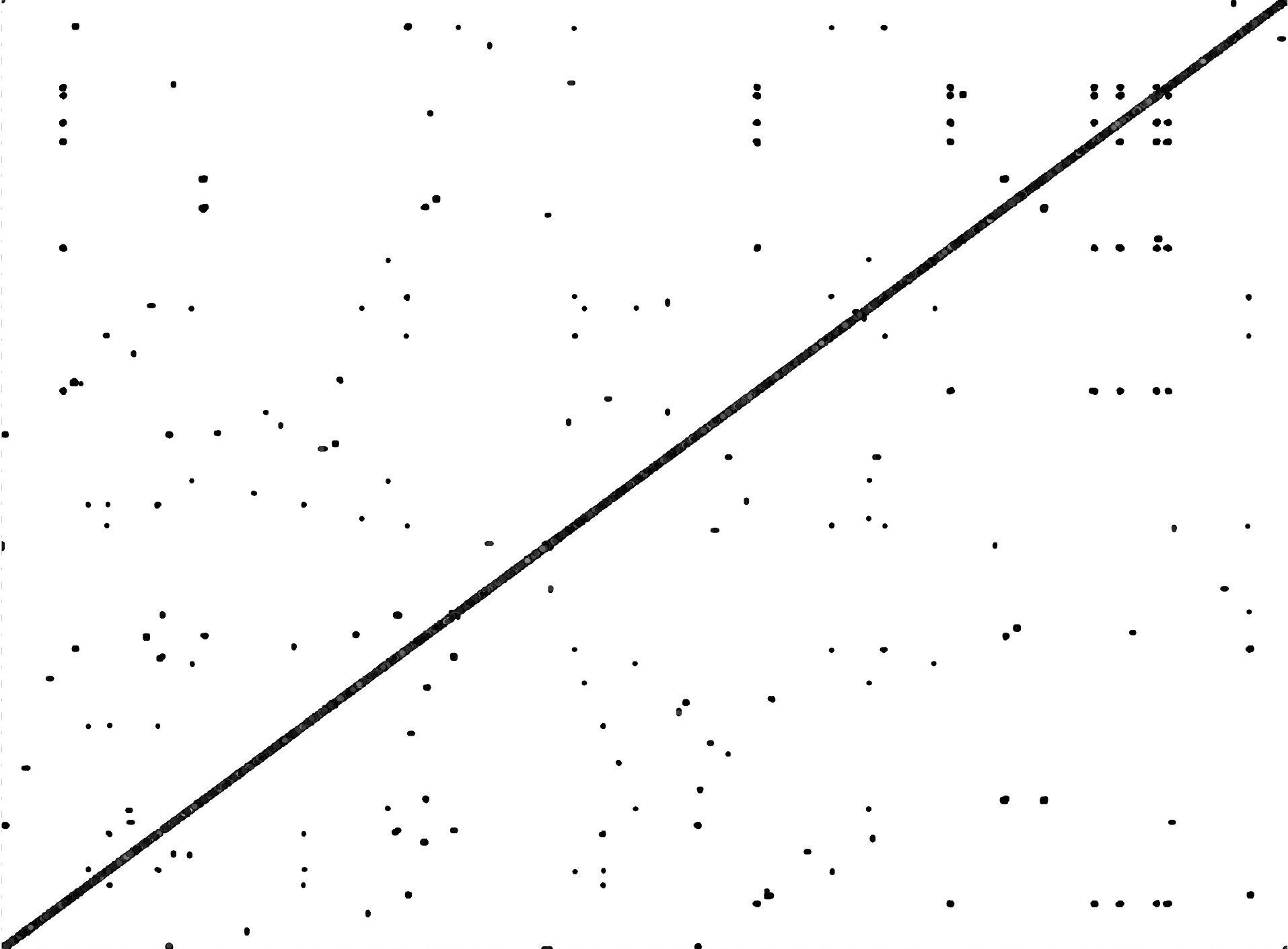}
    &
		\includegraphics[width=0.23\textwidth]{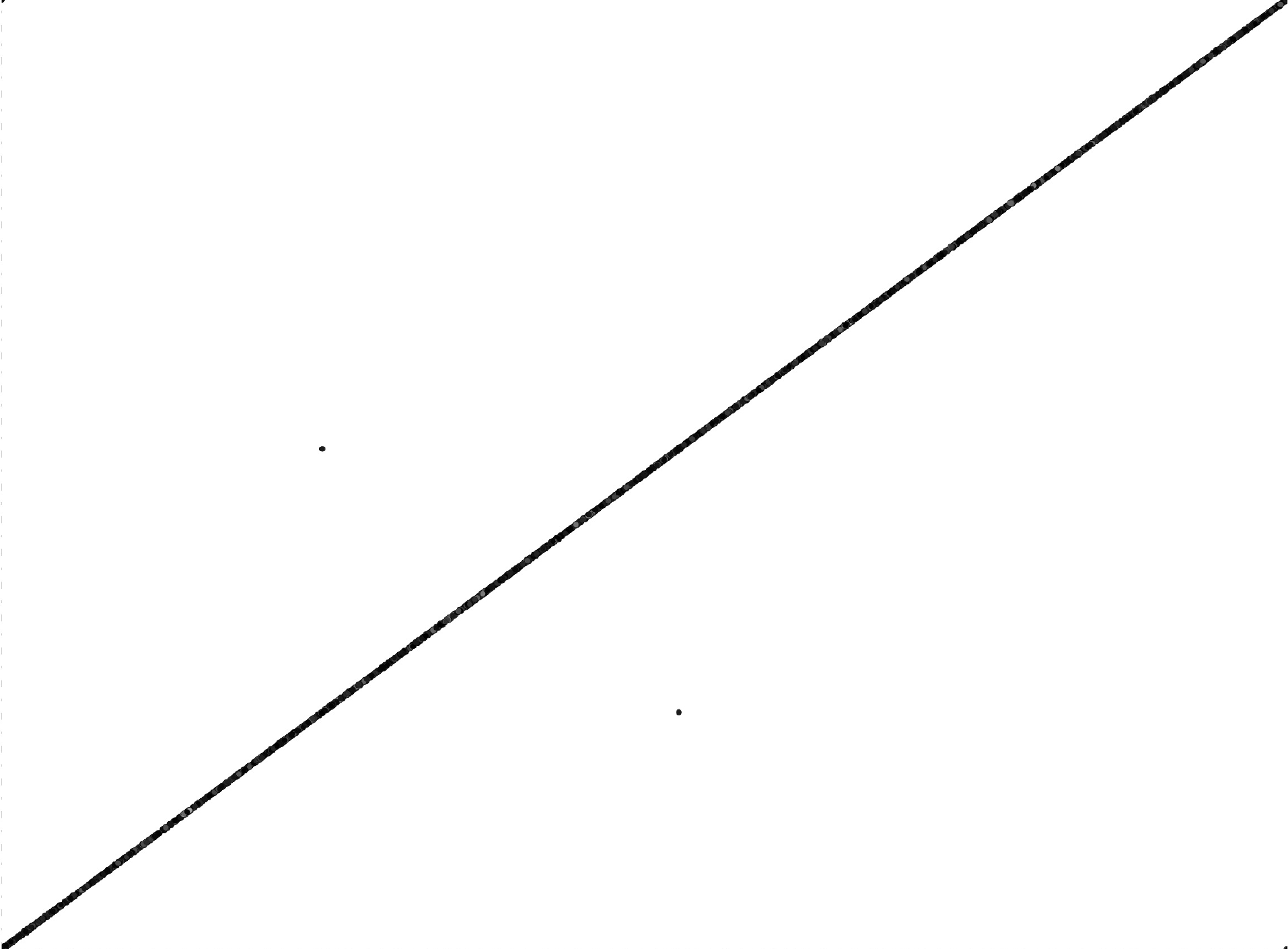}
  \end{tabular}
	\caption{Similarity matrix for \textit{E. coli} ONT sequences before (left) and after (right) thresholding.
  The positions of the reads were obtained by mapping to the reference genome with GraphMap \citep{sovic2016fast}.
	}
	\label{fig:outliers}
\end{figure}

Unfortunately, the correctly ordered (sorted by position of the reads on the backbone sequence) similarity matrix contains outliers outside the main diagonal band (see Figure~\ref{fig:outliers}) that corrupt the ordering.
These outliers are typically caused by either repeated subsequences or sequencing noise (error in the reads and chimeric reads), although errors in the similarity can also be due to hashing approximations made in the overlap algorithm.
We use a threshold on the similarity values and on the length of the overlaps to remove them.
The error-induced overlaps are typically short and yield a low similarity score (\textit{e.g.}, number of shared min-mers),
while repeat-induced overlaps can be as long as the length of the repeated region.
By weighting the similarity, the value associated to repeat-induced overlaps can be lowered.
Weighting can be done with, \textit{e.g}., the \texttt{--weighted} option in MHAP to add a tf-idf style scaling to the MinHash sketch, making repetitive k-mers less likely to cause a match between two sequences, or with default parameters with minimap.
In the Supplementary Material, we describe experiments with real, corrected and simulated reads to assess the characteristics of such overlaps and validate our method.
Supplementary Figure \ref{fig:outliersHisto} shows that although the overlap scores and lengths are lower for outliers than for inliers on average, the distributions of these quantities intersect.
As shown in \ref{fig:simMatSpyNanoSim}, the experiments indicate that all false-overlaps can be removed with a stringent threshold on the overlap length and score.
However, removing all these short or low score overlaps will also remove many true overlaps.
For bacterial genomes, the similarity graph can either remain connected or be broken into several connected components after a threshold-based outlier removal, depending on the initial coverage.
Figure~\ref{fig:simMatSpyNanoSim} illustrates the empirical observation that the coverage needs to be above 60x to keep the graph connected while removing all outliers.
Most outliers can be similarly removed for real and synthetic data from \textit{S. cerevisiae}, although a few outliers, probably harboring telomeric repeats, remain at the ends of chromosomes after thresholding.

There is thus a tradeoff to be reached depending on how many true overlaps one can afford to lose.
With sufficient coverage,
a stringent threshold on overlap score and length
will remove both repeat-induced and error-induced overlaps, while still yielding a connected assembly graph.
Otherwise, aggressive filtering will break the similarity graph into several connected components.
In such a case, since the spectral algorithm only works with a connected similarity graph, we compute the layout and consensus separately in each connected component, resulting in several contigs.
To
set the threshold sufficiently high to remove outliers but small enough to keep the number of contigs minimal, we used a heuristic based on the following empirical observation,
illustrated in Supplementary Figure~\ref{fig:BandwidthHeuristic}.
The presence of outliers in the correctly (based on the positions of the reads) ordered band diagonal matrix imparts an increased bandwidth (maximum distance to the diagonal of non zero entries) on the matrix reordered with the spectral algorithm.

We can therefore run the spectral algorithm, check the bandwidth in the reordered matrix, and increase the threshold if the bandwidth appears too large (typically larger than twice the coverage).

In practice, we chose to set the threshold on the overlap length to 3.5kbp,
and removed the overlaps with the lowest score [in the first 40\%-quantile (respectively 90\% and 95\%) for C$\leq$60X (resp. 60X$\leq$C$\leq$100X and C$\geq$100X)].
As indicated in Algorithm~\ref{alg:assembler}, we let these threshold values increase if indicated by the bandwitdh heuristic.

Finally, we
added a filtering step to remove reads that have non-zero similarity with several sets of reads located in distant parts of the genome, such as chimeric reads.
These reads usually overlap with a first subset of reads at a given position in the genome, and with another distinct subset of reads at another location, with no overlap between these distinct subsets.
We call such reads ``connecting reads'', and they can be detected from the similarity matrix by computing, for each read (index $i$), the set of its neighbors in the graph $\mathcal{N}_{i} =\{ j : A_{ij} > 0 \}$.
The subgraph represented by $A$ restricted to  $\mathcal{N}_{i}$ is either connected (there exists a path between any pair of edges), or split into separate connected components.
In the latter case, we keep the overlaps between read $i$ and its neighbor that belong to only one of these connected components (the largest one).

\begin{algorithm}[tbh]
\footnotesize
\caption{OLC assembly pipeline}\label{alg:assembler}
\begin{algorithmic} [1]
\REQUIRE $n$ long noisy reads
\STATE Compute overlaps with an overlapper (\textit{e.g.} minimap or MHAP)
\STATE Construct similarity matrix $S \in \mathbb{R}^{n \times n}$ from the overlaps
\STATE Remove outliers from $S$ with a threshold on values $S_{ij}$, on overlap length, and removal of connecting reads (as explained in \S\ref{subsec:outliers})
\FORALL{Connected component $A$ of $S$}
	\STATE Reorder $A$ with spectral algorithm (Algorithm \ref{alg:spectral})
	\IF{bandwidth of $A_{reordered}$ $\geq$ 2$\times$ Coverage}
		\STATE{set higher threshold on $A$ and try again}
	\ENDIF
	\STATE Compute layout from the ordering found and overlaps
	\STATE Partition the length of the contig into small windows
	\STATE Compute consensus in each window with SPOA
	\STATE Merge consecutive windows with SPOA
\ENDFOR
\ENSURE Contig consensus sequences
\end{algorithmic}
\end{algorithm}

\end{methods}

\section{Results}\label{sec:results}
\subsection{Data}\label{subsec:data}
We tested this pipeline on
ONT and PacBio data.
The bacterium \textit{ Acinetobacter baylyi ADP1} and the yeast \textit{ Saccharomyces cerevisiae S288C} were sequenced at Genoscope with Oxford Nanopore's MinION device using the R7.3 chemistry, together with an additional dataset of \textit{S. cerevisiae S288C} using the R9 chemistry.
Only the 2D high quality reads were used.
The \textit{S. cerevisiae S288C} ONT sequences were deposited at the European Nucleotide Archive (http://www.ebi.ac.uk/ena) where they can be accessed under Run accessions ERR1539069 to ERR1539080, while \textit{ Acinetobacter baylyi ADP1}  sequences will be made available on https://github.com/antrec/spectrassembler.
We also used the following publicly available data: ONT \textit{ Escherichia coli} by \citet{Loman} (http://bit.ly/loman006 - PCR1 2D pass dataset), and PacBio \textit{ E. coli} K-12 PacBio P6C4, and \textit{ S. cerevisiae W303 P4C2}.
Their key characteristics are given with the assembly results in Table~\ref{tab:allAssembly}, and read length histograms are given in Supplementary Figure~\ref{fig:lengthHistoApp}.
For each dataset, we also used the reads corrected and trimmed by the Canu pipeline as an additional dataset with low error-rate. The results on these corrected datasets are given in Supplementary Figures~\ref{suppfig:bactlayouts} and \ref{suppfig:yeastlayout} and Tables~\ref{tab:correctedAllAssembly}
and \ref{tab:corrMisassemblies}.

\subsection{Layout}
\subsubsection{Bacterial genomes}
\begin{figure}[t]
 	\centering
	\includegraphics[width=0.46\textwidth]{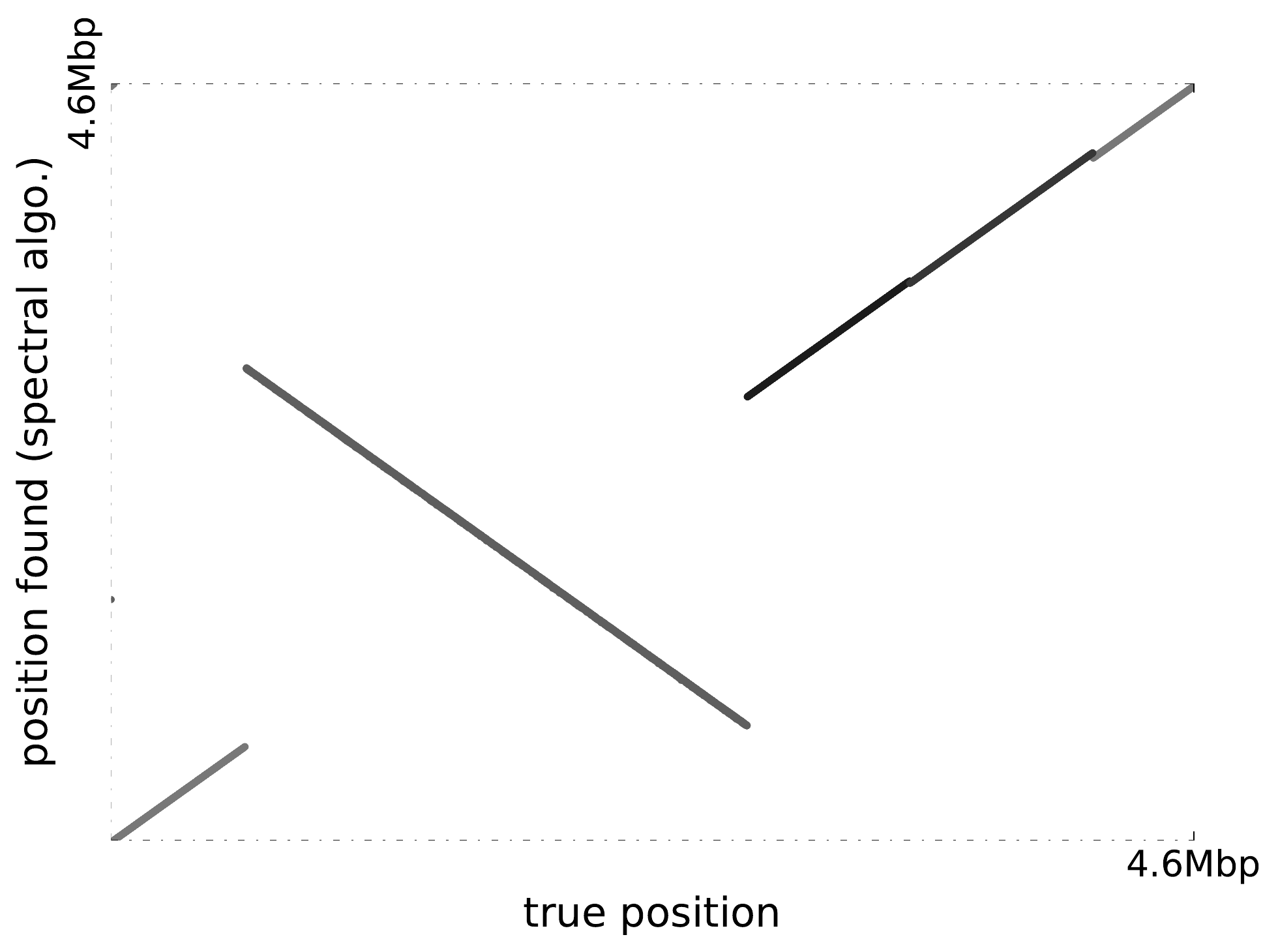}
	\caption{
 	Ordering of the reads computed with the spectral algorithm vs true ordering (obtained by mapping the reads to the reference genome with GraphMap) for the \textit{E. coli} ONT dataset.
  All contigs are  artificially displayed on the same plot for compactness.
  There are two equivalent correct orderings for each contig : (1,2,...,n) and (n, n-1, ..., 1),
  both yielding the same 2-SUM score (\ref{eq:2-sum}) and leading to the same consensus sequence (possibly reverse complemented).
  }
 	\label{fig:layoutBact}
 \end{figure}

\begin{figure}[htb]
  \centering
  \begin{subfigure}[b]{0.25\textwidth}
    \centering \includegraphics[width=\textwidth]{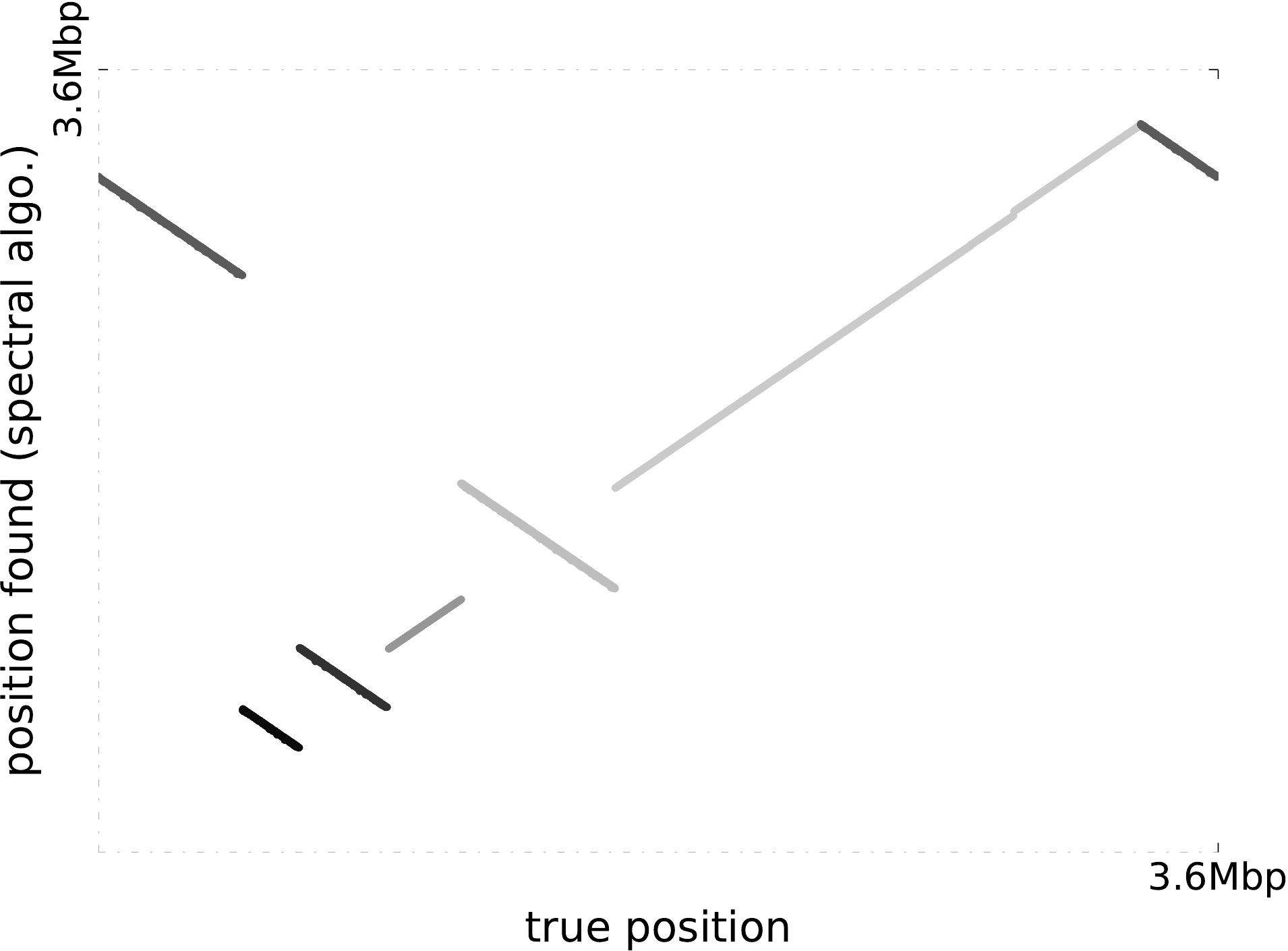}
    \caption{\textit{A. baylyi} ONT}
  \end{subfigure}
  ~
  \begin{subfigure}[b]{0.25\textwidth}
    \centering \includegraphics[width=\textwidth]{figures/oxford_test0_layout.pdf}
    \caption{\textit{E. coli} ONT}
  \end{subfigure}
  ~
  \begin{subfigure}[b]{0.25\textwidth}
    \centering \includegraphics[width=\textwidth]{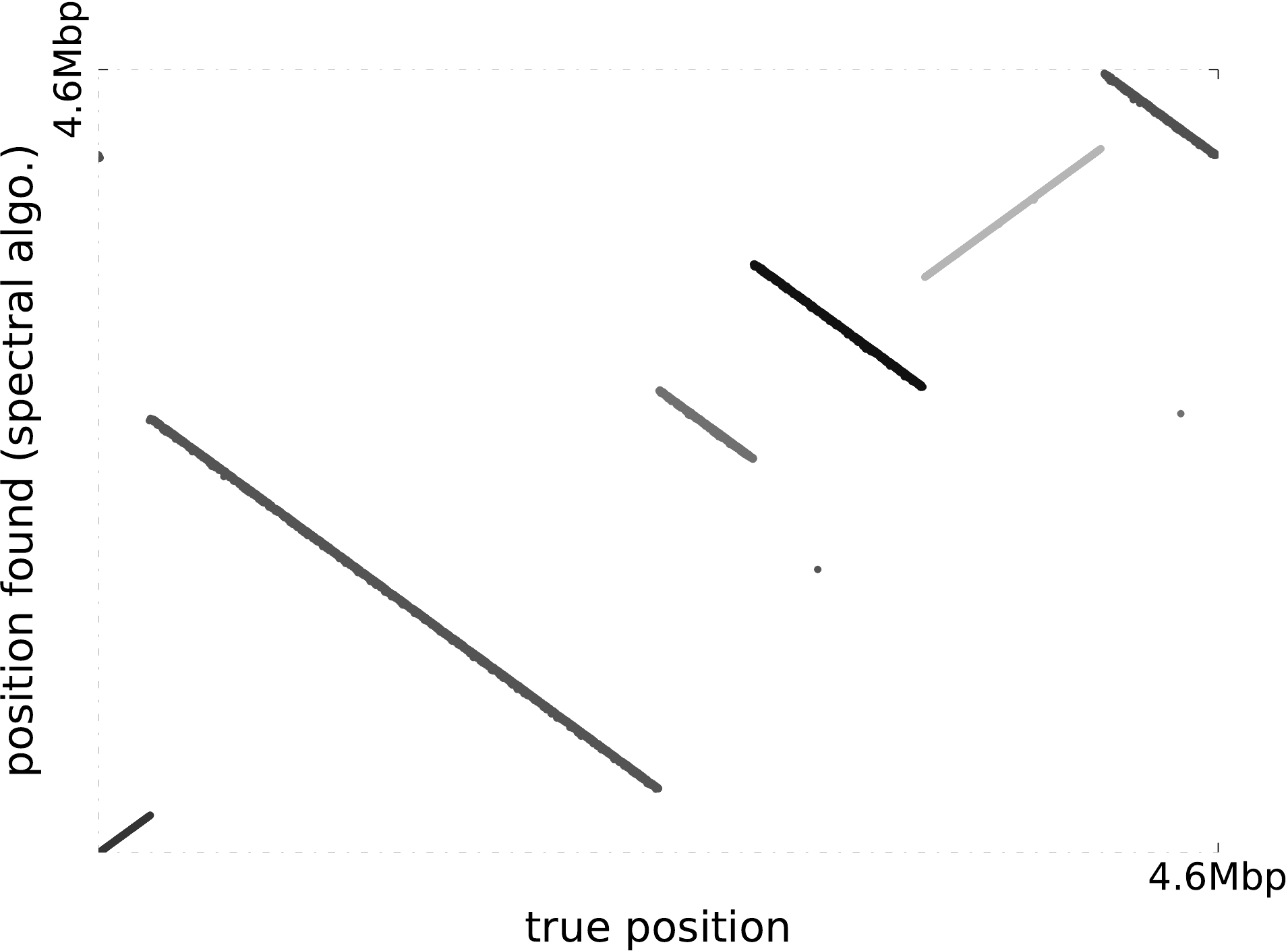}
    \caption{\textit{E. coli} PacBio}
  \end{subfigure}

  \begin{subfigure}[b]{0.25\textwidth}
    \centering \includegraphics[width=\textwidth]{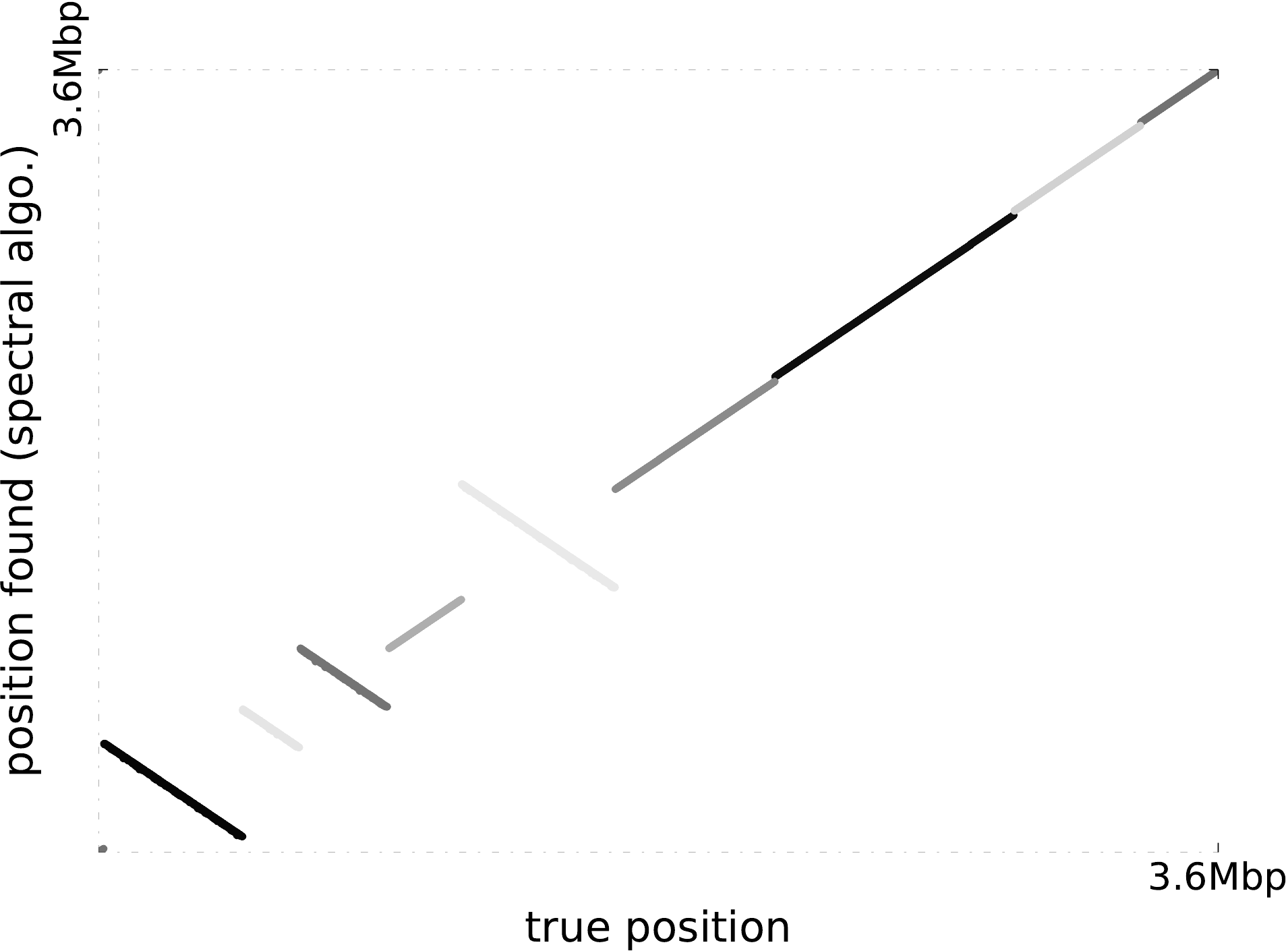}
    \caption{\textit{A. baylyi} ONT corr.}
  \end{subfigure}
  ~
  \begin{subfigure}[b]{0.25\textwidth}
    \centering \includegraphics[width=\textwidth]{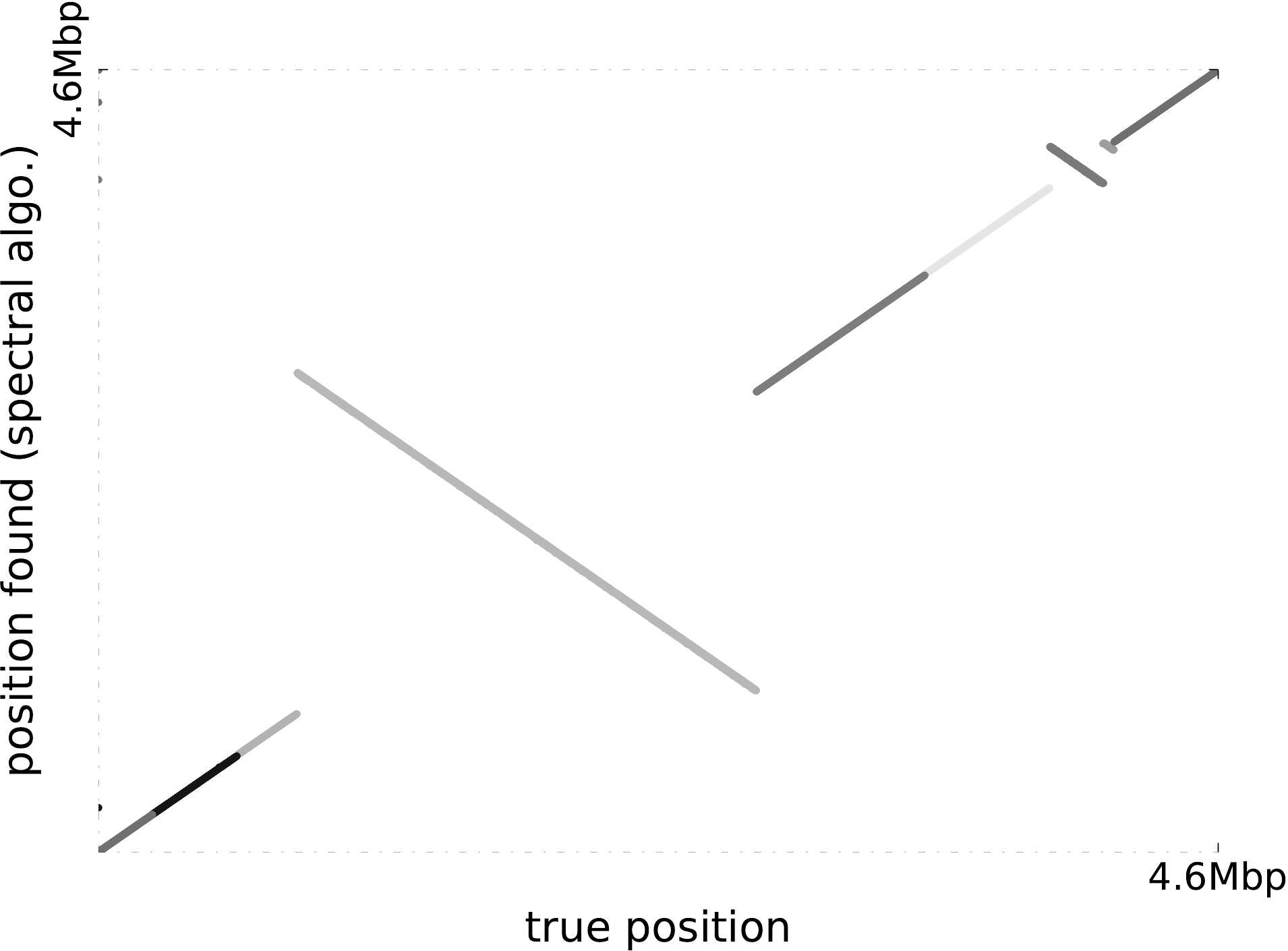}
    \caption{\textit{E. coli} ONT corr.}
  \end{subfigure}
  ~
  \begin{subfigure}[b]{0.25\textwidth}
    \centering \includegraphics[width=\textwidth]{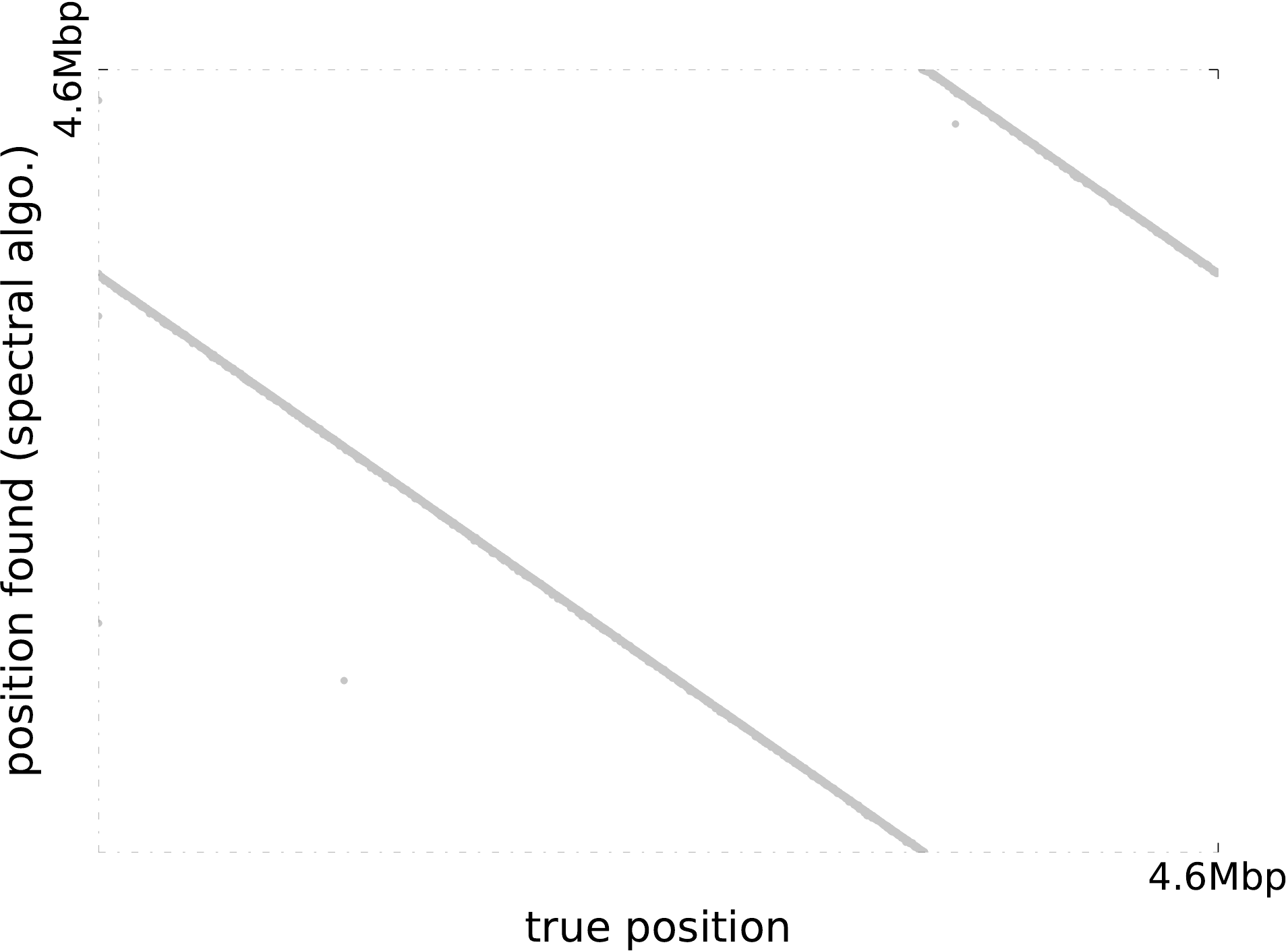}
    \caption{\textit{E. coli} PacBio corr.}
  \end{subfigure}
  \caption{
  Ordering of the reads computed with the spectral algorithm vs true ordering (obtained by mapping the reads to the reference genome with GraphMap) for the original (a-c) and corrected (d-f) bacterial datasets.
  All contigs are  artificially displayed on the same plot for compactness.
}\label{suppfig:bactlayouts}
\end{figure}

minimap was used to compute overlaps between raw reads
(we obtained similar results with MHAP and DALIGNER).
The similarity matrix preprocessed as detailed in Section\ref{subsec:outliers} yielded a few connected components for bacterial genomes.
The reads were successfully ordered in each of these, as one can see in Figure~\ref{fig:layoutBact} for \textit{E. coli}, and in Figure~\ref{suppfig:bactlayouts} for the other datasets.

\subsubsection{Eukaryotic genome}
\begin{figure}[tbh]
	\centering
		\includegraphics[width=0.45\textwidth]{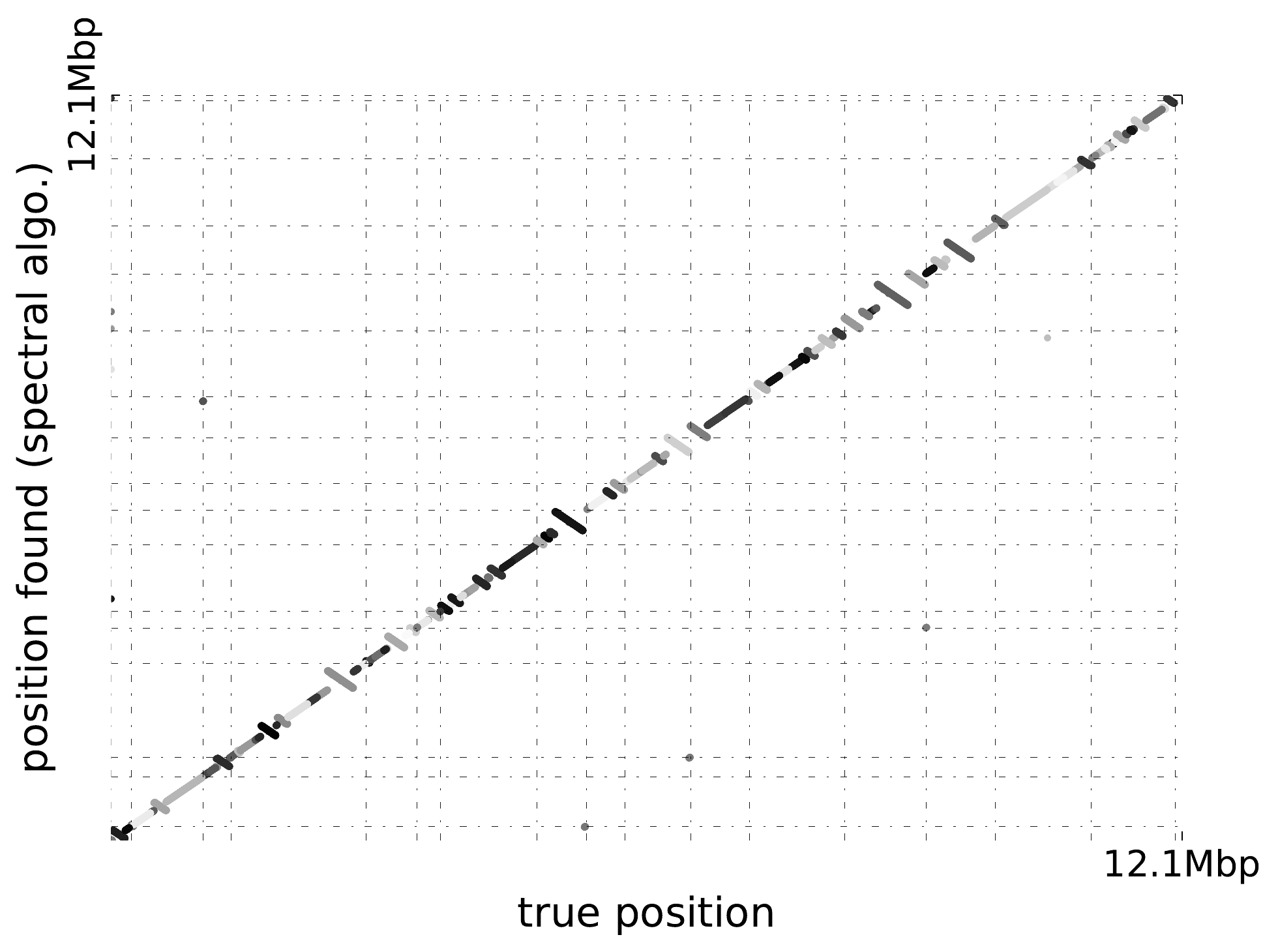}
	\caption{
	Ordering of the \textit{ Saccharomyces cerevisiae} ONT R7.3 reads identified with the spectral algorithm vs true ordering (obtained by mapping the reads to the reference genome with GraphMap and concatenating the ordering found in each chromosome).
  The different chromosomes are separated by grid lines.}
	\label{fig:layoutEuk}
\end{figure}

\begin{figure}[htb]
  \centering
  \begin{subfigure}[b]{0.28\textwidth}
    \centering \includegraphics[width=\textwidth]{figures/yeastR7_test0_layout.pdf}
    \caption{ONT R7.3}
  \end{subfigure}
  ~
  \begin{subfigure}[b]{0.28\textwidth}
    \centering \includegraphics[width=\textwidth]{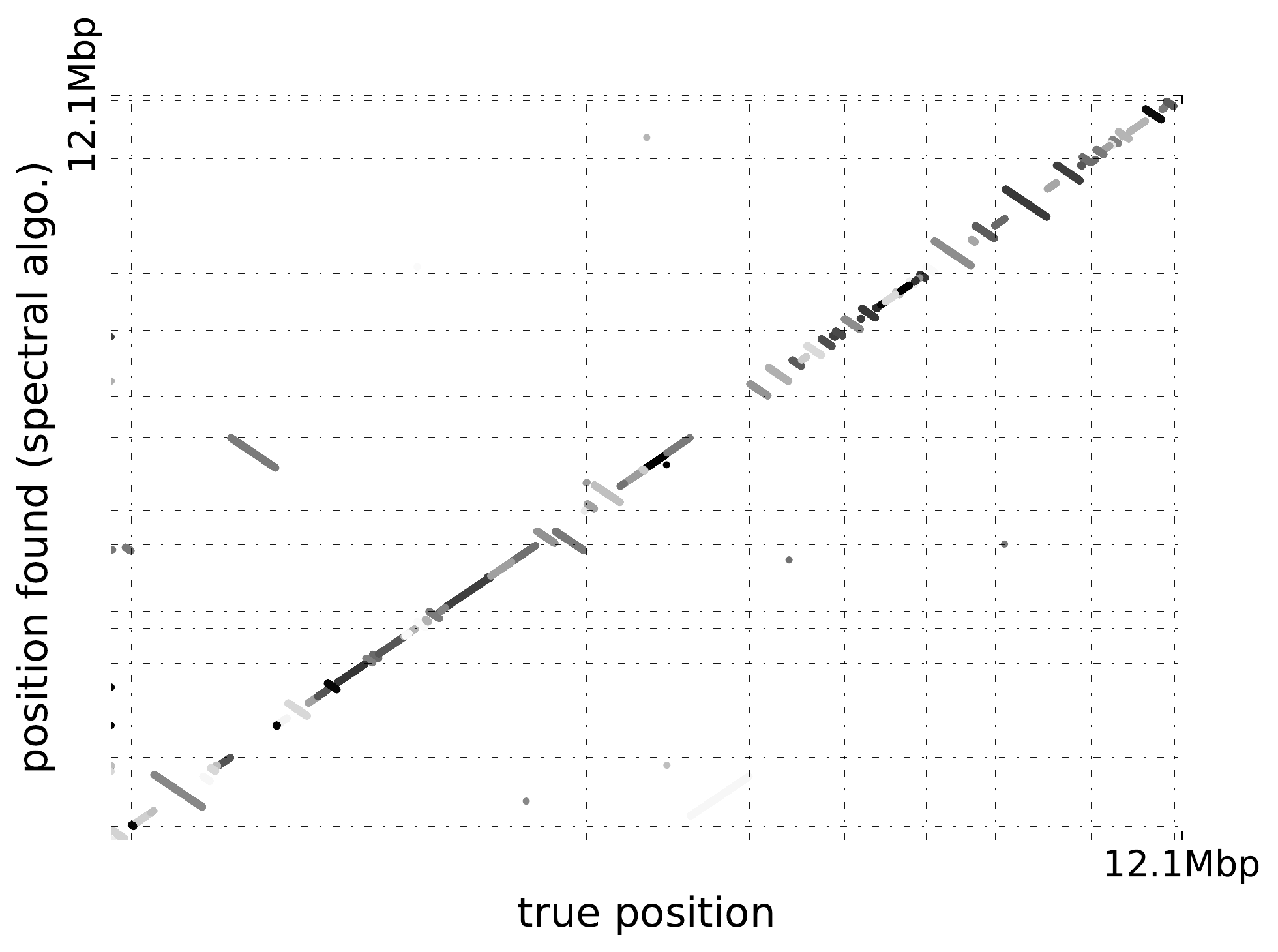}
    \caption{ONT R9}
  \end{subfigure}
  ~
  \begin{subfigure}[b]{0.28\textwidth}
    \centering \includegraphics[width=\textwidth]{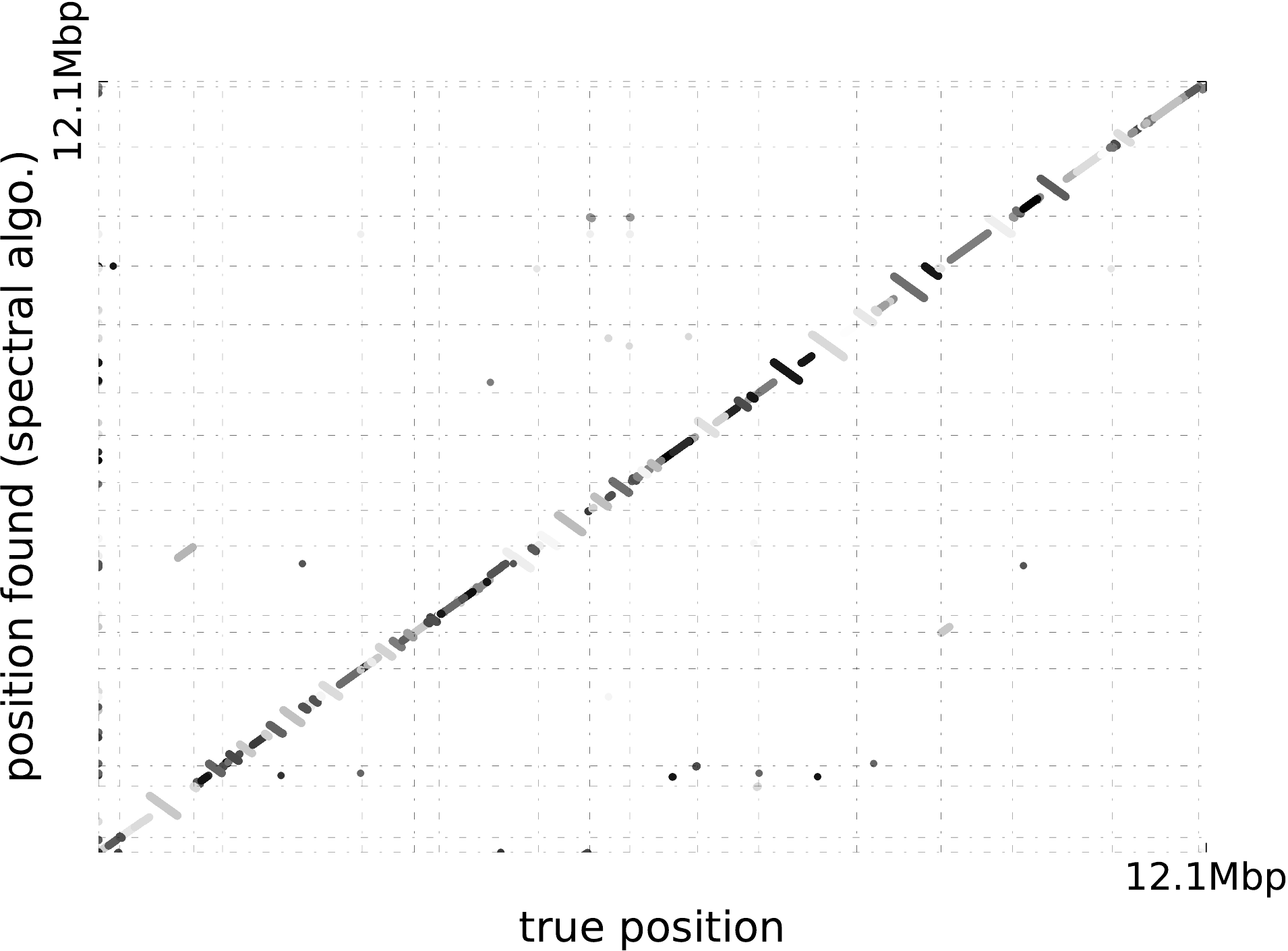}
    \caption{PacBio}
  \end{subfigure}

  \begin{subfigure}[b]{0.28\textwidth}
    \centering \includegraphics[width=\textwidth]{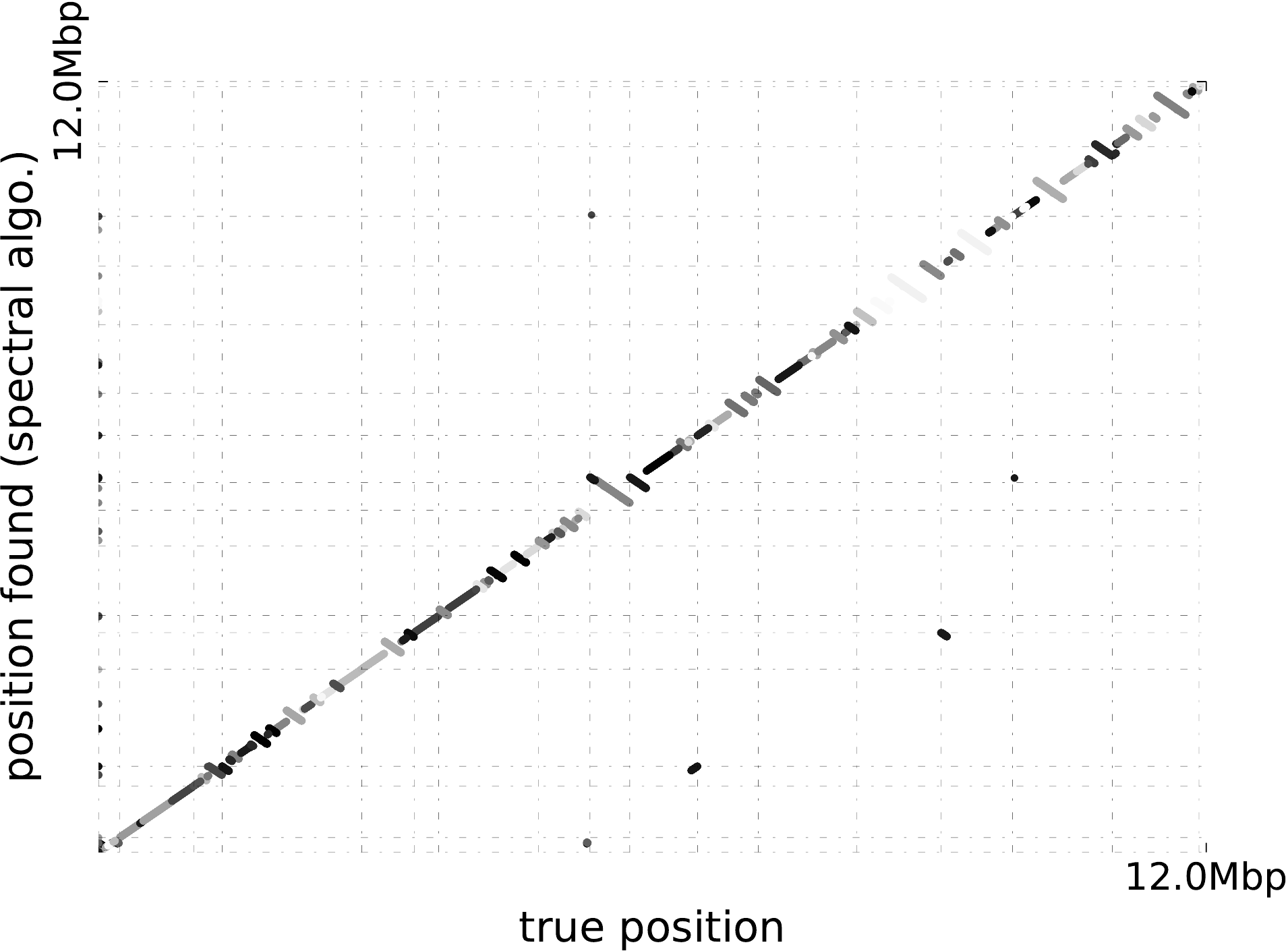}
    \caption{ONT R7.3 corr.}
  \end{subfigure}
  ~
  \begin{subfigure}[b]{0.28\textwidth}
    \centering \includegraphics[width=\textwidth]{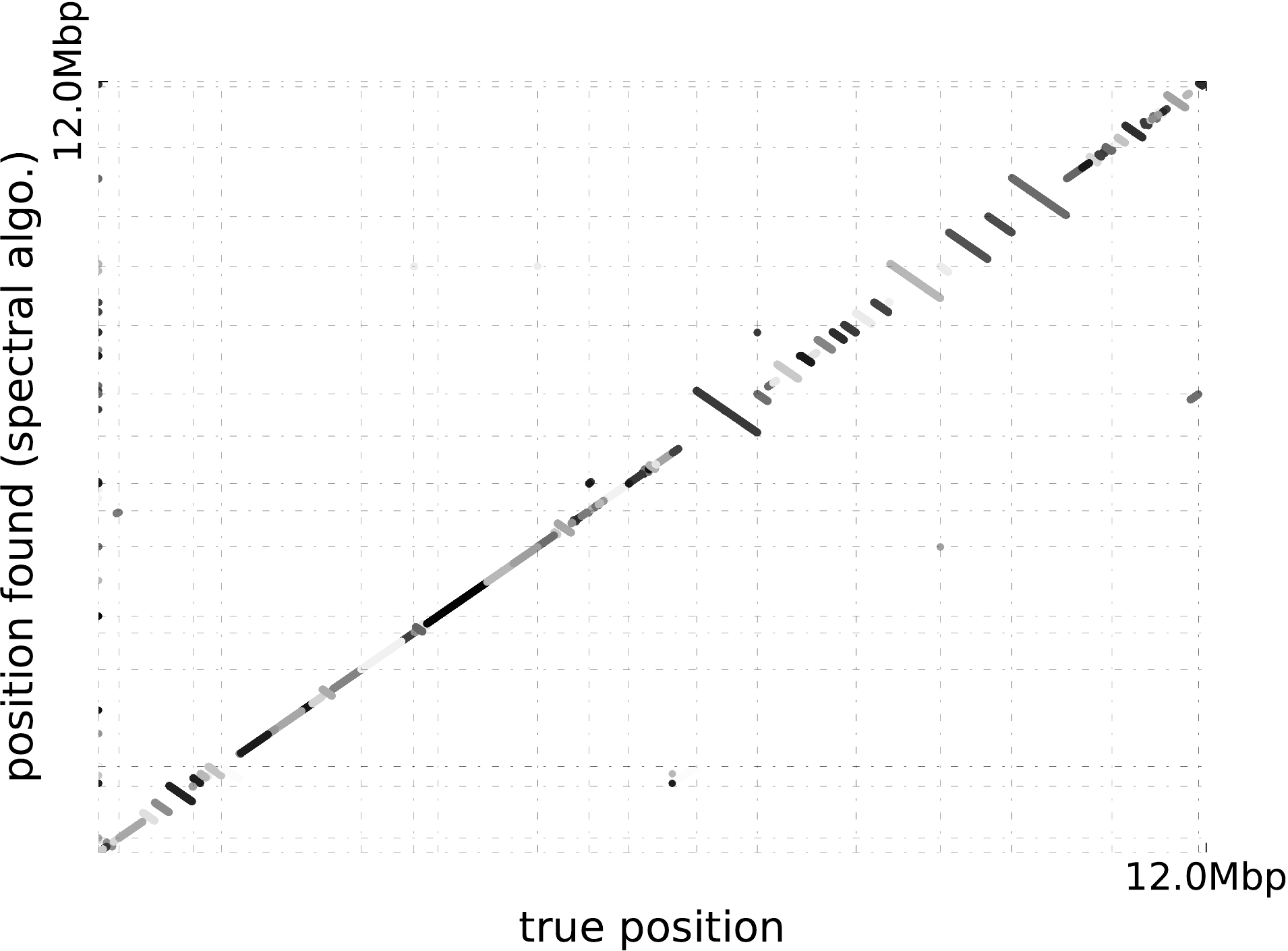}
    \caption{ONT R9 corr.}
  \end{subfigure}
  ~
  \begin{subfigure}[b]{0.28\textwidth}
    \centering \includegraphics[width=\textwidth]{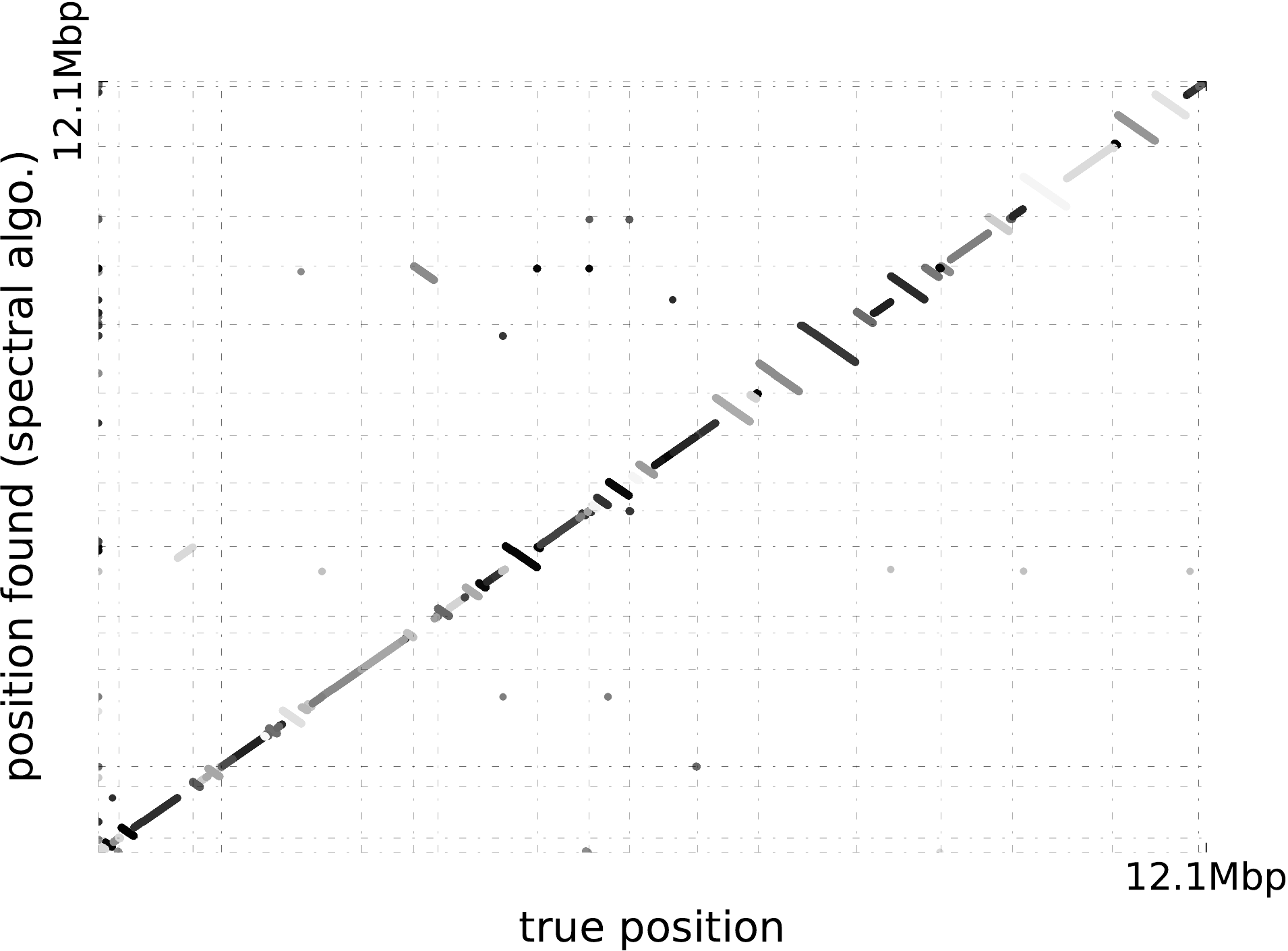}
    \caption{PacBio corr.}
  \end{subfigure}
  \caption{
   Ordering of the reads computed with the spectral algorithm vs true ordering (obtained by mapping the reads to the reference genome with GraphMap) for the original (a-c) and corrected (d-f) yeast (\textit{S. cerevisiae}) datasets.
   All contigs are  artificially displayed on the same plot for compactness.
   The dashed lines represent the boundaries between chromosomes.
   The correction slightly improves the layout for the yeast genomes.
  }\label{suppfig:yeastlayout}
\end{figure}

For the \textit{S. cerevisiae} genome, the threshold on similarity had to be set higher than for bacterial genomes because of a substantially higher number of repetitive regions and false overlaps, leading to a more fragmented assembly.
Most of them are correctly reordered with the spectral algorithm, see Figure~\ref{fig:layoutEuk} and Supplementary Figure~\ref{suppfig:yeastlayout}.

\subsection{Consensus}
\subsubsection{Recovering contiguity}\label{subsec:finisher}
Once the layout was established, the method described above was used to assemble the contigs and generate a consensus sequence.
For the two bacterial genomes, the first round of layout produced a small number of connected components, each of them yielding a contig.
Sufficient overlap was left between the contig sequences to find their layout with a second iteration of the algorithm and produce a single contig spanning the entire genome.
The number of contigs in the yeast assemblies can be reduced similarly.
The fact that the first-pass contigs overlap even though they result from breaking the similarity graph into several connected components might seem counter-intuitive at first sight.
However, note that when cutting an edge $A_{ij}$ results in the creation of two contigs (one containing $i$ and the other $j$), the sequence fragment at the origin of the overlap between the two reads is still there on both contigs to yield an overlap between them in the second iteration.
Alternatively, we found the following method useful to link the contigs' ends:
1. extract the ends of the contig sequences, 2. compute their overlap with minimap,
3. propagate the overlaps to the contig sequences,
4. use miniasm with all pre-selection parameters and thresholds off, to just concatenate the contigs (see Supplementary Material \S\ref{subsec:Implementation}).

\subsubsection{Consensus quality evaluation}
\begin{figure}[tb]
	\centering
		\includegraphics[width=0.46\textwidth]{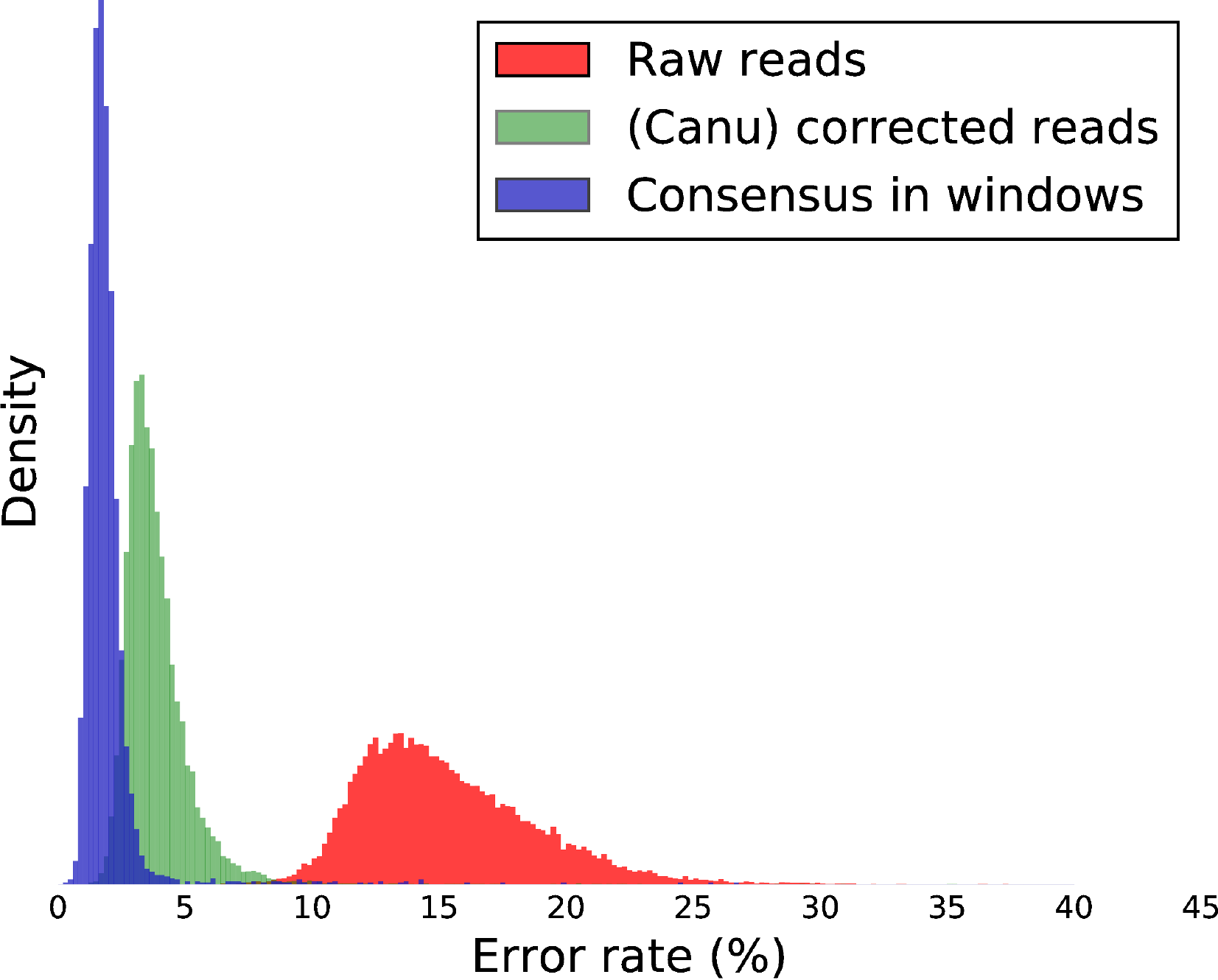}
	\caption{
  Error rate of consensus window sequences, compared to the raw and corrected (with the Canu correction and trimming modules) reads for the \textit{A. baylyi} ONT dataset.
  The error rates were computed by mapping the sequences to the \textit{A. baylyi} reference genome.
  Histograms for the other datasets are available in Supplementary Figure~\ref{fig:errorHistoApp}.}
	\label{fig:qual}
\end{figure}

\begin{figure}[htb]
  \centering
  \begin{subfigure}[b]{0.27\textwidth}
    \centering \includegraphics[width=\textwidth]{figures/a_baylyi_test0_err_hist3.pdf}
    \caption{\textit{A. baylyi} ONT}
  \end{subfigure}
  ~
  \begin{subfigure}[b]{0.27\textwidth}
    \centering \includegraphics[width=\textwidth]{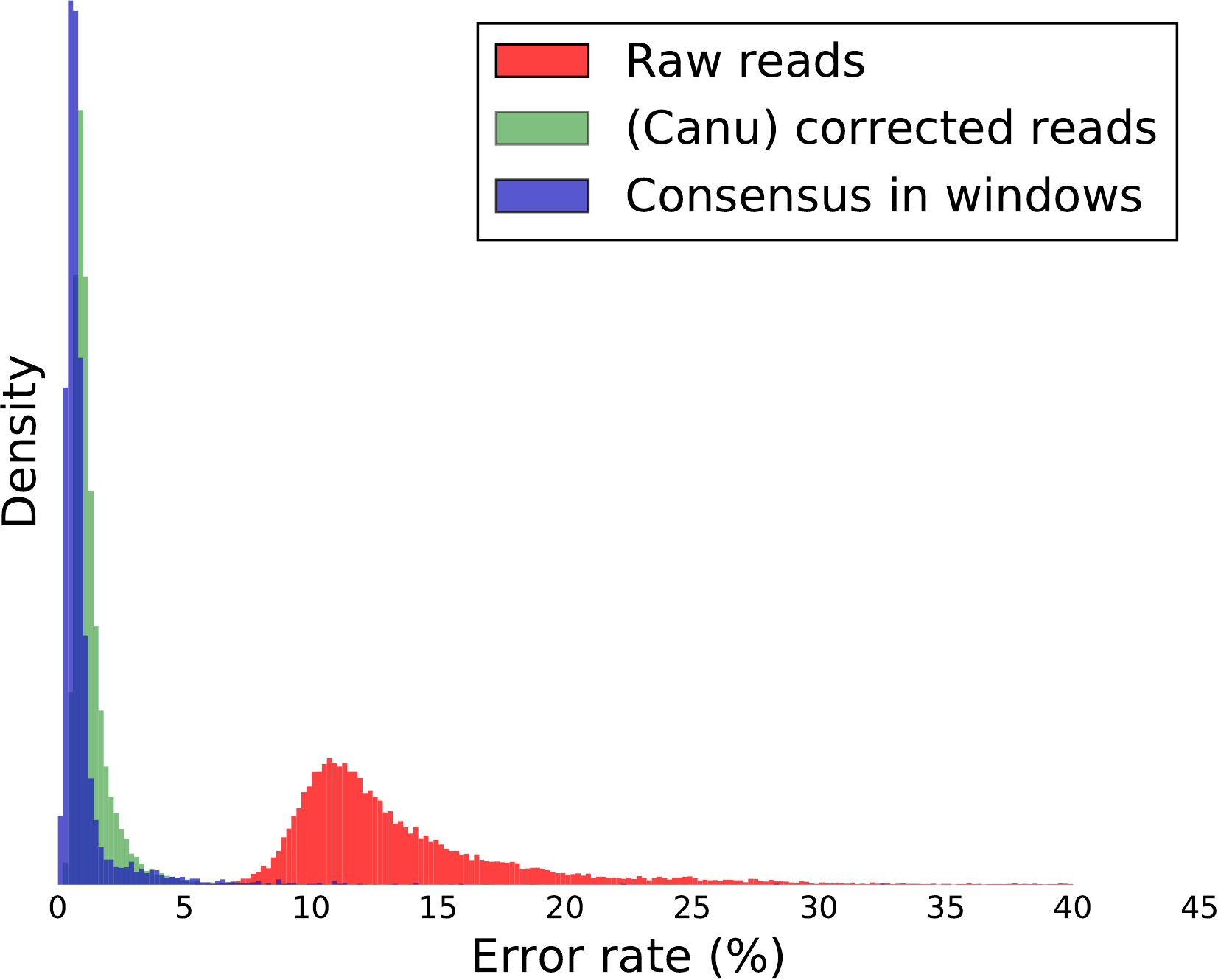}
    \caption{\textit{E. coli} ONT}
  \end{subfigure}
  ~
  \begin{subfigure}[b]{0.27\textwidth}
    \centering \includegraphics[width=\textwidth]{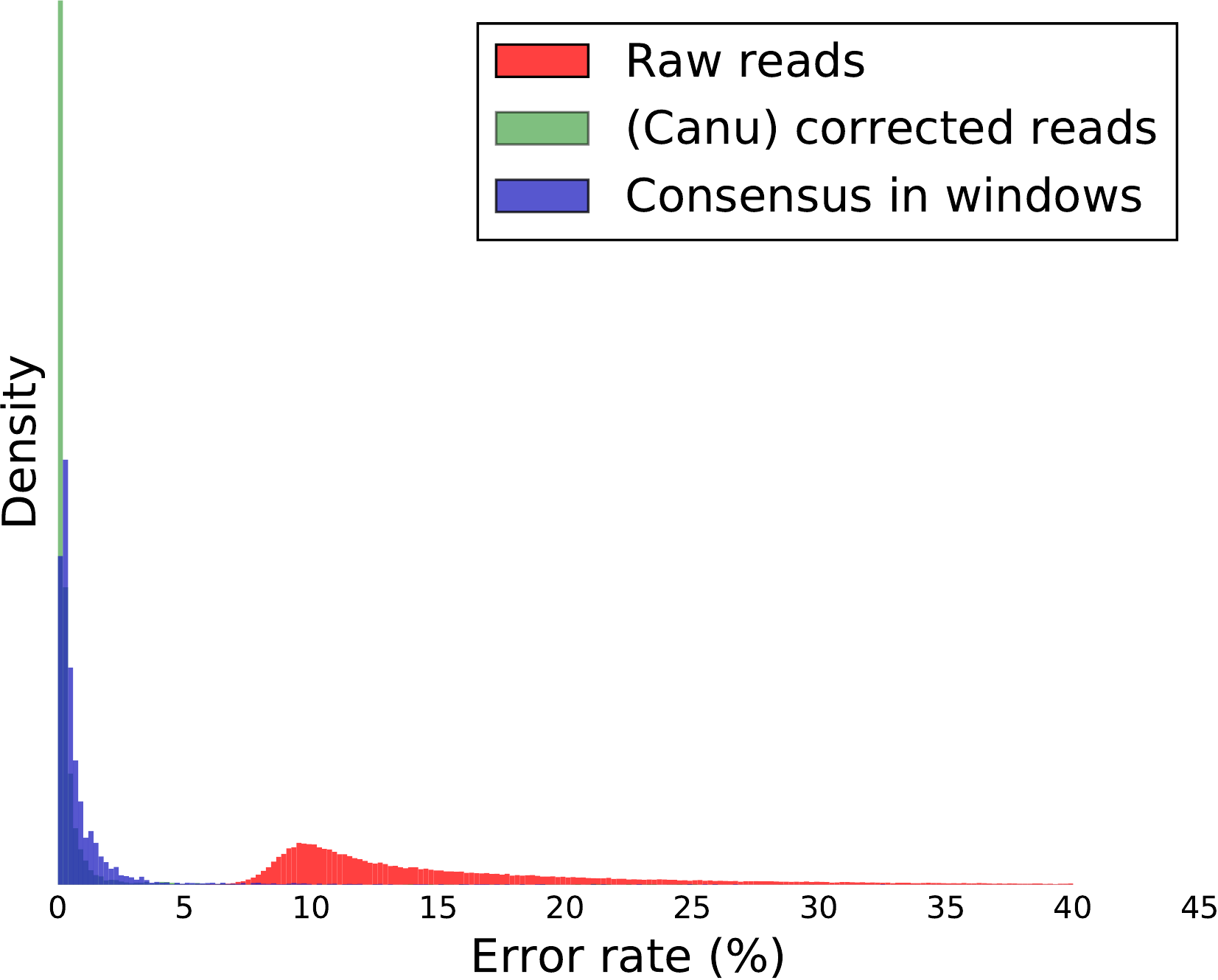}
    \caption{\textit{E. coli} PacBio}
  \end{subfigure}

  \begin{subfigure}[b]{0.27\textwidth}
    \centering \includegraphics[width=\textwidth]{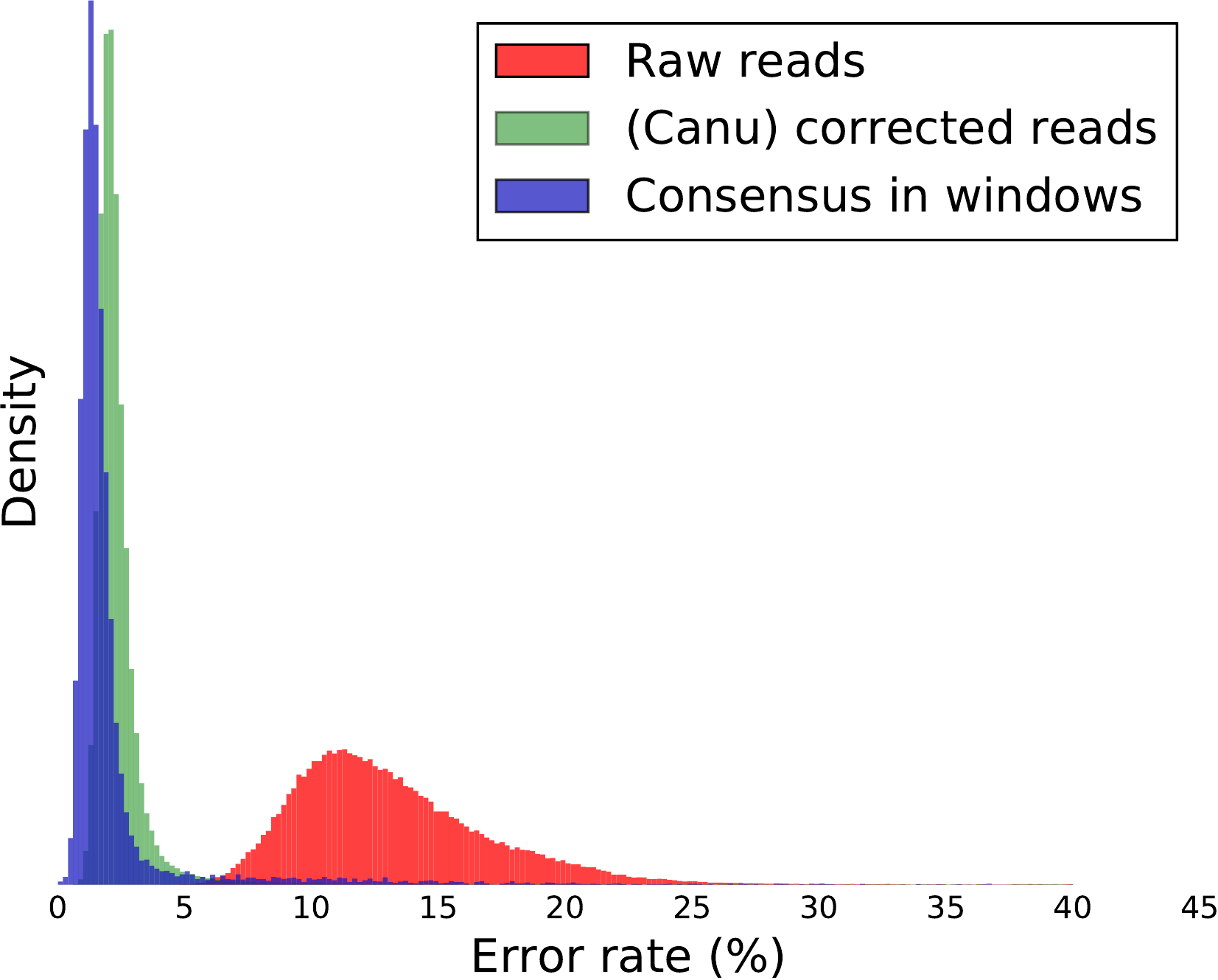}
    \caption{\textit{S. cerevisiae} ONT R7.3}
  \end{subfigure}
  ~
  \begin{subfigure}[b]{0.27\textwidth}
    \centering \includegraphics[width=\textwidth]{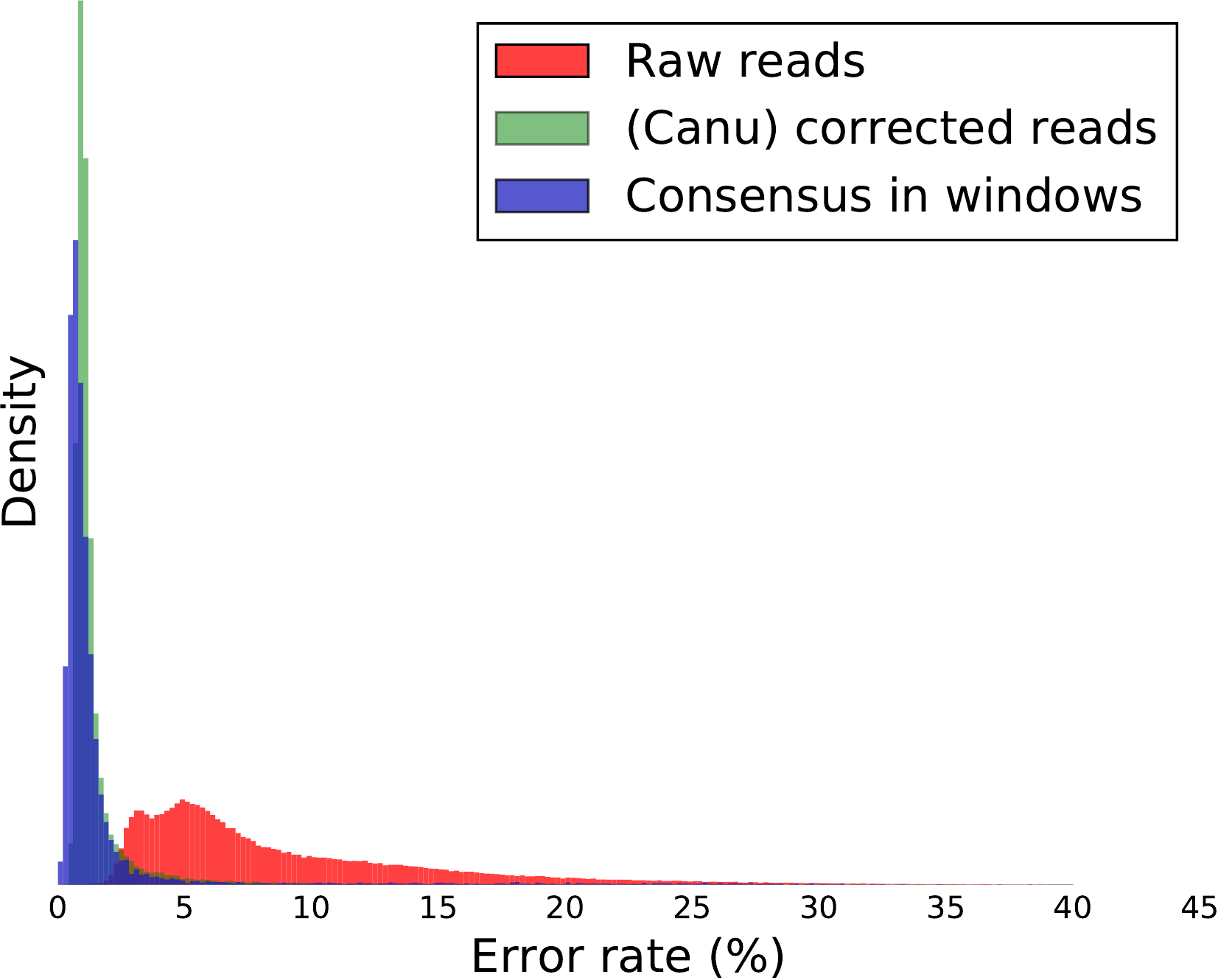}
    \caption{\textit{S. cerevisiae} ONT R9}
  \end{subfigure}
  ~
  \begin{subfigure}[b]{0.27\textwidth}
    \centering \includegraphics[width=\textwidth]{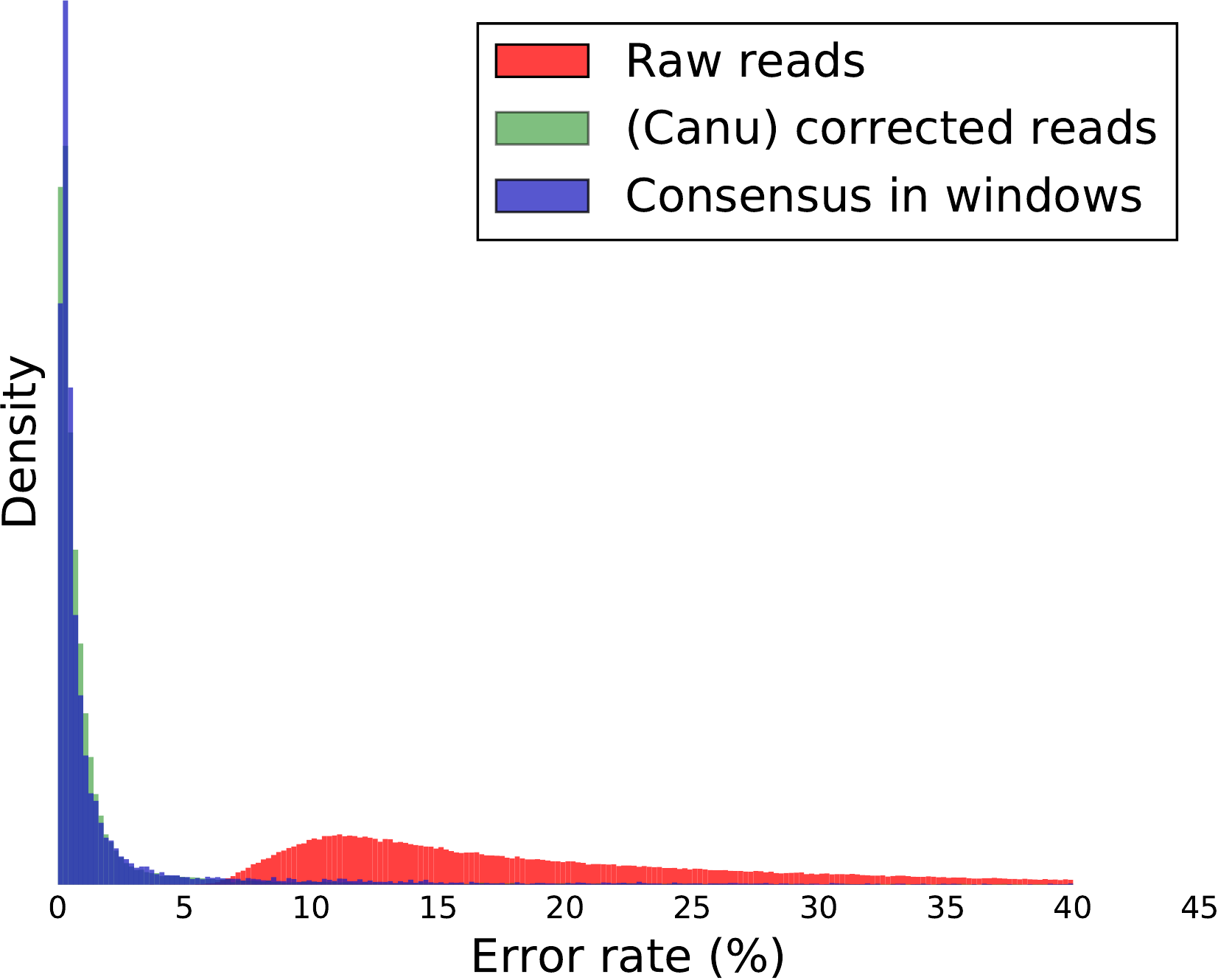}
    \caption{\textit{S. cerevisiae} PacBio}
  \end{subfigure}
  \caption{
    Error-rates in consensus windows, raw reads and corrected reads for the six real datasets.
    With ONT R7.3 data, the consensus produced by our pipeline appears more accurate than via the correction module of Canu,
    while the contrary is true for PacBio data.
}\label{fig:errorHistoApp}
\end{figure}

\begin{figure}[htb]
  \centering
  \begin{subfigure}[b]{0.3\textwidth}
    \centering \includegraphics[width=\textwidth]{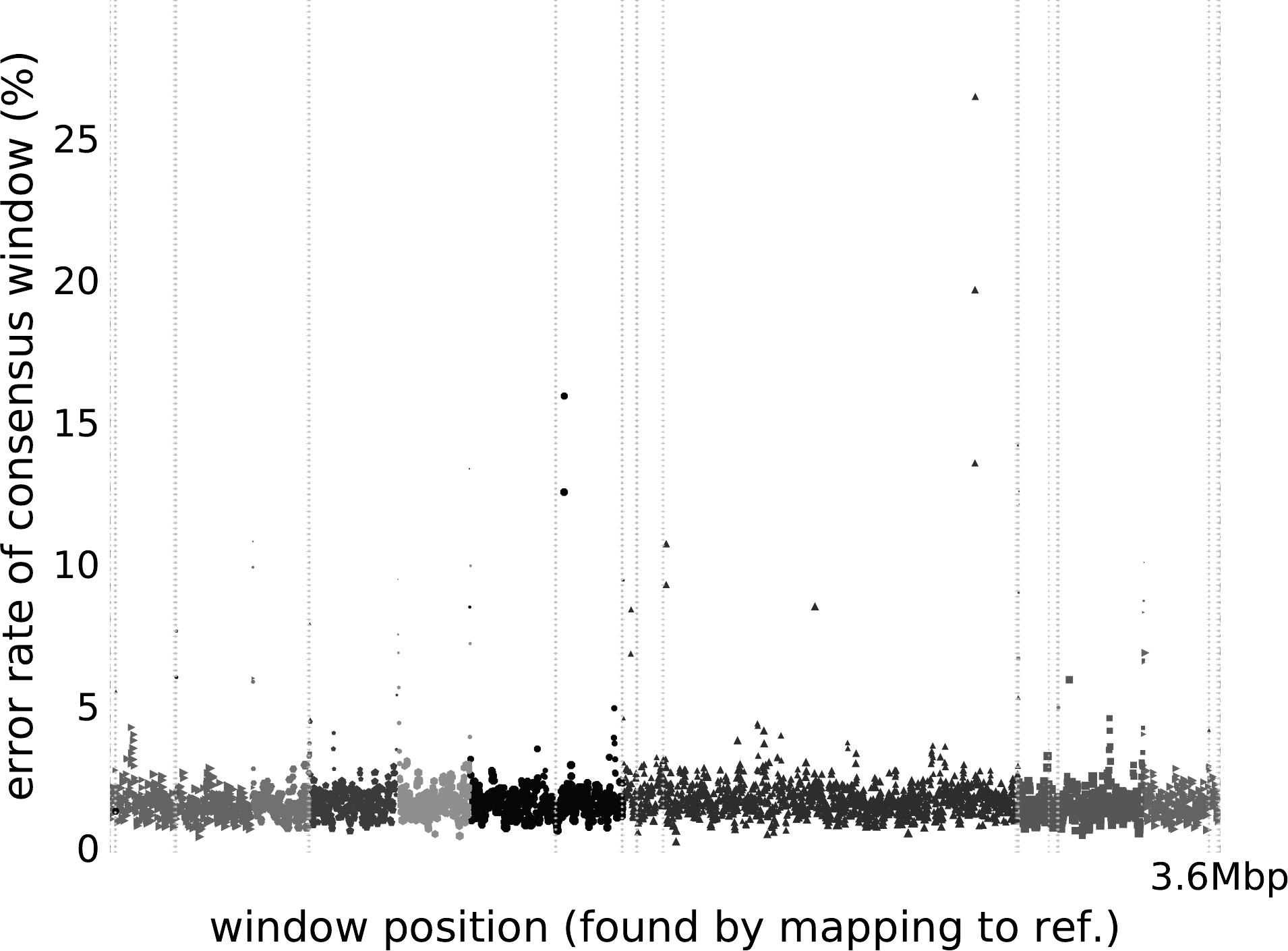}
    \caption{\textit{A. baylyi} ONT}
  \end{subfigure}
  ~
  \begin{subfigure}[b]{0.3\textwidth}
    \centering \includegraphics[width=\textwidth]{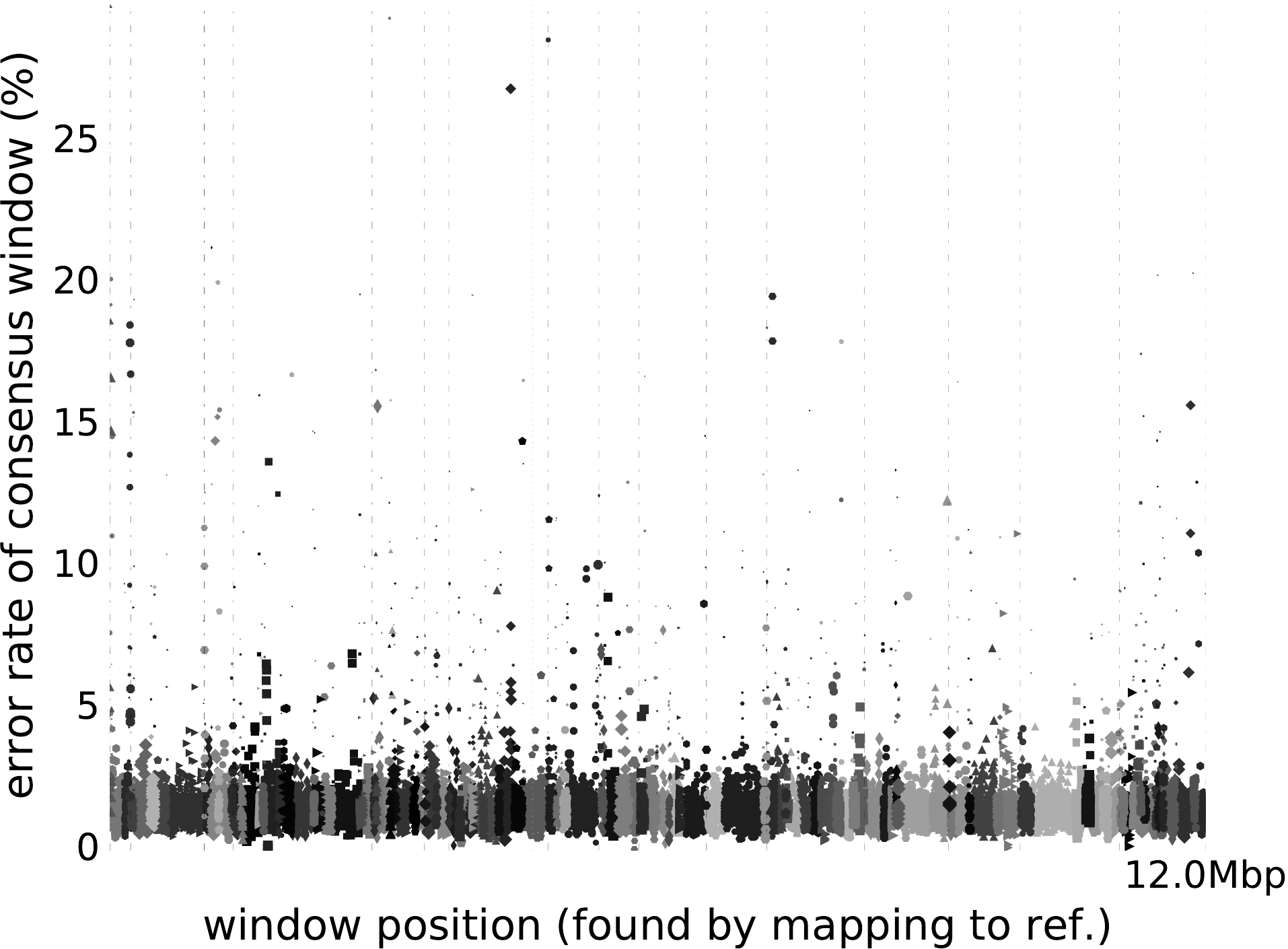}
    \caption{\textit{S. cerevisiae} ONT R7.3}
  \end{subfigure}

  \begin{subfigure}[b]{0.3\textwidth}
    \centering \includegraphics[width=\textwidth]{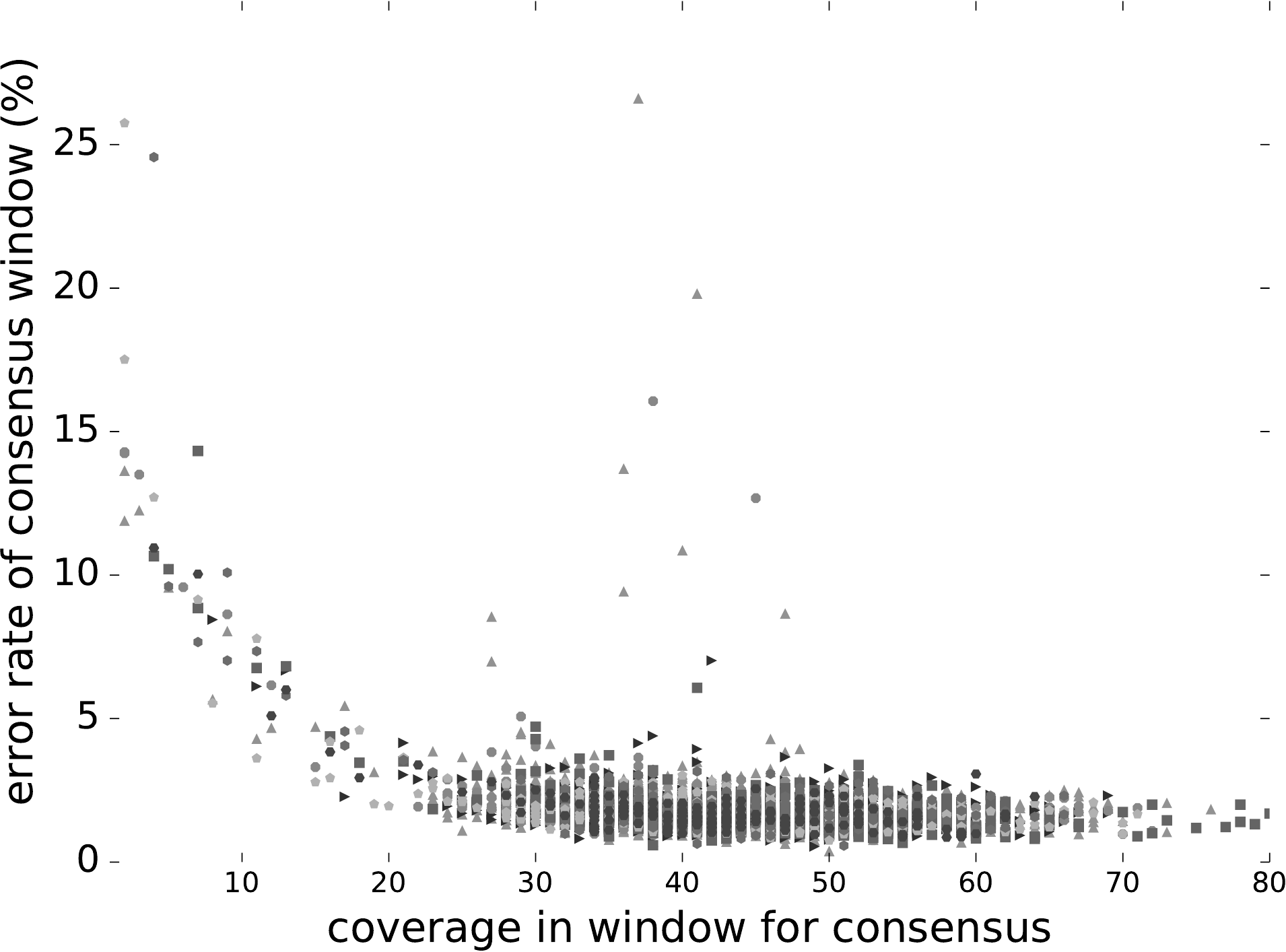}
    \caption{\textit{A. baylyi} ONT}
  \end{subfigure}
  ~
  \begin{subfigure}[b]{0.3\textwidth}
    \centering \includegraphics[width=\textwidth]{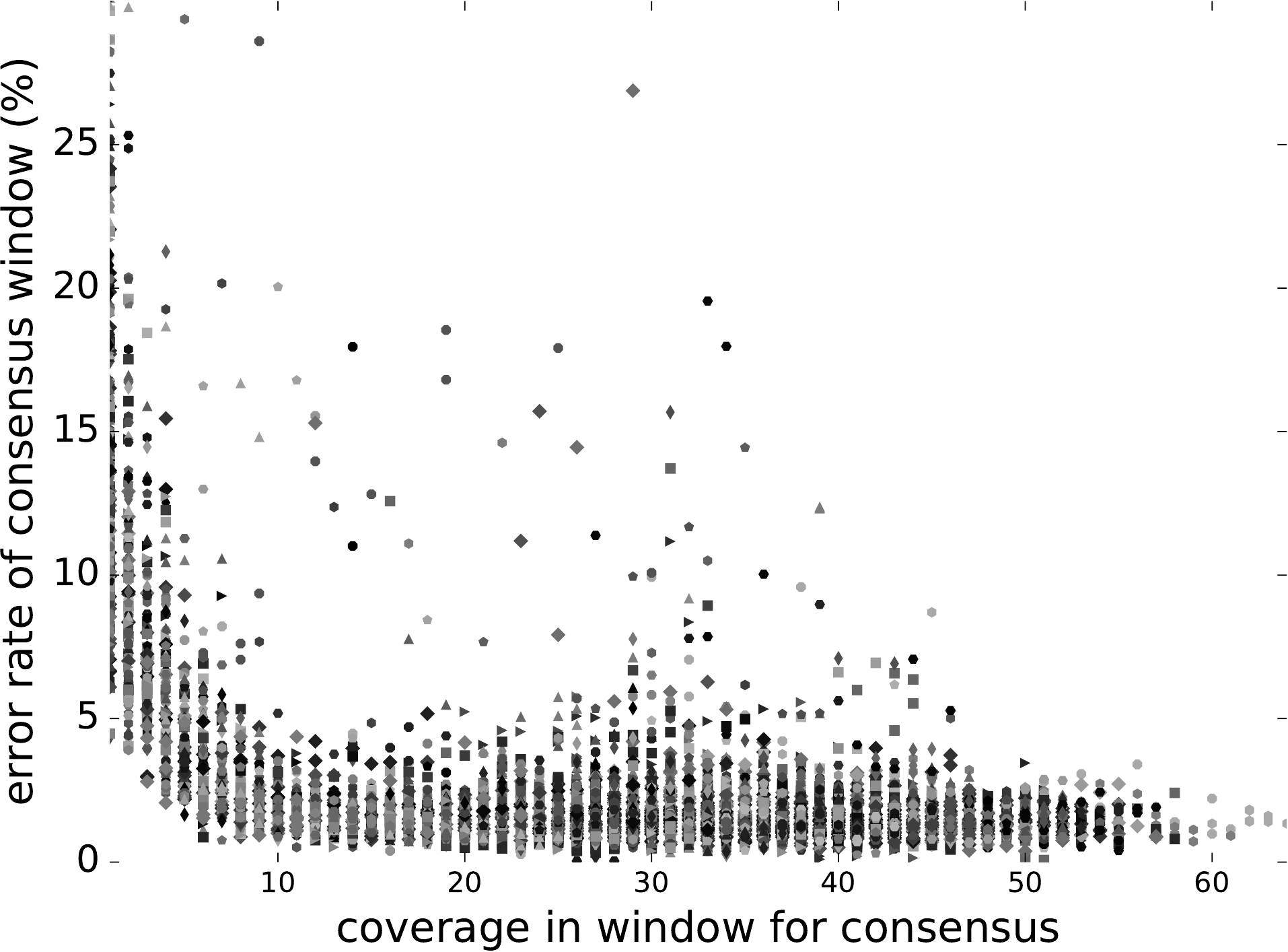}
    \caption{\textit{S. cerevisiae} ONT R7.3}
  \end{subfigure}
  \caption{
  Error-rates in consensus windows \textit{versus} position of the windows on the reference genome (a,b).
  The dashed lines represent the location of repeats for \textit{A. baylyi}, and the separation between chromosomes for \textit{S. cerevisiae}.
  The size of each scatter marker is proportional to the coverage of the window.
  The (c,d) panel represents the error-rates in consensus windows \textit{versus} the coverage of the windows.
  The error-rate was computed with the errorrates.py script from samtools, using the mapping obtained from GraphMap.
  Most of the windows with a high error rate are positioned at the ends of the contigs they belong to.
  We also observed that repeats are often positioned at the edge between two contigs, though this does not seem to be the determinant factor.
  The bottom plots represent the error-rate in the windows against their estimated coverage,
  defined as the total length of sequences used to perform the multiple alignment in the window normalized by the length of the consensus sequence.
  Overall, one can see that the windows with high error rate are the ones with low coverage.
  Nevertheless, especially for the yeast genomes, there are also several
  windows with high values for both error-rate and coverage.
  Manual inspection of these reveals that they usually do not span repeated regions, but their high error-rates arise from imperfections in the layout.
}\label{fig:accuracyWindows}
\end{figure}

We first investigated the quality of the consensus sequences derived in each window. Figures~\ref{fig:qual} and \ref{fig:errorHistoApp} highlight the correcting effect of the consensus.
Supplementary Figure~\ref{fig:accuracyWindows} suggests that the error-rate in the consensus windows depends mainly on the local coverage.
We then compared our results to those obtained with other long reads assemblers :  Miniasm, Canu and Racon \citep{racon2016}.
Racon takes a draft assembly, the raw reads, and a mapping of the reads to the draft assembly as input. We used it with the draft assembly produced by Miniasm (as done by \citet{racon2016}). We label this method ``Miniasm+Racon'' in our results.
We also used Racon with the draft assembly derived by our method (``Spectral+Racon'' method), using Minimap to map the raw reads to the draft assemblies before using Racon.
A summary of assembly reports generated with DNAdiff \citep{MUMmer} and QUAST \citep{GurevichQUAST} are given in Table~\ref{tab:allAssembly} and Supplementary Table~\ref{tab:rawMisassemblies}.
Briefly, the assemblies displayed between $98\%$ and $99\%$ average identity to their reference genome, with errors mostly consisting in deletions.
Misassemblies were rare in reconstructed bacterial genomes but more frequent in assembled yeast genomes, where they mostly consisted in translocations and relocations caused by either deletions and/or misplaced reads in the layout.

\begin{table*}[tb]
\centering
\processtable{Assembly results of the spectral method, compared to Miniasm, Canu and Racon, across the different datasets}
{\footnotesize\label{tab:allAssembly}
\begin{tabular}{cc K{1.91cm}K{1.91cm}K{1.91cm}K{1.91cm}K{1.91cm}K{1.91cm}}
  \toprule
   & & Miniasm & Spectral & Canu & Miniasm+Racon & Miniasm+Racon (2 iter.) & Spectral+Racon\\
  \hline
  \multirow{8}*{\parbox{1.cm}{\textit{A. baylyi} ONT R7.3 28x}}& Ref. genome size [bp] & 3598621 & 3598621 & 3598621 & 3598621 & 3598621 & 3598621\\
  & Total bases [bp] & 3531295 & 3551582 & 3513432 & 3564823 & 3566438 & 3551094\\
  & Ref. chromosomes [\#] & 1 & 1 & 1 & 1 & 1 & 1\\
  & Contigs [\#] & 5 & 1 (7) & 1 & 5 & 5 & 1 (7)\\
  & Aln. bases ref [bp] & 3445457(95.74\%) & 3596249(99.93\%) & 3595082(99.90\%) & 3596858(99.95\%) & 3596854(99.95\%) & 3598181(99.99\%)\\
  & Aln. bases query [bp] & 3379002(95.69\%) & 3549290(99.94\%) & 3513081(99.99\%) & 3564455(99.99\%) & 3566021(99.99\%) & 3550742(99.99\%)\\
  & Misassemblies [\#] & 0 & 0 & 2 & 2 & 2 & 0\\
  & Avg. identity & 87.31 & 98.17 & 97.59 & 98.18 & 98.36 & \textbf{98.42}\\\hline
  \multirow{8}*{\parbox{1.cm}{\textit{E. coli} ONT R7.3 30x}}& Ref. genome size [bp] & 4641652 & 4641652 & 4641652 & 4641652 & 4641652 & 4641652\\
  & Total bases [bp] & 4759346 & 4662043 & 4625543 & 4647066 & 4643235 & 4629112\\
  & Ref. chromosomes [\#] & 1 & 1 & 1 & 1 & 1 & 1\\
  & Contigs [\#] & 3 & 1 (4) & 2 & 3 & 3 & 1 (4)\\
  & Aln. bases ref [bp] & 4355121(93.83\%) & 4612515(99.37\%) & 4638255(99.93\%) & 4640127(99.97\%) & 4640127(99.97\%) & 4641457(100.00\%)\\
  & Aln. bases query [bp] & 4432658(93.14\%) & 4623823(99.18\%) & 4625535(100.00\%) & 4642837(99.91\%) & 4639816(99.93\%) & 4628962(100.00\%)\\
  & Misassemblies [\#] & 0 & 2 & 8 & 3 & 3 & 2\\
  & Avg. identity & 89.28 & 98.80 & 99.40 & 99.31 & 99.45 & \textbf{99.46}\\\hline
  \multirow{8}*{\parbox{1.cm}{\textit{S. cerevisiae} ONT R7.3 68x}}& Ref. genome size [bp] & 12157105 & 12157105 & 12157105 & 12157105 & 12157105 & 12157105\\
  & Total bases [bp] & 11813544 & 12213218 & 12142953 & 11926664 & 11926191 & 12167363\\
  & Ref. chromosomes [\#] & 17 & 17 & 17 & 17 & 17 & 17\\
  & Contigs [\#] & 29 & 71 (127) & 36 & 29 & 29 & 71 (127)\\
  & Aln. bases ref [bp] & 11566318(95.14\%) & 12043050(99.06\%) & 12086977(99.42\%) & 12084923(99.41\%) & 12086556(99.42\%) & 12061384(99.21\%)\\
  & Aln. bases query [bp] & 11236806(95.12\%) & 12134480(99.36\%) & 12089056(99.56\%) & 11923058(99.97\%) & 11918621(99.94\%) & 12135284(99.74\%)\\
  & Misassemblies [\#] & 0 & 7 & 34 & 18 & 19 & 11\\
  & Avg. identity & 89.00 & 98.00 & 98.33 & 98.49 & \textbf{98.63} & 98.61\\\hline
  \multirow{8}*{\parbox{1.cm}{\textit{S. cerevisiae} ONT R9 86x}}& Ref. genome size [bp] & 12157105 & 12157105 & 12157105 & 12157105 & 12157105 & 12157105\\
  & Total bases [bp] & 11734150 & 11795644 & 12217497 & 12128279 & 12129086 & 11750114\\
  & Ref. chromosomes [\#] & 17 & 17 & 17 & 17 & 17 & 17\\
  & Contigs [\#] & 30 & 48 (85) & 26 & 30 & 29 & 48 (85)\\
  & Aln. bases ref [bp] & 11947453(98.28\%) & 11607131(95.48\%) & 12126980(99.75\%) & 12126663(99.75\%) & 12127467(99.76\%) & 11695983(96.21\%)\\
  & Aln. bases query [bp] & 11549494(98.43\%) & 11668882(98.93\%) & 12179843(99.69\%) & 12118506(99.92\%) & 12121202(99.93\%) & 11717047(99.72\%)\\
  & Misassemblies [\#] & 0 & 23 & 39 & 18 & 19 & 36\\
  & Avg. identity & 93.55 & 98.81 & 99.02 & 99.16 & \textbf{99.20} & 99.10\\\hline
  \multirow{8}*{\parbox{1.cm}{\textit{E. coli} PacBio 161x}}& Ref. genome size [bp] & 4641652 & 4641652 & 4641652 & 4641652 & 4641652 & 4641652\\
  & Total bases [bp] & 4845211 & 4731239 & 4670125 & 4653228 & 4645420 & 4674460\\
  & Ref. chromosomes [\#] & 1 & 1 & 1 & 1 & 1 & 1\\
  & Contigs [\#] & 1 & 2 (6)& 1 & 1 & 1 & 2 (6)\\
  & Aln. bases ref [bp] & 4437473(95.60\%) & 4617713(99.48\%) & 4641652(100.00\%) & 4641551(100.00\%) & 4641500(100.00\%) & 4641652(100.00\%)\\
  & Aln. bases query [bp] & 4601587(94.97\%) & 4705704(99.46\%) & 4670125(100.00\%) & 4653140(100.00\%) & 4645420(100.00\%) & 4673065(99.97\%)\\
  & Misassemblies [\#] & 0 & 5 & 4 & 4 & 4 & 4\\
  & Avg. identity & 89.13 & 98.63 & \textbf{99.99} & 99.64 & 99.91 & 99.87\\\hline
  \multirow{8}*{\parbox{1.cm}{\textit{S. cerevisiae} PacBio 127x}}& Ref. genome size [bp] & 12157105 & 12157105 & 12157105 & 12157105 & 12157105 & 12157105\\
  & Total bases [bp] & 12266420 & 12839034 & 12346258 & 12070971 & 12052148 & 12695031\\
  & Ref. chromosomes [\#] & 17 & 17 & 17 & 17 & 17 & 17\\
  & Contigs [\#] & 30 & 90 (136) & 29 & 30 & 30 & 90 (136)\\
  & Aln. bases ref [bp] & 11250453(92.54\%) & 11917823(98.03\%) & 12091868(99.46\%) & 12023040(98.90\%) & 12024968(98.91\%) & 12002816(98.73\%)\\
  & Aln. bases query [bp] & 11396172(92.91\%) & 12456415(97.02\%) & 12304982(99.67\%) & 12045088(99.79\%) & 12027812(99.80\%) & 12485128(98.35\%)\\
  & Misassemblies [\#] & 0 & 57 & 76 & 61 & 59 & 68\\
  & Avg. identity & 88.29 & 98.41 & \textbf{99.87} & 99.43 & 99.72 & 99.54\\
  \botrule
\end{tabular}
}{
For the spectral method, we give the results after contig merging (see \S\ref{subsec:finisher}); the number of contigs before this post-processing is given between parentheses.
Racon's use here can be seen as a polishing phase for the sequences outputted by the spectral method and Miniasm.
To keep both assemblers on an equal footing, we compared Spectral+Racon to two iterations of Miniasm+Racon (since one pass of Miniasm does not implement any consensus).
The best results in terms of average identity are highlighted in bold (but other metrics should also be used to compare the assemblies).
Canu clearly outperforms the spectral method on PacBio data, while both assemblers yield comparable results on the ONT datasets.

}
\end{table*}

\subsubsection{Optical mapping}
After the first iteration of the bacterial genome assembly pipeline, overlaps between the first-pass contigs were sufficient to find their layout.
It should be anticipated however that not all overlaps might be apparent in some cases, e.g. if too many reads were removed during the preprocessing step. One attractive option is to use optical mapping \citep{RefOM} to layout the contigs. We had such an optical map available for the \textit{A. baylyi} genome, and implemented the algorithm of \citet{Nagarajan08} to map the contigs to the restriction map, which led to the same layout as the one identified from our two-round assemblies (data not shown), thus providing a ``consistency check'' for the layout.
We suggest in Supplementary Figure~\ref{fig:opticalMap} and Table~\ref{tbl:EukSepChr} that optical maps could be particularly valuable for the ordering of contigs from more structurally complex eukaryotic genomes such as \textit{S. cerevisiae}.

\section{Discussion}\label{sec:discuss}
We have shown that seriation based layout algorithms can be successfully applied to \textit{de novo} genome assembly problems, at least for genomes harboring a limited number of repeats.

In a similar vein to the recent report about the miniasm assembly engine \citep{Li:Miniasm}, our work confirms that the layout of long reads can be found without prior error correction, using only overlap information generated from raw reads by tools such as minimap, MHAP or DALIGNER.
However, unlike miniasm, which does not derive a consensus but instead concatenates the reads into a full sequence, we take advantage of read coverage to produce contigs with a consensus quality on par with that achieved by assembly pipelines executing dedicated error-correction steps.
The results of Table~\ref{tab:allAssembly} appear promising. For example, our assembler combined with Racon yields among the highest average identities with the reference for the ONT datasets.
In terms of speed however, our pipeline is clearly outperformed by Miniasm, but also by Miniasm+Racon, the latter improving overall accuracy.
Still, compared to approaches implementing error correction steps, we gain significant speed-ups by highly localizing the error correction and consensus generation processes, which is made possible by knowledge of the layout.
We believe that tools such as Miniasm and Racon are implemented in a much more efficient way than our own, but the layout method itself is efficient (see Supplementary Table~\ref{supptab:runtime}) and is known to be scalable as it relies on the same algorithmic core as Google's PageRank.

The main limitation of our layout algorithm is its sensitivity to outliers in the similarity matrix, hence the need to remove them in a pre-processing phase. Higher coverage and quality of the input reads, both expected in the near future, would likely improve the robustness of our pipeline.
Still, for eukaryotic genomes, we found that some outliers require additional information to be resolved (see Supplementary Figure\ref{fig:simMatSpyNanoSim}), which could be provided in the future by extracting topological information from the assembly graph.

In the meantime, our pipeline behaves like a draft generating assembler for prokaryotic genomes, and a first-pass unitigger for eukaryotic genomes.
Importantly, the overall approach is modular and can integrate other algorithms to increase layout robustness or consensus quality, as illustrated here by the integration of Racon as an optional polishing module.

Our original contribution here consists in the layout computation. The spectral OLC assembler we built on top of it could be enhanced in many ways. We have shown that the spectral algorithm is suited to find the layout for bacterial genomes, even though there is room left for performance improvements on repeat-rich eukaryotic genomes.

For these eukaryotic genomes, it could make sense to use the spectral algorithm jointly with other assembly engines (\textit{e.g.} Miniasm or Canu), to check the consistency of connected components before they are assembled.
Our consensus generation method is coarse-grained for now and does not take into account statistical properties of ONT sequencing errors.
Nevertheless, the three components (O, L and C) of the method being independent, an external and more refined consensus generation process could readily be plugged after the overlap and layout computations to further improve results and increase accuracy.

\section*{Acknowledgement}
TB would like to thank Genoscope's sequencing (Laboratoire de S\'equen\c{c}age) and bioinformatics (Laboratoire d'Informatique Scientifique) teams for sharing some \textit{ Acinetobacter baylyi} ADP1 and \textit{ Sacharomyces cerevisiae} S288C MinION data, and is grateful to Oxford Nanopore Technologies Ltd for granting Genoscope access to its MinION device via the MinION Access Programme.

AA and AR would like to acknowledge support from the European Research Council (project SIPA). The authors would also like to acknowledge support from the chaire {\em \'Economie des nouvelles donn\'ees}, the {\em data science} joint research initiative with the {\em fonds AXA pour la recherche} and a gift from Soci\'et\'e G\'en\'erale Cross Asset Quantitative Research.

\bibliography{biblio}

\clearpage

\beginsupplement
\onecolumn

\section{Supplementary Material}

\begin{table*}[bh]
\centering
\processtable{Running time for the different methods on the datasets presented in Section\ref{subsec:data}}
{\footnotesize\label{supptab:runtime}
\begin{tabular}{c K{2.2cm} K{2.cm} K{1.7cm}K{1.7cm}K{1.7cm}K{1.7cm}K{1.7cm}}
  \toprule

  & & \textbf{Spectral Layout} & \textbf{Spectral (full, +Minimap)}  & \textbf{Canu}  & \textbf{Minimap + Miniasm}  & \textbf{Racon after Miniasm}  & \textbf{Racon after Spectral} \\
  \hline
  \multirow{2}*{\parbox{1.7cm}{\textit{A. baylyi} ONT R7.3 28x}}& Runtime [h:mm:ss] & 0:00:23 (0:00:59) & 0:12:52 & 0:25:55 & 0:00:28 & 0:01:54 & 0:01:48 \\
  & Max mem [Gb] & 1.966 & 1.966 & 3.827 & 1.499 & 0.756 & 0.484 \\
  \hline
  \multirow{2}*{\parbox{1.7cm}{\textit{E. coli} ONT R7.3 30x}}& Runtime [h:mm:ss] & 0:00:41 (0:01:25) & 0:16:15 & 0:28:40 & 0:00:13 & 0:04:36 & 0:02:14 \\
  & Max mem [Gb] & 1.216 & 1.216 & 4.655 & 2.099 & 0.879 & 0.645 \\
  \hline
  \multirow{2}*{\parbox{1.7cm}{\textit{S. cerevisiae} ONT R7.3 68x}}& Runtime [h:mm:ss] & 0:01:41 (0:07:60) & 1:41:20 & 4:33:08 & 0:01:17 & 0:21:11 & 0:21:32 \\
  & Max mem [Gb] & 12.208 & 12.208 & 4.015 & 8.506 & 2.376 & 2.325 \\
  \hline
  \multirow{2}*{\parbox{1.7cm}{\textit{S. cerevisiae} ONT R9 86x}}& Runtime [h:mm:ss] & 0:03:38 (0:09:28) & 2:26:44 & 7:15:41 & 0:02:14 & 0:23:09 & 0:22:03 \\
  & Max mem [Gb] & 32.928 & 32.928 & 3.986 & 12.397 & 2.966 & 2.775 \\
  \hline
  \multirow{2}*{\parbox{1.7cm}{\textit{E. coli} PacBio 161x}}& Runtime [h:mm:ss] & 0:05:19 (0:05:44) & 1:32:13 & 0:51:32 & 0:01:16 & 0:16:51 & 0:18:18 \\
  & Max mem [Gb] & 21.650 & 21.650 & 3.770 & 9.969 & 8.082 & 4.619 \\
  \hline
  \multirow{2}*{\parbox{1.7cm}{\textit{S. cerevisiae} PacBio 127x}}& Runtime [h:mm:ss] & 0:03:11 (0:07:01) & 2:59:41 & 1:50:23 & 0:02:10 & 0:20:54 & 0:23:32 \\
  & Max mem [Gb] & 32.184 & 32.184 & 3.810 & 16.881 & 4.290 & 4.307 \\

  \botrule
\end{tabular}
}{
Run-time and peak memory for the previously compared methods, when run on a 24 cores Intel Xeon E5-2640 2.50GHz node.
Runtime and Max mem correspond to the wall-clock and maximum resident set size fields of the unix /usr/bin/time -v command.
The first column (Spectral Layout) displays the running time of the layout phase of our method in the following way: time to reorder contigs with the spectral algorithm (total time to get fine-grained layout); the total time for the layout (including the fine-grained computation of the position of the reads on a backbone sequence) is given between parentheses next to the time for the ordering.
The second column gives the runtime for our full pipeline, including running minimap to obtain the overlaps.
The runtime for Racon includes the time to map the reads to the backbone sequence with Minimap and to run Racon for the consensus (Racon requires a backbone sequence, obtained either with Miniasm or Spectral in the present experiments).
Indeed, the Racon pipeline maps the reads to a draft sequence to get the layout and then computes consensus sequences in windows across the genome.
Our pipeline instead directly computes the layout and then generates consensus sequences in windows across the genome (the latter task being embarassingly parallel).
Canu is faster than our method on the PacBio datasets (probably at least because because we did not adapt our pipeline (as Canu does) to the much higher coverage, nor to the higher fraction of chimeric reads typical of PacBio data), but not on the ONT datasets.
The memory for the spectral method can be allocated among several cores.
}
\end{table*}
\begin{figure}[bh]
 \centering
 \includegraphics[width=0.5\textwidth]{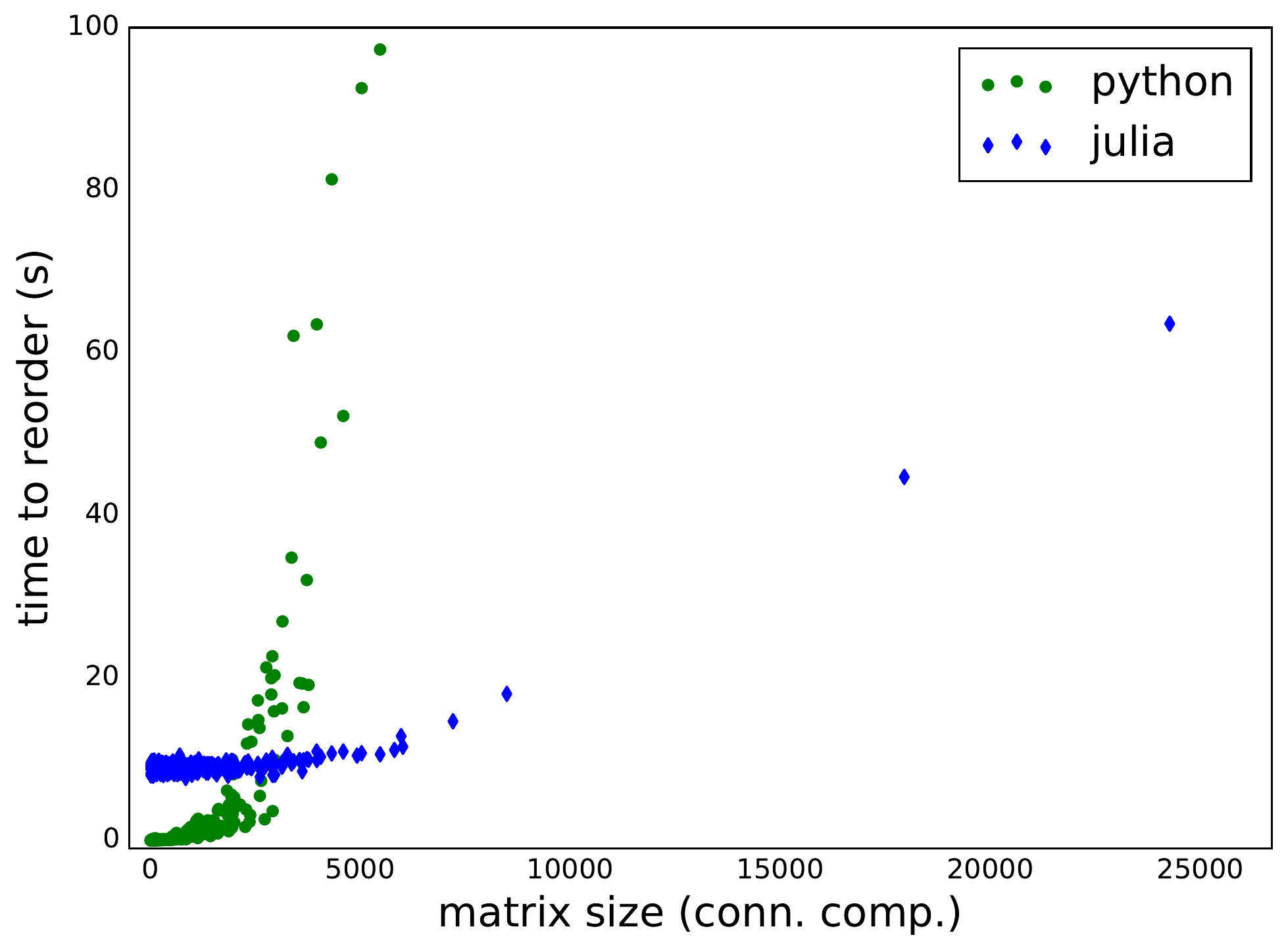}
  \caption{
 Runtime of the spectral ordering algorithm in connected components of different sizes (across all datasets), with two solvers for the eigenvalues computations (scipy.sparse.eigsh and the eigs function from Julia \citep{BEKS14}).
 We implemented a call to Julia for matrices of size larger than 3000 in the code since its eigenvector computation scales better for large matrices (probably due to the fact that the matrices are passed by reference in Julia) but has a non-negligible overhead for small matrices.
 }
 \label{suppfig:runtimeReordering}
\end{figure}

\begin{figure}[htb]
  \centering
  \begin{subfigure}[b]{0.27\textwidth}
    \centering \includegraphics[width=\textwidth]{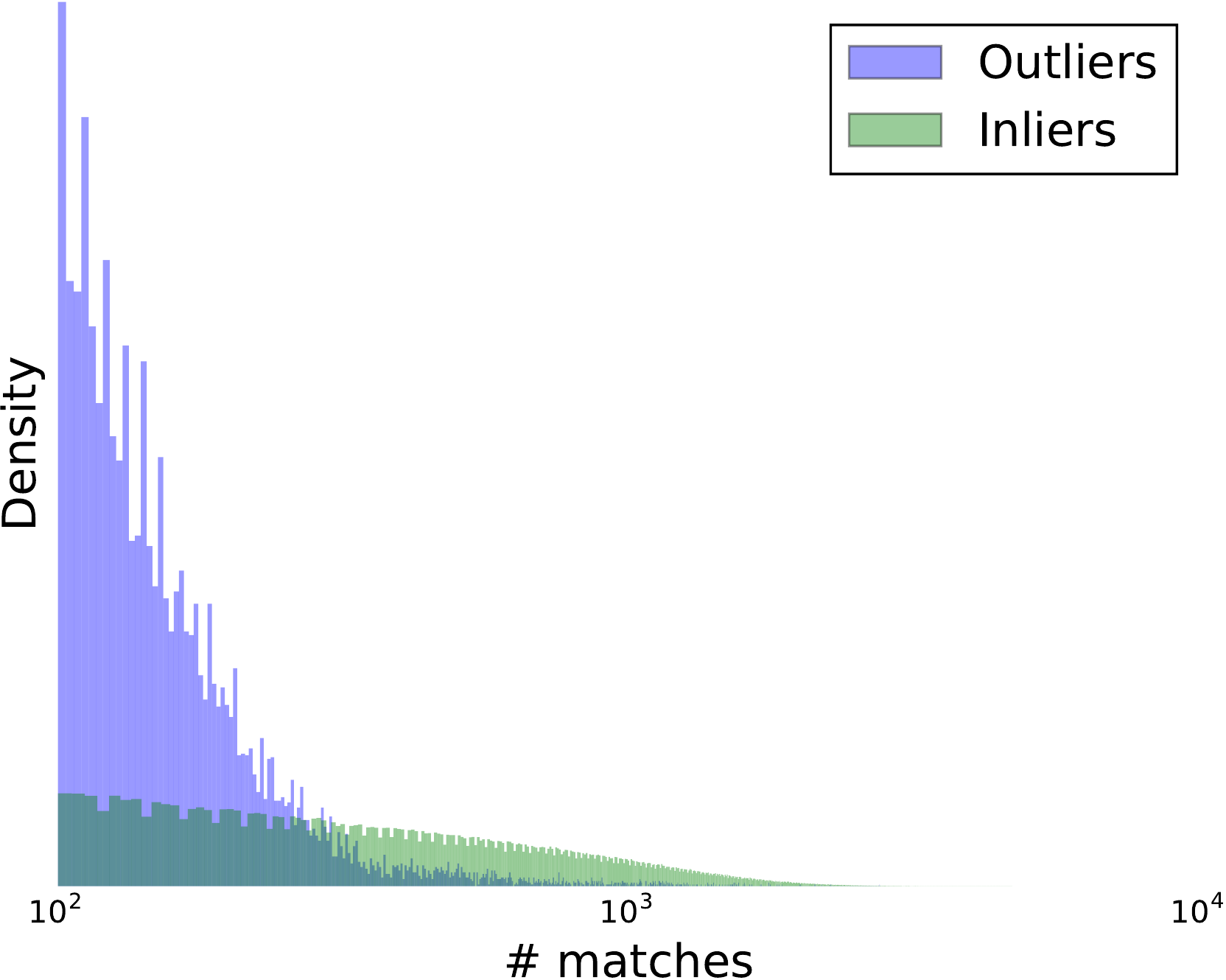}
    \caption{\textit{A. baylyi} ONT}
  \end{subfigure}
  ~
  \begin{subfigure}[b]{0.27\textwidth}
    \centering \includegraphics[width=\textwidth]{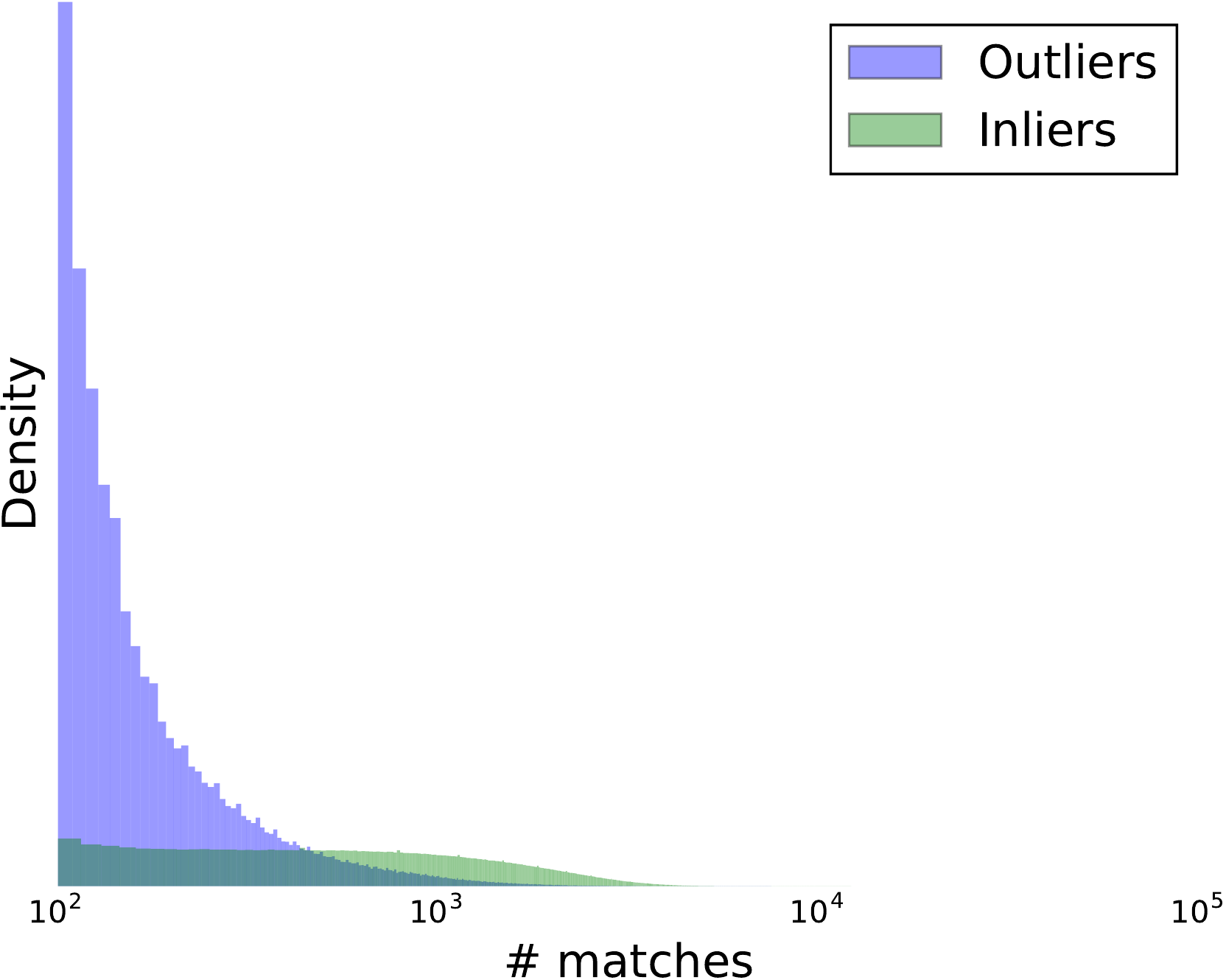}
    \caption{\textit{S. cerevisiae} ONT R7.3}
  \end{subfigure}
  ~
  \begin{subfigure}[b]{0.27\textwidth}
    \centering \includegraphics[width=\textwidth]{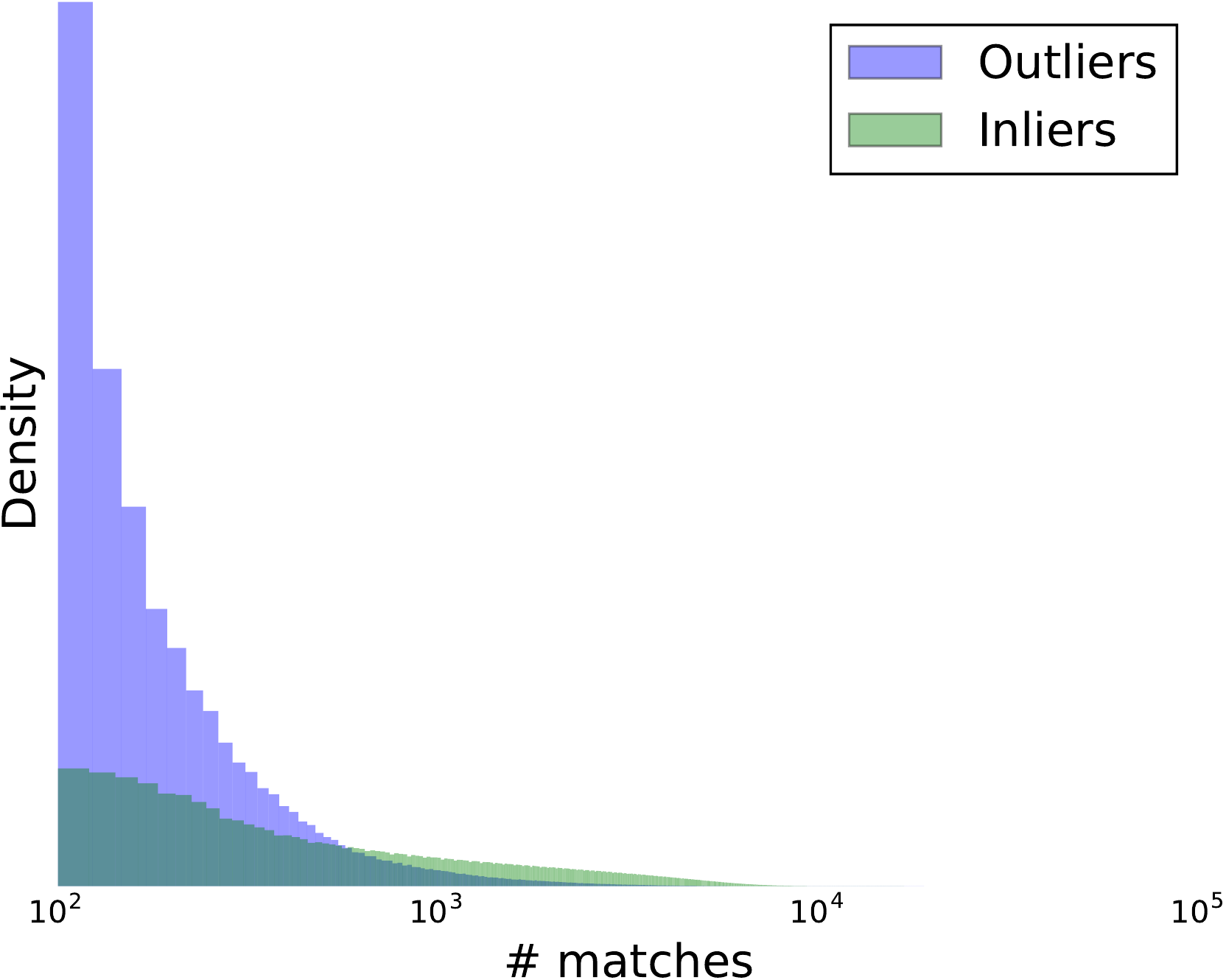}
    \caption{\textit{S. cerevisiae} ONT R9}
  \end{subfigure}

  \begin{subfigure}[b]{0.27\textwidth}
    \centering \includegraphics[width=\textwidth]{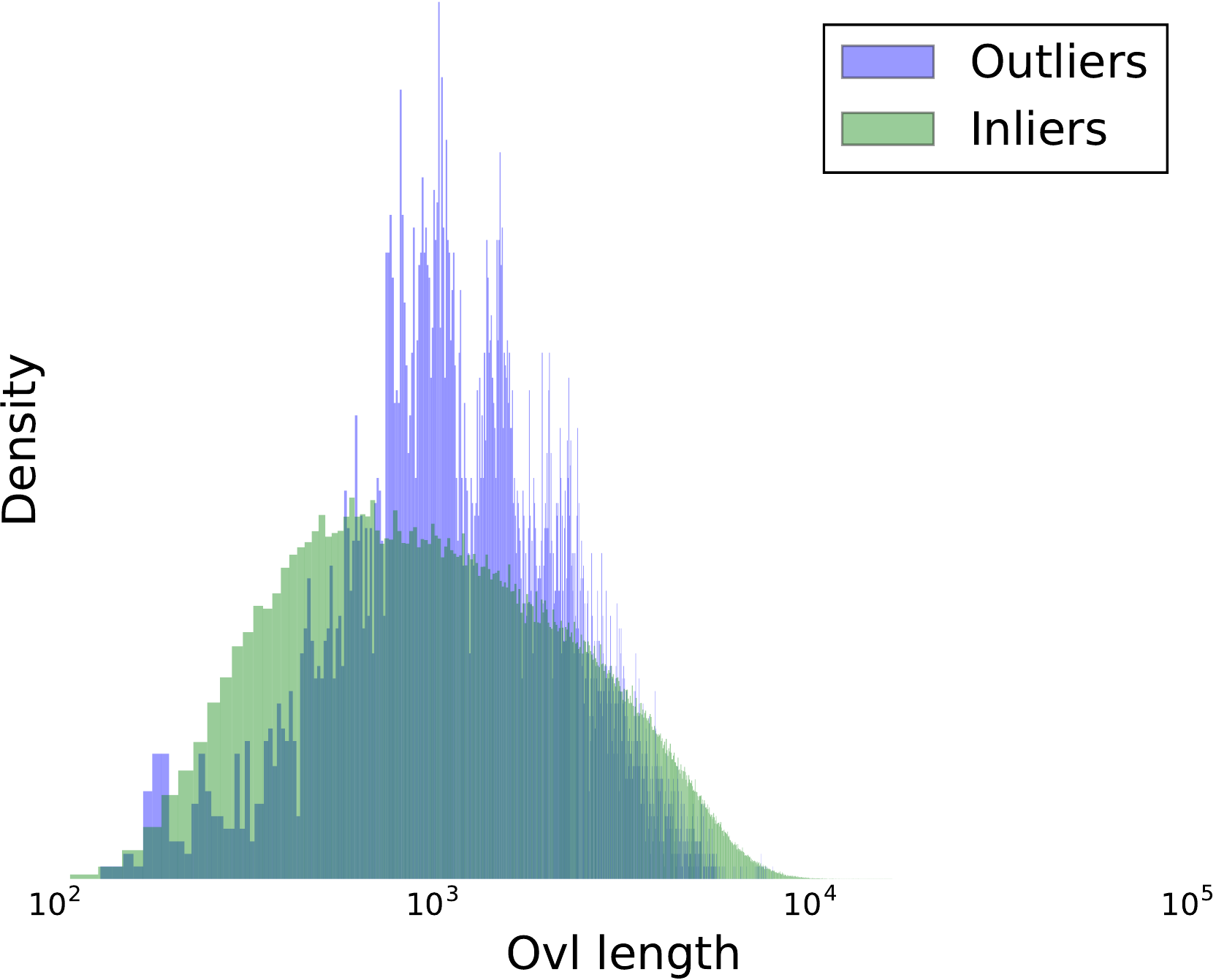}
    \caption{\textit{A. baylyi} ONT}
  \end{subfigure}
  ~
  \begin{subfigure}[b]{0.27\textwidth}
    \centering \includegraphics[width=\textwidth]{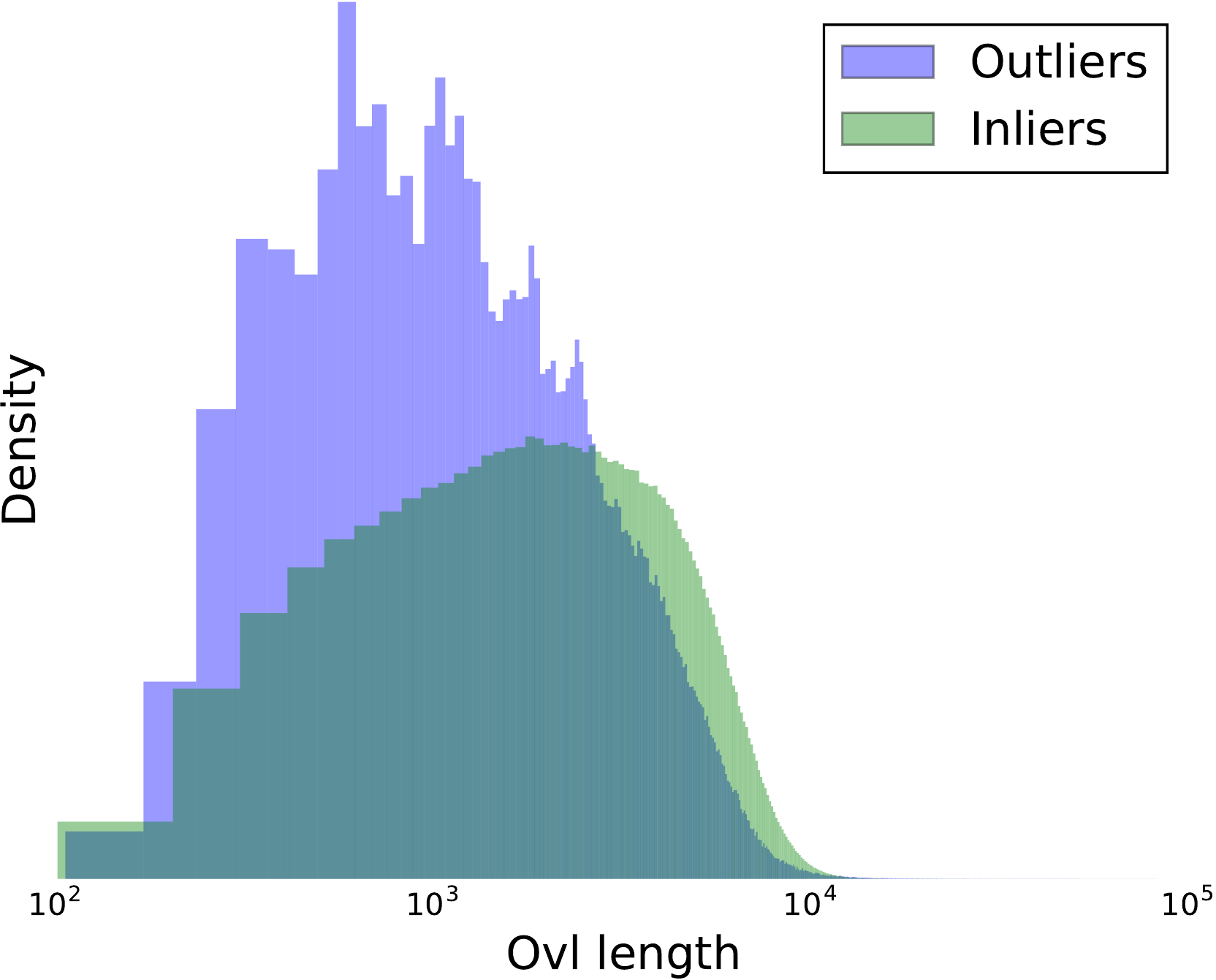}
    \caption{\textit{S. cerevisiae} ONT R7.3}
  \end{subfigure}
  ~
  \begin{subfigure}[b]{0.27\textwidth}
    \centering \includegraphics[width=\textwidth]{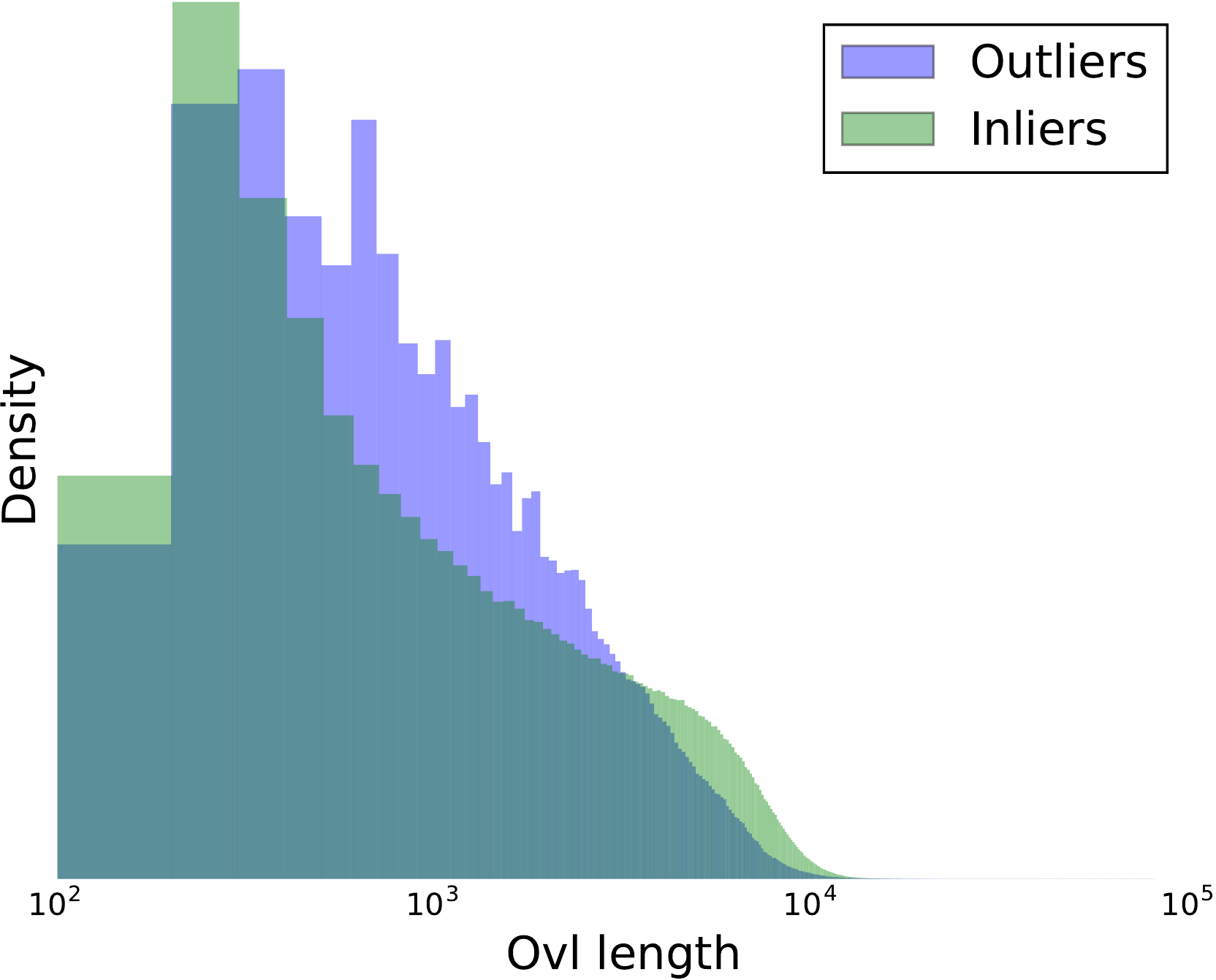}
    \caption{\textit{S. cerevisiae} ONT R9}
  \end{subfigure}
  \caption{
  Histograms of overlap scores [number of matches from minimap] (a-c) and overlap lengths (d-f) for the ONT datasets, for outliers (blue) and inliers (green).
  The x-axis is in log scale.
  The mapping of the reads to the reference genome with GraphMap was used to label inliers and outliers.
}\label{fig:outliersHisto}
\end{figure}

\begin{figure}[hb]
  \centering
  \begin{subfigure}[b]{0.3\textwidth}
    \centering \includegraphics[width=\textwidth]{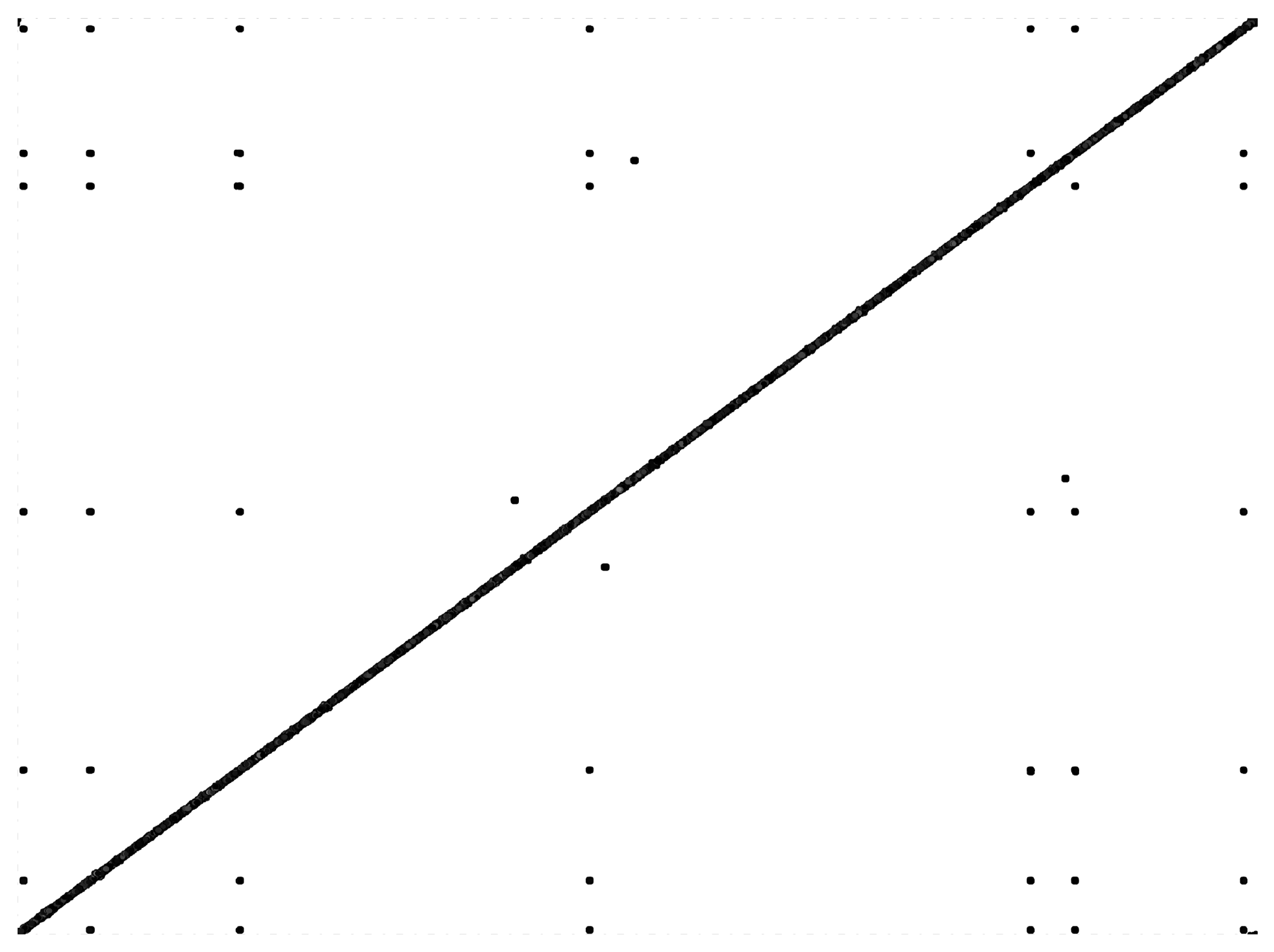}
    \caption{\textit{A. baylyi} simu. perfect 104X}\label{subfig:abaylyiNanosimPerfect05}
  \end{subfigure}
  ~
  \begin{subfigure}[b]{0.3\textwidth}
    \centering \includegraphics[width=\textwidth]{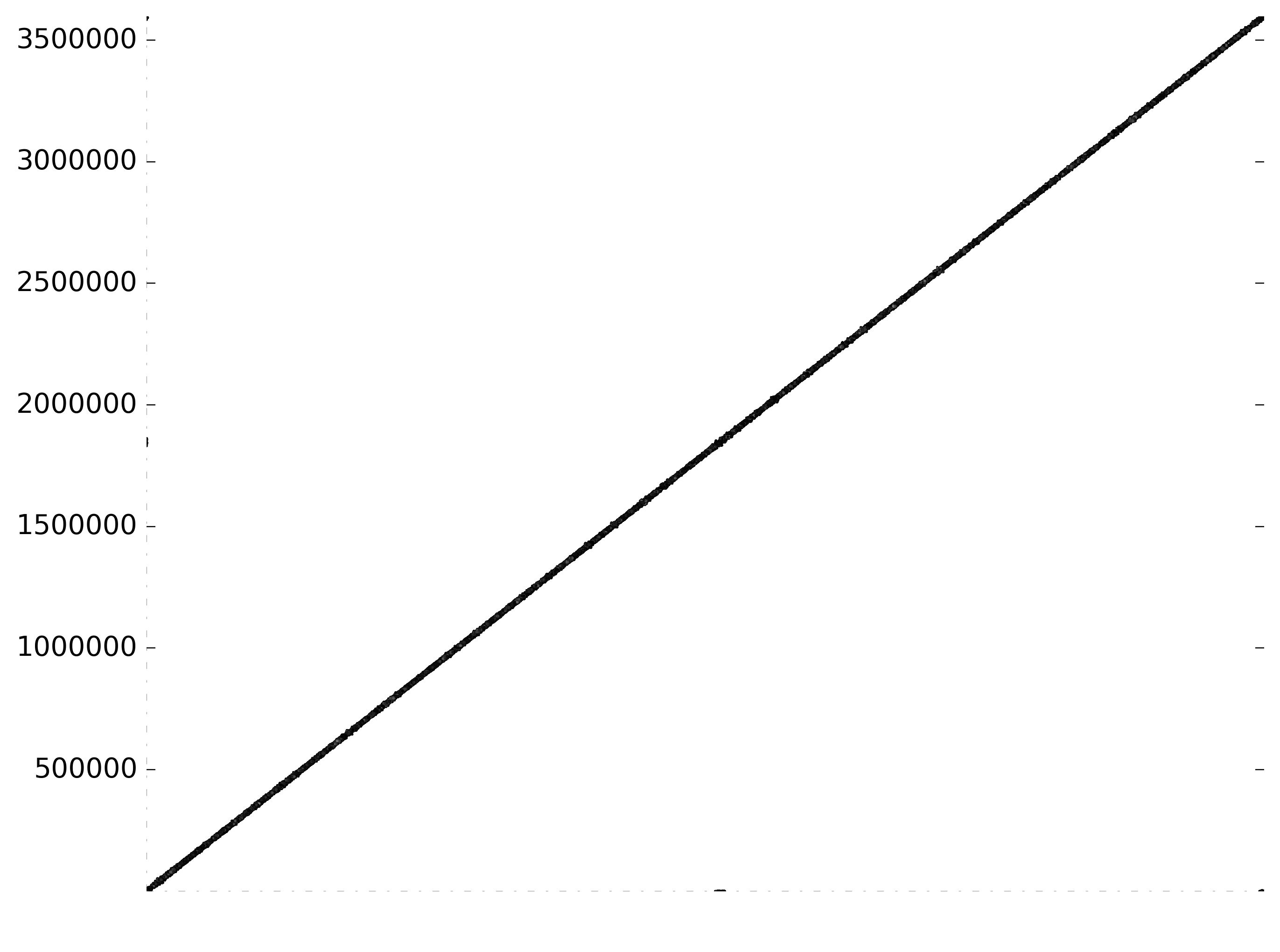}
    \caption{\textit{A. baylyi} simu. raw 104X}\label{subfig:abaylyiNanosimRaw05}
  \end{subfigure}
  ~
  \begin{subfigure}[b]{0.3\textwidth}
    \centering \includegraphics[width=\textwidth]{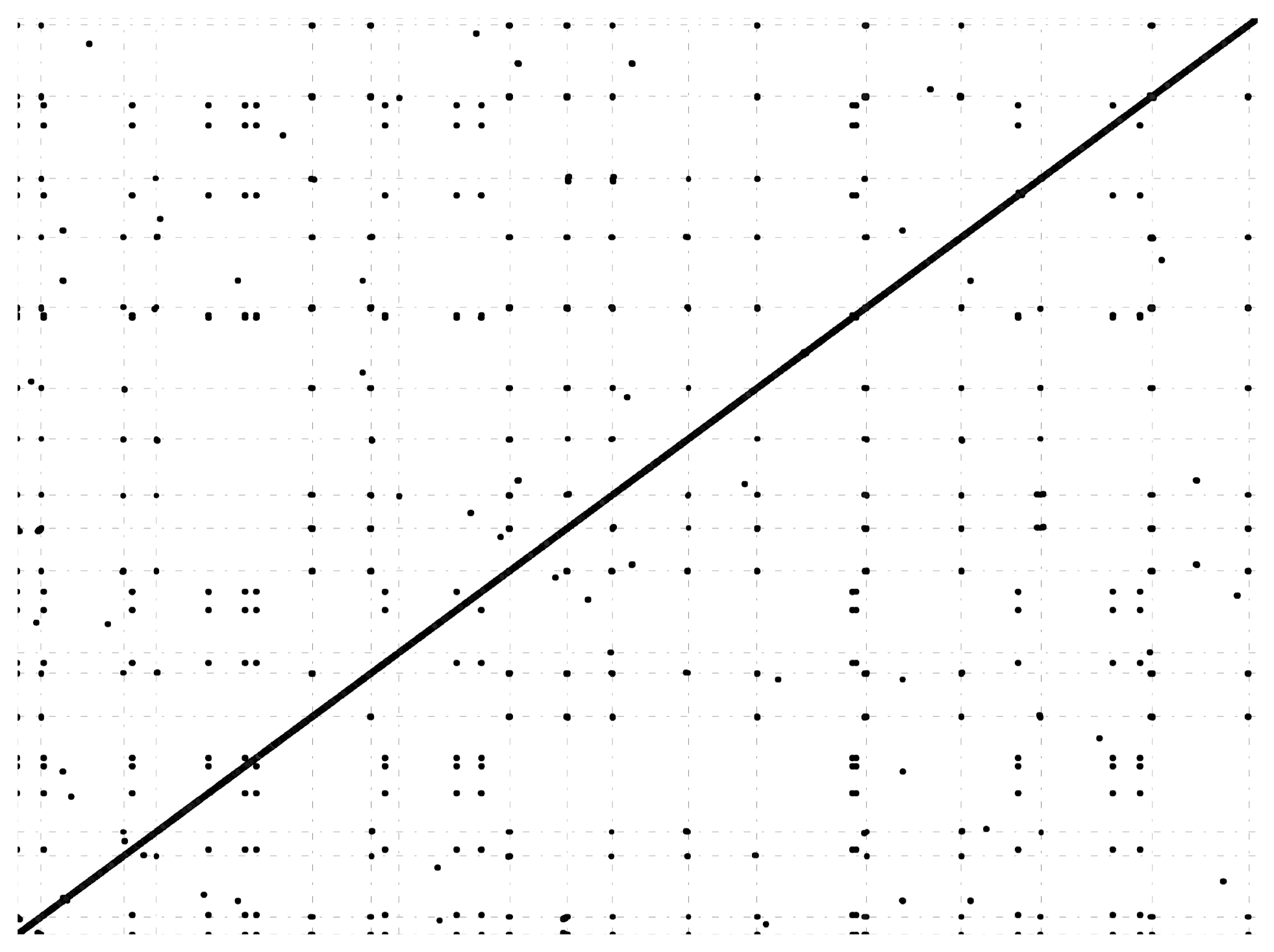}
    \caption{\textit{S. cerevisiae} simu. perfect 87x}\label{subfig:yeastNanosimPerfect05}
  \end{subfigure}

  \begin{subfigure}[b]{0.3\textwidth}
    \centering \includegraphics[width=\textwidth]{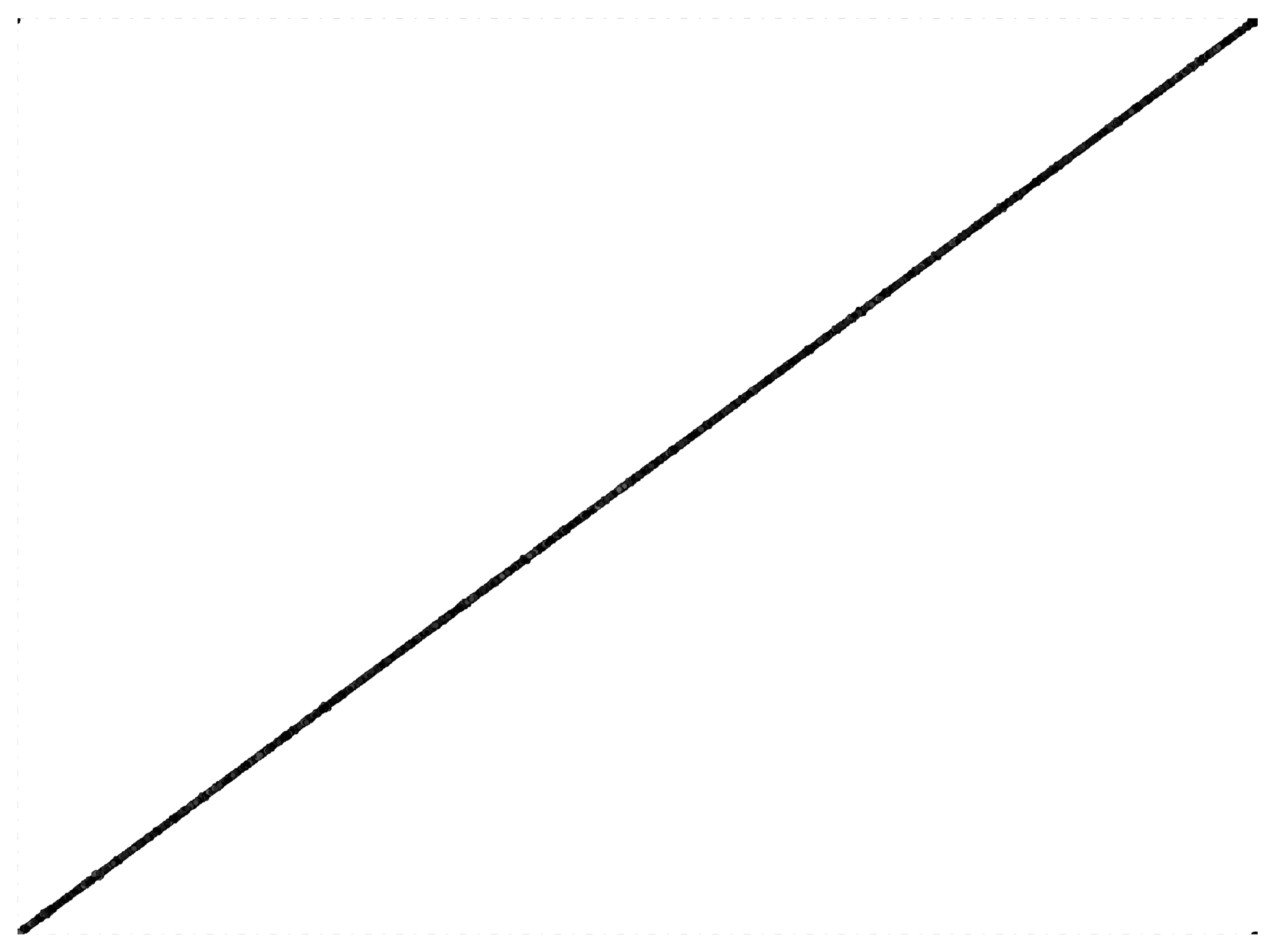}
    \caption{\textit{A. baylyi} sim. perfect 104X}\label{subfig:abaylyiNanosimPerfect09}
  \end{subfigure}
  ~
  \begin{subfigure}[b]{0.3\textwidth}
    \centering \includegraphics[width=\textwidth]{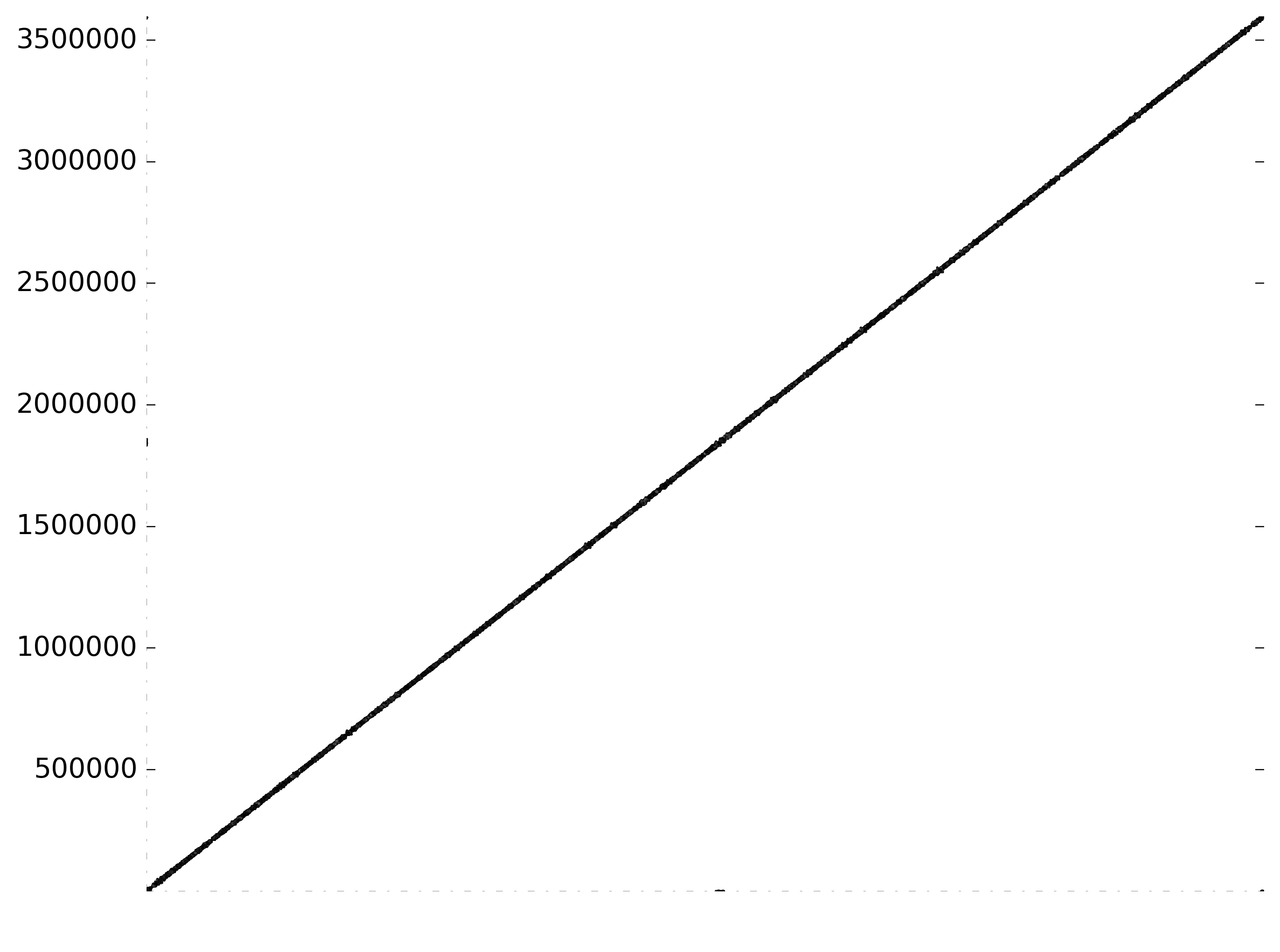}
    \caption{\textit{A. baylyi} simu. raw 104X}\label{subfig:abaylyiNanosimRaw09}
  \end{subfigure}
  ~
  \begin{subfigure}[b]{0.3\textwidth}
    \centering \includegraphics[width=\textwidth]{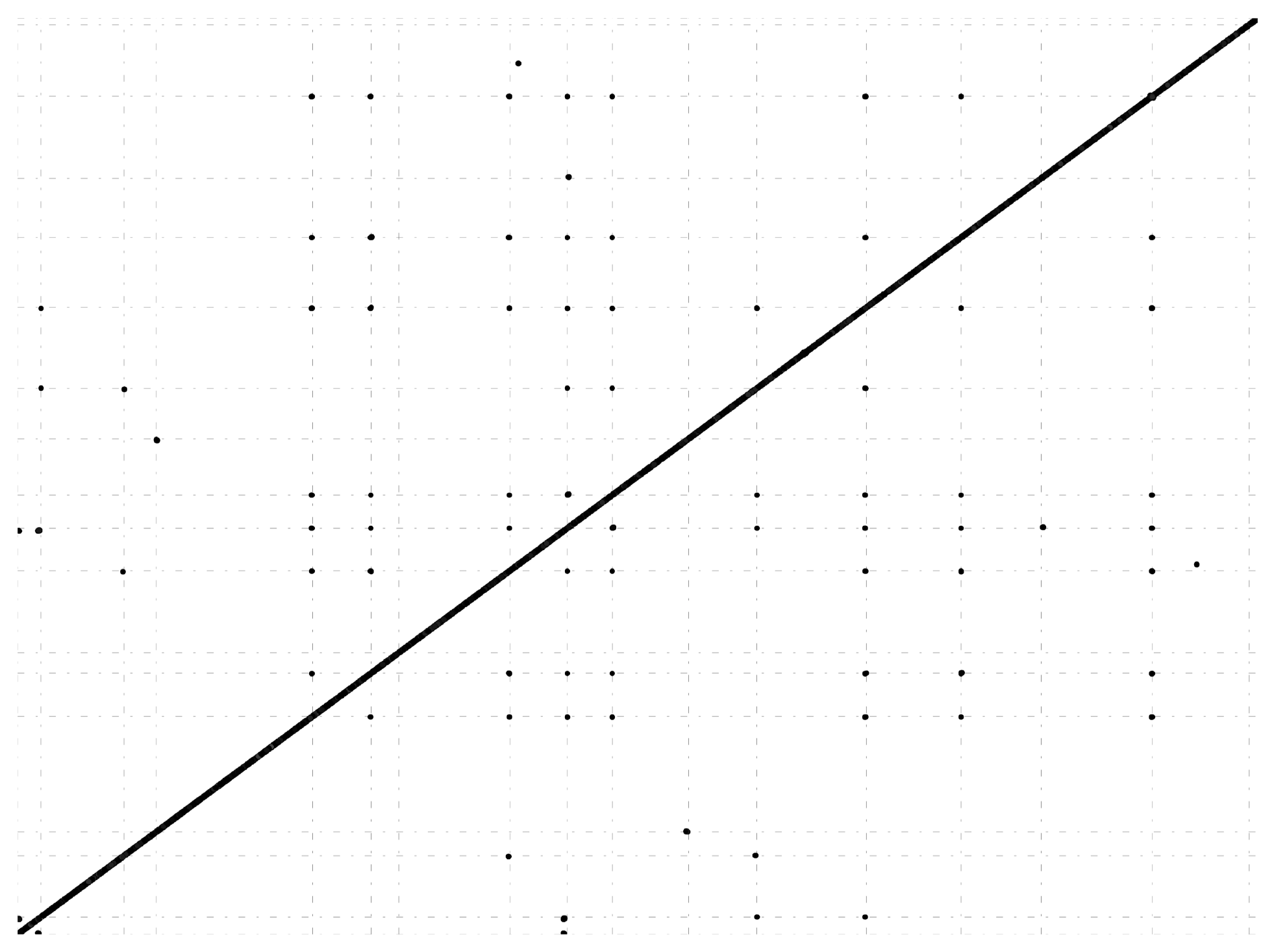}
    \caption{\textit{S. cerevisiae} simu. perfect 87x}\label{subfig:yeastNanosimPerfect09}
  \end{subfigure}
  \caption{
  Ordered similarity matrices for simulated datasets after removing 50\% of the overlaps (a-c) or 90\% (d-f) to illustrate the outlier removal by thresholding on the overlap score.
  The reads were simulated with NanoSim \citep{yang2016nanosim}, from the \textit{A. baylyi} ONT R7.3 and \textit{S. cerevisiae} ONT R9 datasets.
  Subfigures~\ref{subfig:abaylyiNanosimPerfect05} and \ref{subfig:abaylyiNanosimPerfect09}
  (respectively \ref{subfig:yeastNanosimPerfect05} and \ref{subfig:yeastNanosimPerfect09}) represent the similarity for reads generated with NanoSim from the \textit{A. baylyi} ONT R7.3 (respectively \textit{S. cerevisiae} ONT R9) dataset with option --perfect,
  which means these synthetic reads follow the same length distribution than the original dataset, but have no errors, and have the coverage specified above.
  The matrices \ref{subfig:abaylyiNanosimRaw05} and \ref{subfig:abaylyiNanosimRaw09} were generated from the \textit{A. baylyi} ONT R7.3 dataset without the --perfect option, which means they have the same length and error distribution than the original data, but with higher coverage.
  For perfect and noisy synthetic \textit{A. baylyi} reads and with sufficient coverage, all outliers could be removed by thresholding while keeping a connected similarity graph (all matrices in the Figure are connected).
  On the other hand, the similarity matrix generated with \textit{S. cerevisiae} perfect reads still harbors a few outliers after removing 90\% of the overlaps (with lowest score).
  When increasing the threshold value, the connectivity within some individual chromosomes will be broken before all outliers have been removed. Additional structural information (as used in Canu or Miniasm) will be required to resolve repeats in such situations.
}\label{fig:simMatSpyNanoSim}
\end{figure}

\begin{figure}[htb]
  \centering
  \begin{subfigure}[b]{0.3\textwidth}
    \centering \includegraphics[width=\textwidth]{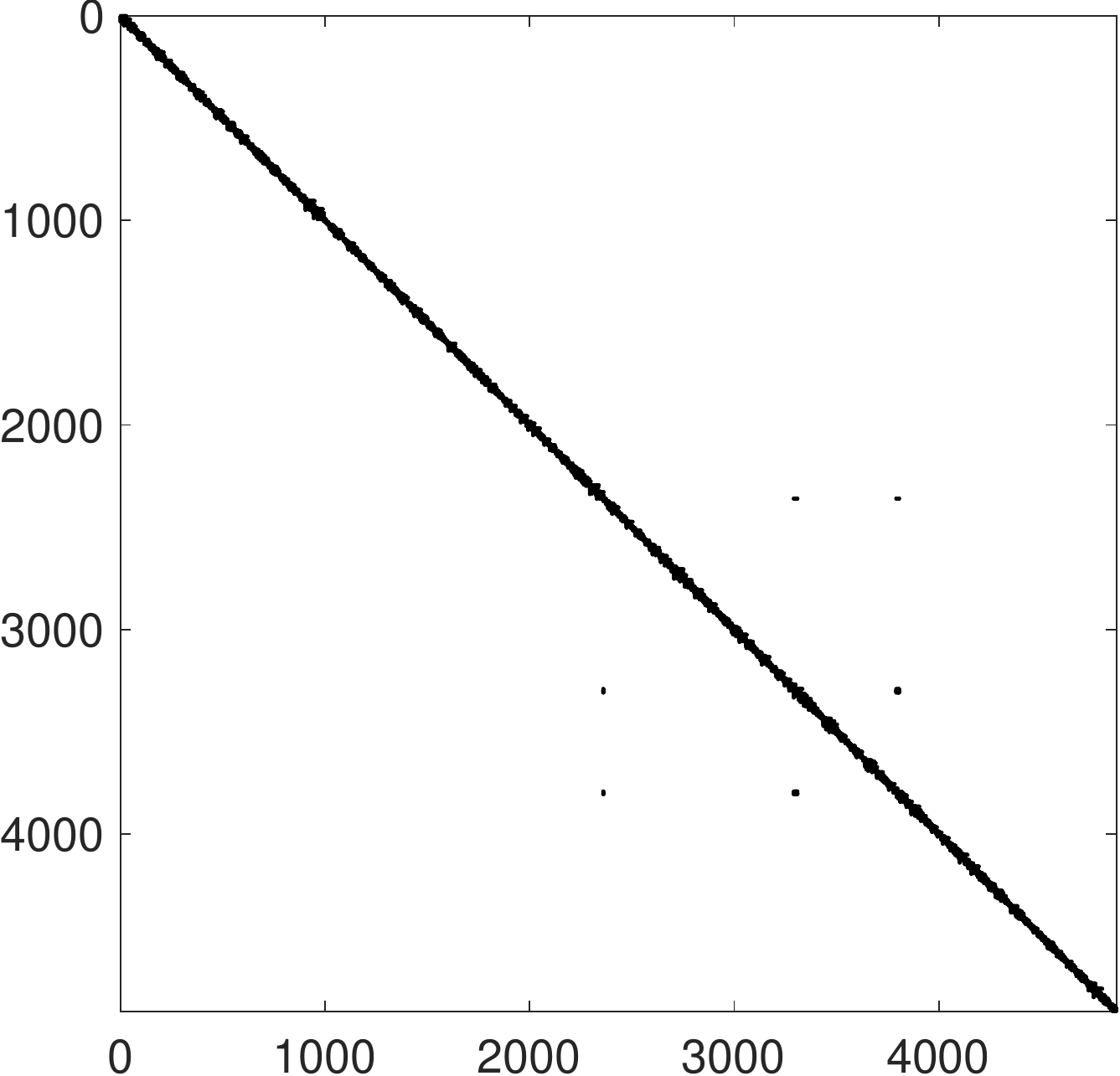}
    \caption{}\label{subfig:trueordmat}
  \end{subfigure}
  ~
  \begin{subfigure}[b]{0.3\textwidth}
    \centering \includegraphics[width=\textwidth]{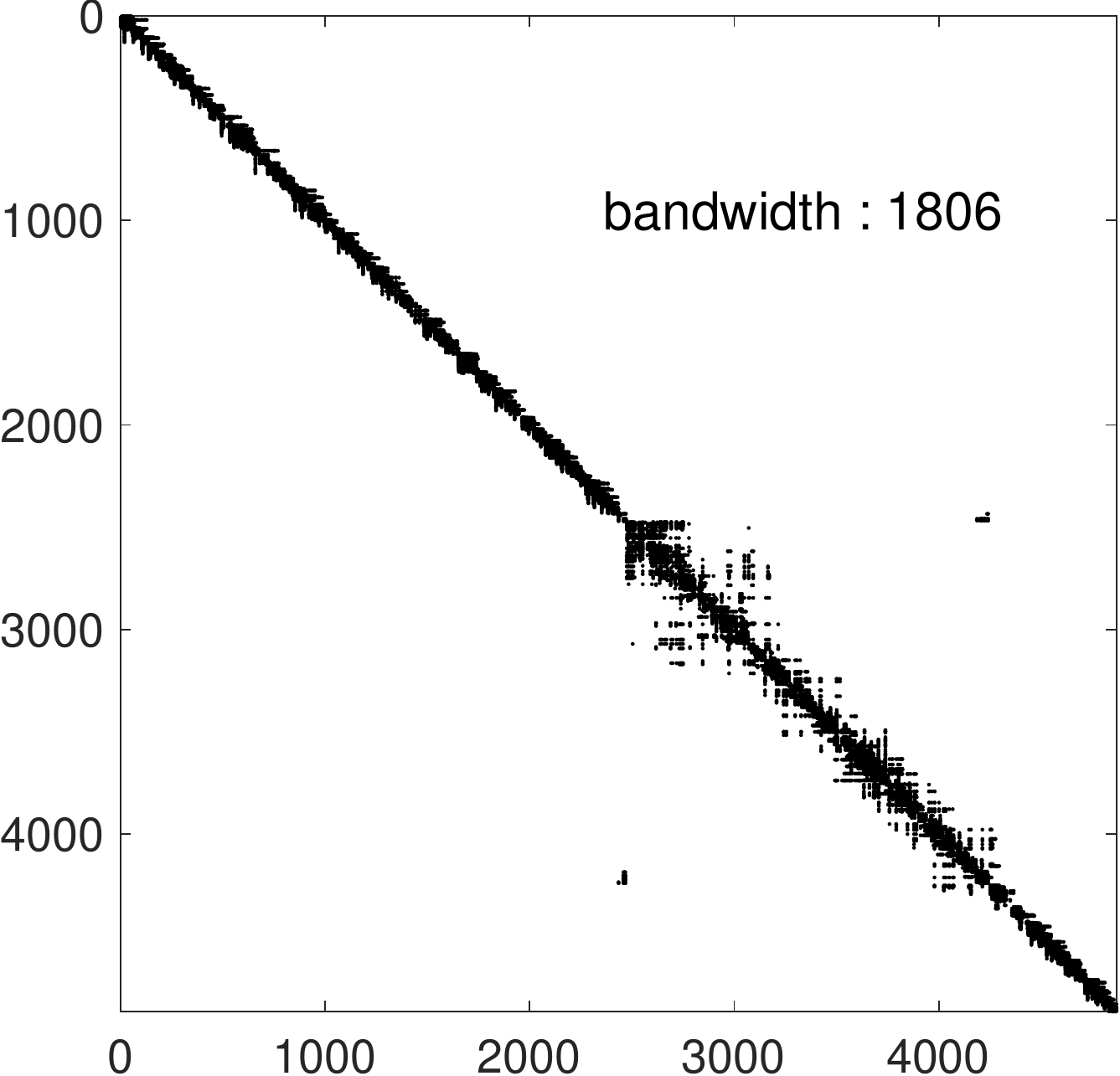}
    \caption{}\label{subfig:wrongord}
  \end{subfigure}
  \caption{
  Similarity matrices containing outliers, displayed with true ordering (obtained by mapping the reads to the reference genome with GraphMap) and generated with a subset of \textit{A. baylyi} ONT NanoSim perfect reads \ref{subfig:trueordmat},
  and the same matrix incorrectly reordered with the spectral algorithm \ref{subfig:wrongord}.
  The bandwidth is about 50 times as large as in the absence of outliers.
  This significant gap (an order of magnitude difference) between the bandwidth of the matrix reordered with the spectral algorithm depending on whether the original matrix (ordered by increasing position of the reads) contained outliers (\textit{i.e.}, is band-diagonal) or not motivated the development of the heuristic for assessing the ordering found by the spectral algorithm, as explained in \S2.3.
  However, this heuristic is not applicable when the size of the similarity matrix is small, e.g., if the matrix is of size 100, the bandwidth cannot exceed 100 and the use of the heuristic is precluded.
}\label{fig:BandwidthHeuristic}
\end{figure}

\begin{figure}[htb]
  \centering
  \begin{subfigure}[b]{0.27\textwidth}
    \centering \includegraphics[width=\textwidth]{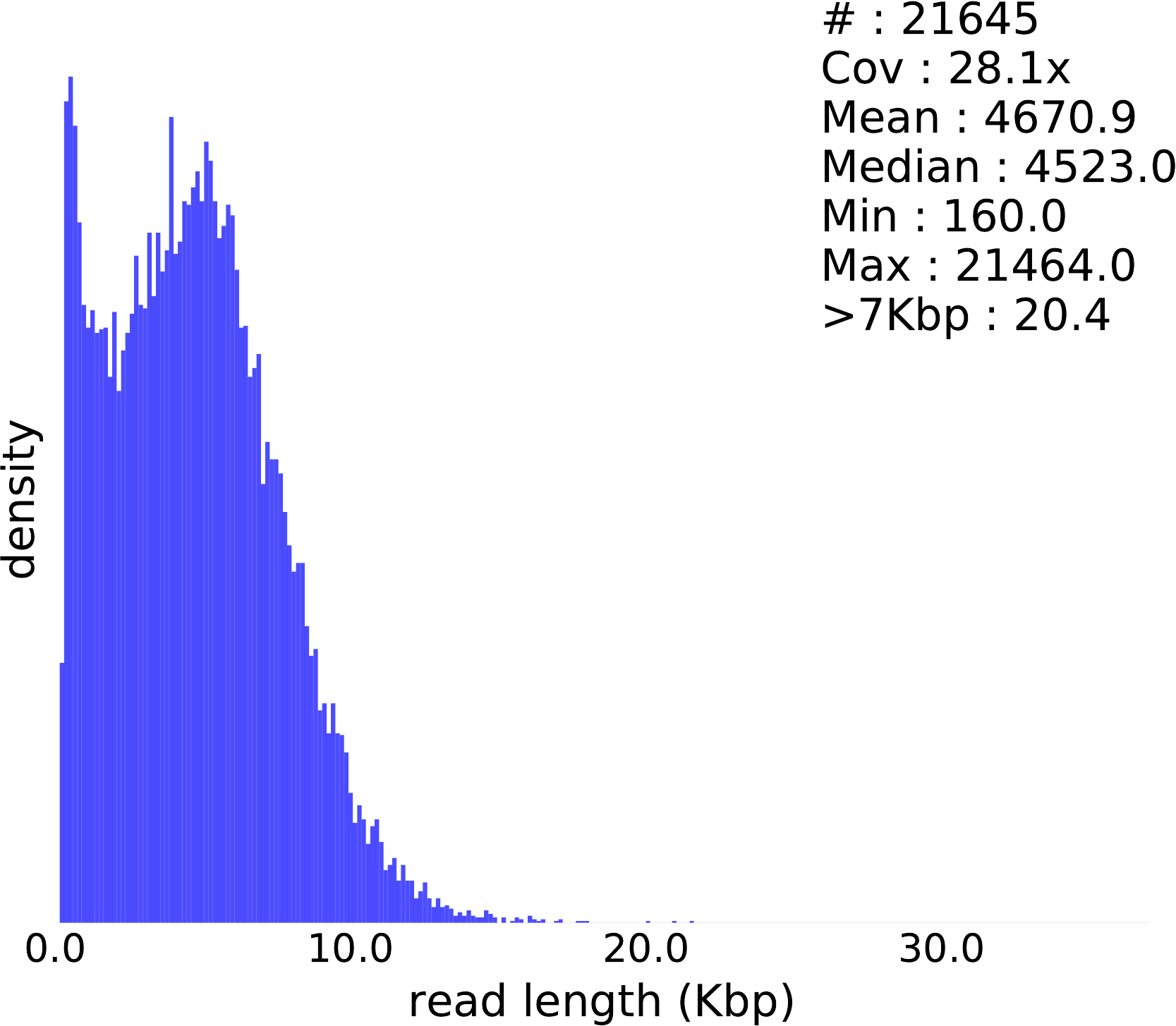}
    \caption{\textit{A. baylyi} ONT}
  \end{subfigure}
  ~
  \begin{subfigure}[b]{0.27\textwidth}
    \centering \includegraphics[width=\textwidth]{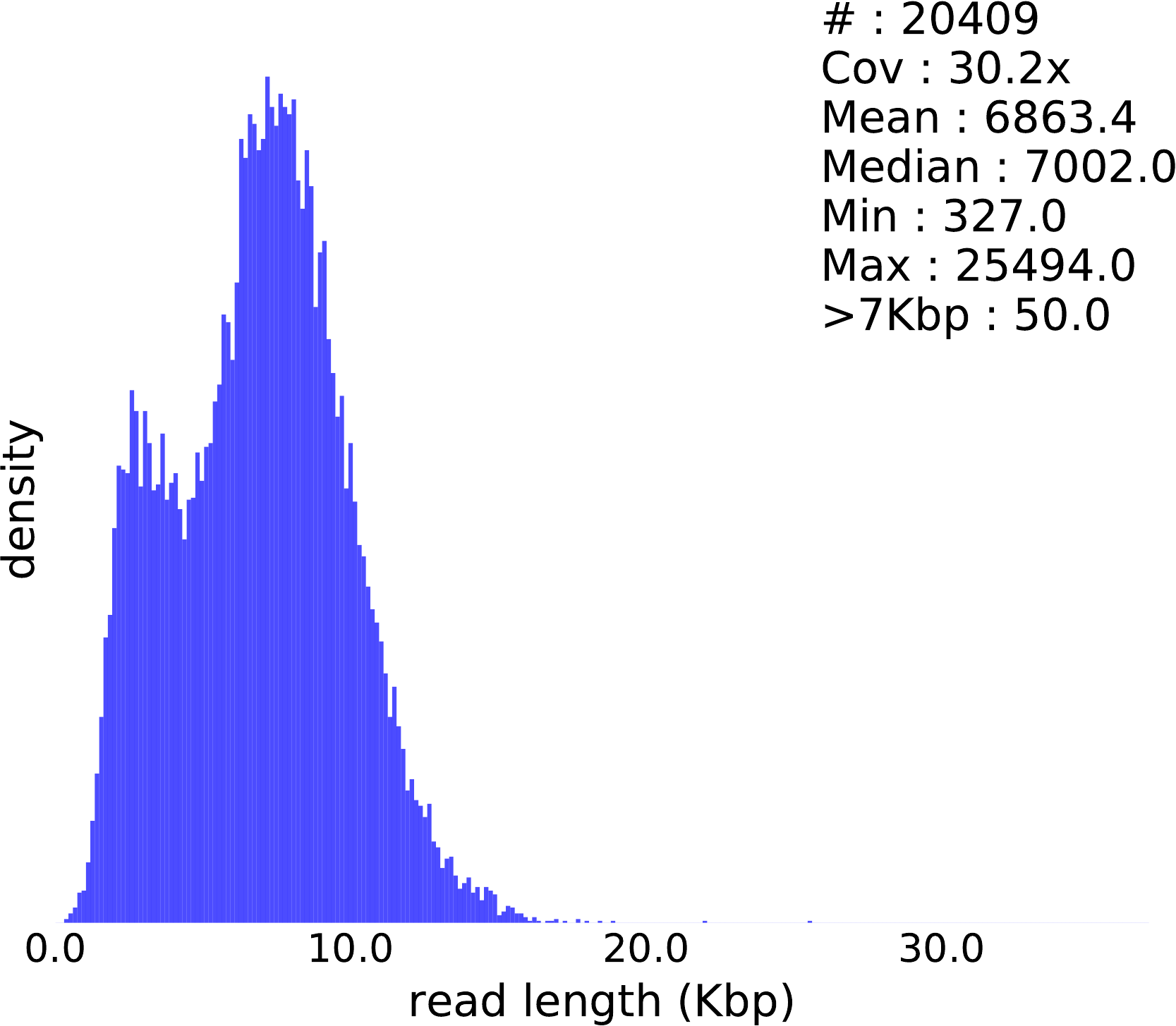}
    \caption{\textit{E. coli} ONT}
  \end{subfigure}
  ~
  \begin{subfigure}[b]{0.27\textwidth}
    \centering \includegraphics[width=\textwidth]{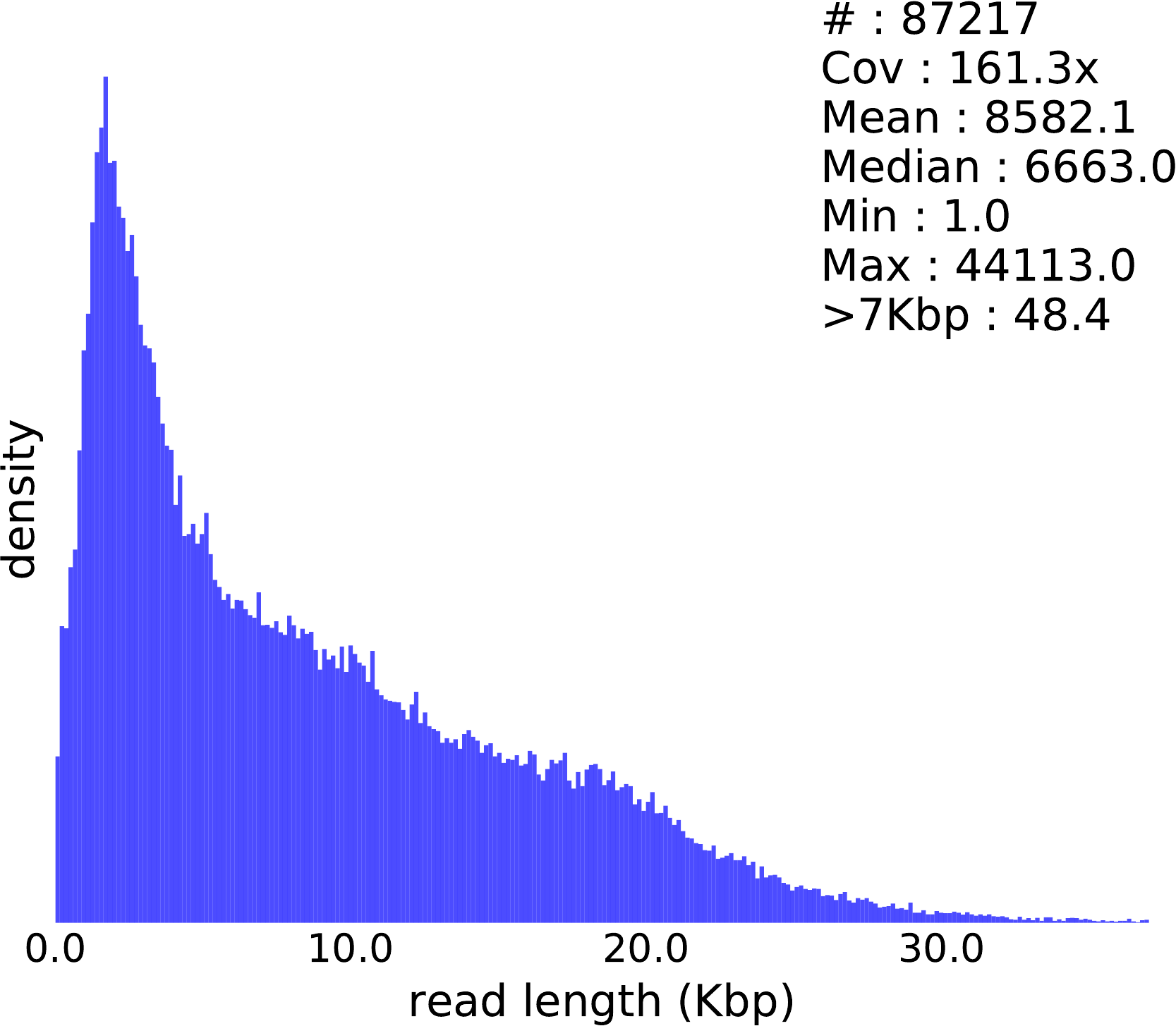}
    \caption{\textit{E. coli} PacBio}
  \end{subfigure}

  \begin{subfigure}[b]{0.27\textwidth}
    \centering \includegraphics[width=\textwidth]{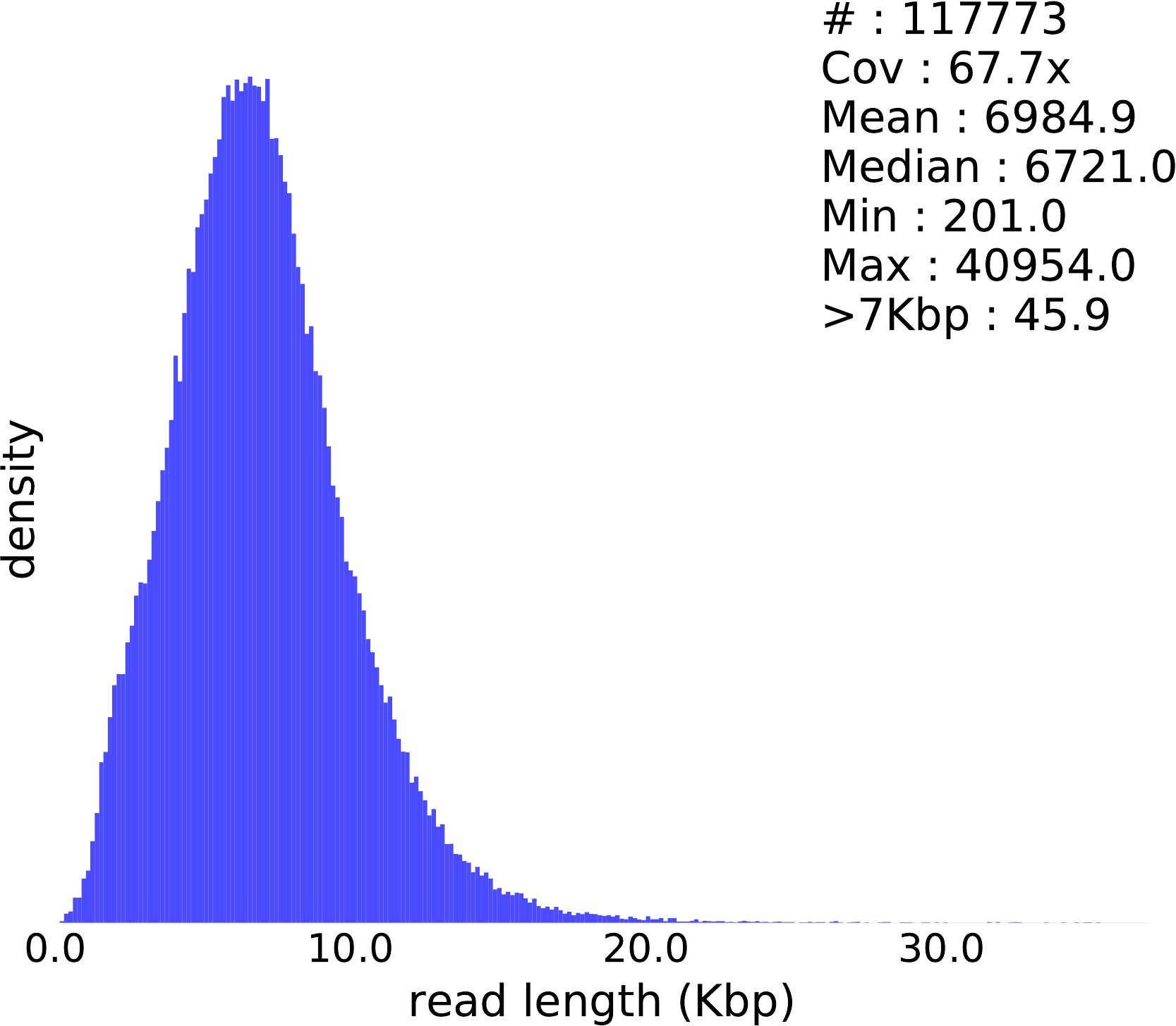}
    \caption{\textit{S. cerevisiae} ONT R7.3}
  \end{subfigure}
  ~
  \begin{subfigure}[b]{0.27\textwidth}
    \centering \includegraphics[width=\textwidth]{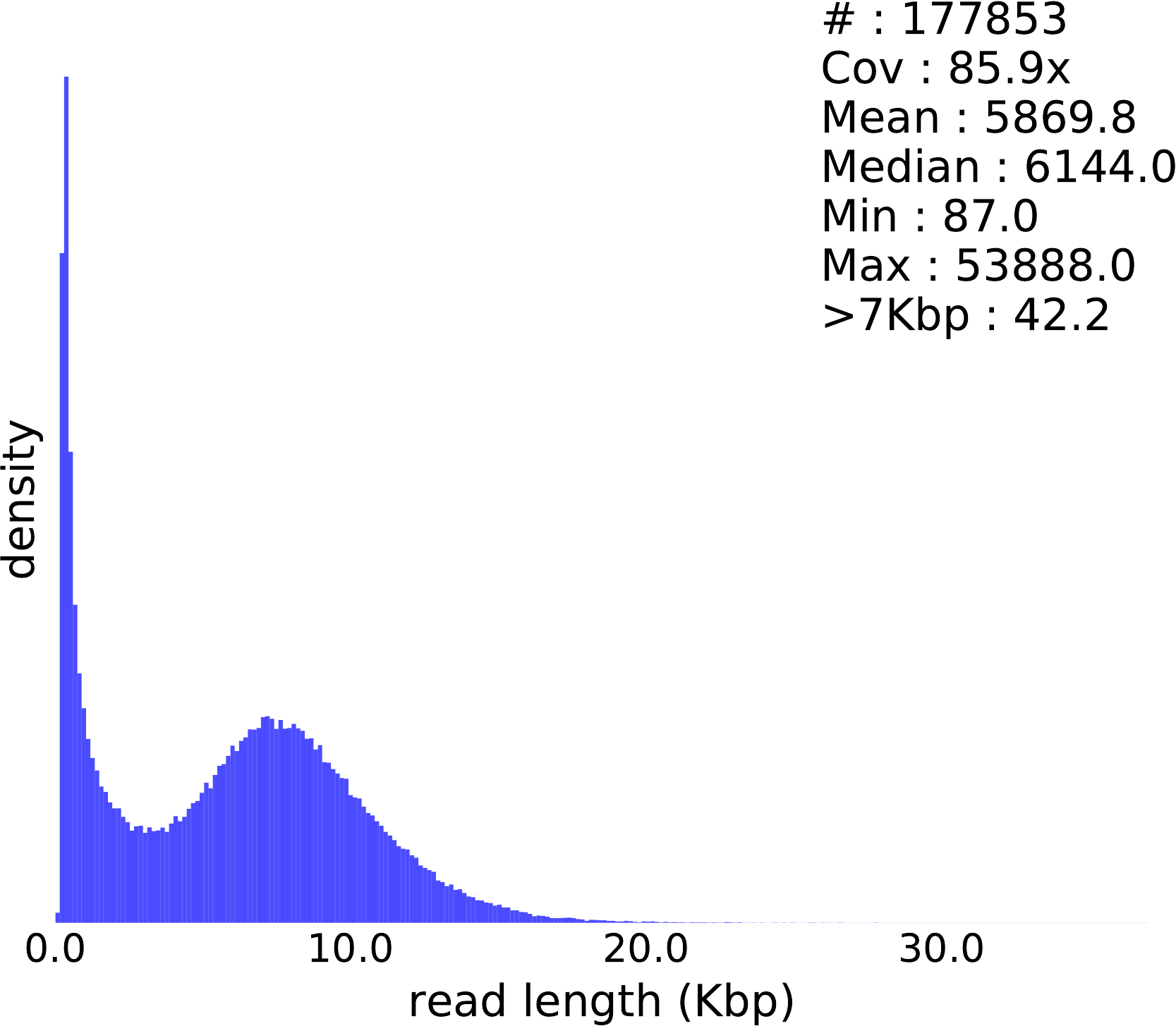}
    \caption{\textit{S. cerevisiae} ONT R9}
  \end{subfigure}
  ~
  \begin{subfigure}[b]{0.27\textwidth}
    \centering \includegraphics[width=\textwidth]{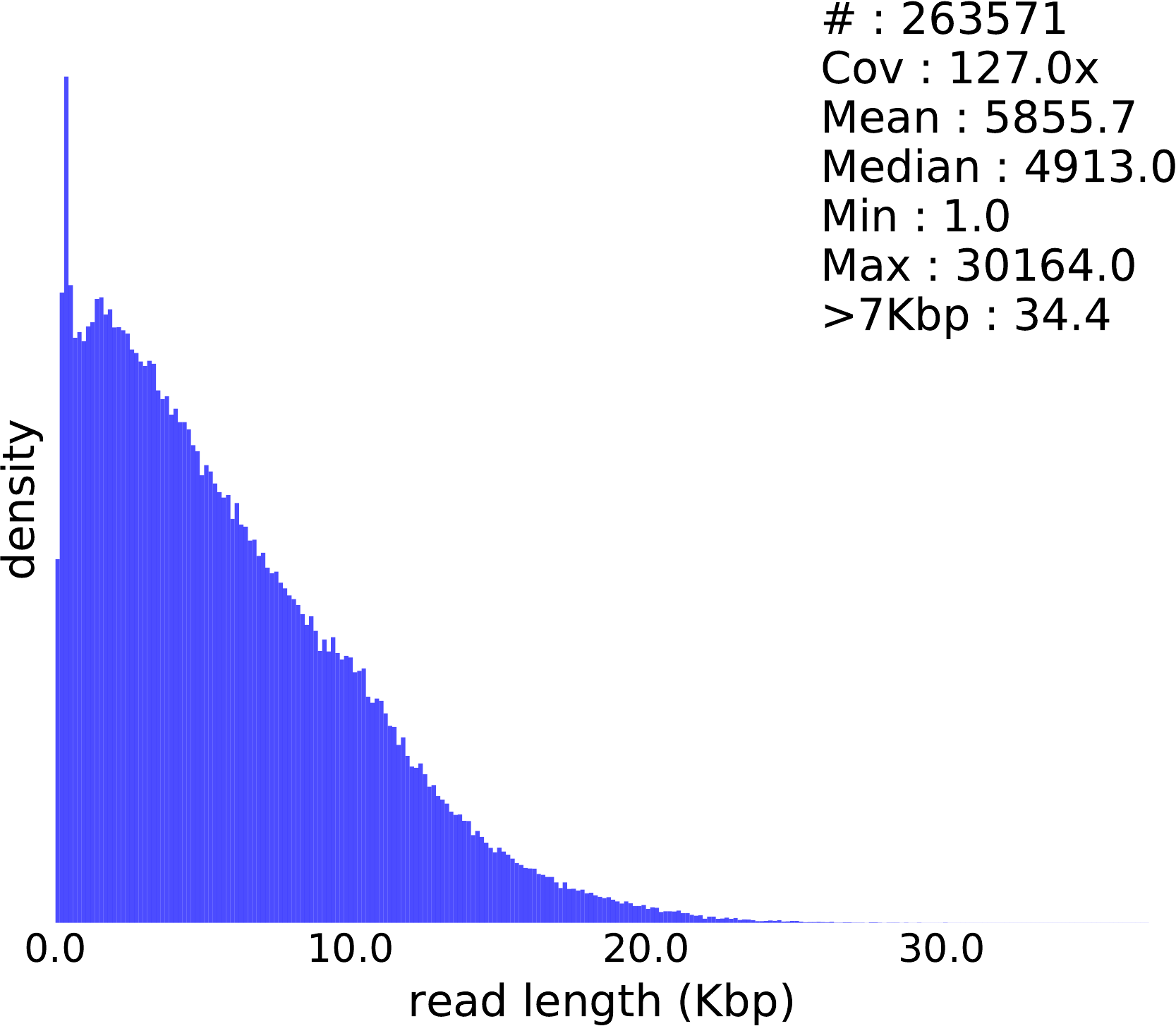}
    \caption{\textit{S. cerevisiae} PacBio}
  \end{subfigure}
  \caption{
    Read length histograms of the raw datasets.
}\label{fig:lengthHistoApp}
\end{figure}

\begin{table}[tbh]
\centering
\processtable{Assembly results of several assemblers across the datasets \textbf{corrected} with Canu}
{\footnotesize\label{tab:correctedAllAssembly}
\begin{tabular}{cc K{1.915cm}K{1.915cm}K{1.915cm}K{1.915cm}K{1.915cm}K{1.915cm}}

  \toprule
   & & Miniasm & Spectral & Canu & Miniasm+Racon & Miniasm+Racon (2 iter.) & Spectral+Racon\\
  \hline
  \multirow{8}*{\parbox{1.cm}{\textit{A. baylyi} ONT R7.3 28x (26x)}}& Ref. genome size [bp] & 3598621 & 3598621 & 3598621 & 3598621 & 3598621 & 3598621\\
  & Total bases [bp] & 3493724 & 3523055 & 3516777 & 3540178 & 3540766 & 3522315\\
  & Ref. chromosomes [\#] & 1 & 1 & 1 & 1 & 1 & 1\\
  & Contigs [\#] & 5 & 2 (9) & 2 & 5 & 5 & 2 (9)\\
  & Aln. bases ref [bp] & 3594663(99.89\%) & 3596069(99.93\%) & 3595264(99.91\%) & 3595193(99.90\%) & 3595193(99.90\%) & 3596269(99.93\%)\\
  & Aln. bases query [bp] & 3492976(99.98\%) & 3522804(99.99\%) & 3516440(99.99\%) & 3539856(99.99\%) & 3540444(99.99\%) & 3522311(100.00\%)\\
  & Misassemblies [\#] & 2 & 1 & 2 & 2 & 2 & 1\\
  & Avg. identity & 96.40 & \textbf{97.87} & 97.61 & 97.79 & 97.85 & 97.86\\\hline
  \multirow{8}*{\parbox{1.cm}{\textit{E. coli} ONT R7.3 30x (27x)}}& Ref. genome size [bp] & 4641652 & 4641652 & 4641652 & 4641652 & 4641652 & 4641652\\
  & Total bases [bp] & 4597538 & 4613973 & 4627578 & 4617120 & 4617100 & 4613521\\
  & Ref. chromosomes [\#] & 1 & 1 & 1 & 1 & 1 & 1\\
  & Contigs [\#] & 3 & 1 (8) & 2 & 3 & 3 & 1 (8)\\
  & Aln. bases ref [bp] & 4639179(99.95\%) & 4639815(99.96\%) & 4639396(99.95\%) & 4639355(99.95\%) & 4639355(99.95\%) & 4639420(99.95\%)\\
  & Aln. bases query [bp] & 4597389(100.00\%) & 4613972(100.00\%) & 4627577(100.00\%) & 4617119(100.00\%) & 4617099(100.00\%) & 4613520(100.00\%)\\
  & Misassemblies [\#] & 2 & 2 & 4 & 2 & 2 & 2\\
  & Avg. identity & 98.89 & \textbf{99.43} & 99.41 & 99.42 & \textbf{99.43} & \textbf{99.43}\\\hline
  \multirow{8}*{\parbox{1.cm}{\textit{S. cerevisiae} ONT R7.3 68x (38x)}}& Ref. genome size [bp] & 12157105 & 12157105 & 12157105 & 12157105 & 12157105 & 12157105\\
  & Total bases [bp] & 11814836 & 11959669 & 12112186 & 11877015 & 11876882 & 11949674\\
  & Ref. chromosomes [\#] & 17 & 17 & 17 & 17 & 17 & 17\\
  & Contigs [\#] & 29 & 67 (126) & 37 & 28 & 28 & 67 (126)\\
  & Aln. bases ref [bp] & 12061456(99.21\%) & 11963869(98.41\%) & 12068379(99.27\%) & 12062161(99.22\%) & 12061809(99.22\%) & 11969742(98.46\%)\\
  & Aln. bases query [bp] & 11814252(100.00\%) & 11930637(99.76\%) & 12069253(99.65\%) & 11876268(99.99\%) & 11876225(99.99\%) & 11925068(99.79\%)\\
  & Misassemblies [\#] & 19 & 22 & 26 & 20 & 20 & 24\\
  & Avg. identity & 97.81 & 98.32 & 98.36 & \textbf{98.39} & \textbf{98.39} & 98.38\\\hline
  \multirow{8}*{\parbox{1.cm}{\textit{S. cerevisiae} ONT R9 86x (40x)}}& Ref. genome size [bp] & 12157105 & 12157105 & 12157105 & 12157105 & 12157105 & 12157105\\
  & Total bases [bp] & 11946760 & 12081487 & 12184545 & 11970672 & 11970529 & 12061759\\
  & Ref. chromosomes [\#] & 17 & 17 & 17 & 17 & 17 & 17\\
  & Contigs [\#] & 21 & 65 (108) & 30 & 20 & 20 & 65 (108) \\
  & Aln. bases ref [bp] & 12055448(99.16\%) & 11851023(97.48\%) & 12110461(99.62\%) & 12056562(99.17\%) & 12056734(99.17\%) & 11879607(97.72\%)\\
  & Aln. bases query [bp] & 11944969(99.99\%) & 12043650(99.69\%) & 12184122(100.00\%) & 11970041(99.99\%) & 11969729(99.99\%) & 12040521(99.82\%)\\
  & Misassemblies [\#] & 21 & 32 & 26 & 22 & 22 & 38\\
  & Avg. identity & 98.83 & 98.90 & \textbf{99.06} & \textbf{99.06} & 99.05 & 99.04\\\hline
  \multirow{8}*{\parbox{1.cm}{\textit{E. coli} PacBio 161x (38x)}}& Ref. genome size [bp] & 4641652 & 4641652 & 4641652 & 4641652 & 4641652 & 4641652\\
  & Total bases [bp] & 4642736 & 4663427 & 4670125 & 4642423 & 4642443 & 4662179\\
  & Ref. chromosomes [\#] & 1 & 1 & 1 & 1 & 1 & 1\\
  & Contigs [\#] & 1 & 1 (1) & 1 & 1 & 1 & 1 (1)\\
  & Aln. bases ref [bp] & 4639048(99.94\%) & 4640514(99.98\%) & 4641652(100.00\%) & 4641623(100.00\%) & 4641616(100.00\%) & 4641652(100.00\%)\\
  & Aln. bases query [bp] & 4639955(99.94\%) & 4662891(99.99\%) & 4670125(100.00\%) & 4642423(100.00\%) & 4642443(100.00\%) & 4662172(100.00\%)\\
  & Misassemblies [\#] & 2 & 4 & 4 & 4 & 4 & 4\\
  & Avg. identity & 99.59 & 99.97 & \textbf{99.99} & \textbf{99.99} & \textbf{99.99} & \textbf{99.99}\\\hline
  \multirow{8}*{\parbox{1.cm}{\textit{S. cerevisiae} PacBio 127x (37x)}}& Ref. genome size [bp] & 12157105 & 12157105 & 12157105 & 12157105 & 12157105 & 12157105\\
  & Total bases [bp] & 12174558 & 12232964 & 12346261 & 12194786 & 12193481 & 12217702\\
  & Ref. chromosomes [\#] & 17 & 17 & 17 & 17 & 17 & 17\\
  & Contigs [\#] & 26 & 55 (86) & 29 & 26 & 26 & 55 (86)\\
  & Aln. bases ref [bp] & 12036689(99.01\%) & 12008560(98.78\%) & 12091871(99.46\%) & 12042104(99.05\%) & 12041381(99.05\%) & 12018488(98.86\%)\\
  & Aln. bases query [bp] & 12151704(99.81\%) & 12179852(99.57\%) & 12304982(99.67\%) & 12177020(99.85\%) & 12175701(99.85\%) & 12172316(99.63\%)\\
  & Misassemblies [\#] & 74 & 75 & 76 & 76 & 76 & 80\\
  & Avg. identity & 99.22 & 99.78 & 99.87 & \textbf{99.88} & \textbf{99.88} & 99.86\\
  \botrule
\end{tabular}
}{These corrected datasets were obtained by running Canu with the saveReadCorrections=True option on the datasets presented in \ref{subsec:data}.
Canu includes correction and trimming, resulting in a removal of short reads and a lower coverage than in the original raw data.
However, it is the coverage of the raw dataset which is relevant since higher coverage in the latter will result in longer reads in the corrected data, even though the coverage in all corrected datasets are roughly below 40x.
We indicate the coverage of the corrected datasets in parentheses next to the coverage of the original dataset.
For the spectral method, we give the results after the contig merging step (see \ref{subsec:finisher}). The number of contigs before this post-processing is given between parentheses.
Unlike with raw data, the polishing effect of adding Racon to our pipeline is not significant.
All methods have comparable results on the corrected datasets.
The best result in terms of average identity only is indicated in bold (but other metrics should also be used to compare the assemblies).

}
\end{table}

\begin{table}[tbh]
\centering
\processtable{Misassemblies report of the different assemblers across the various datasets}
{\footnotesize\label{tab:rawMisassemblies}
\begin{tabular}{cc K{1.7cm}K{1.7cm}K{1.7cm}K{1.7cm}K{1.7cm}K{1.7cm}}
  \toprule
   & & Miniasm & Spectral & Canu & Miniasm+Racon & Miniasm+Racon (2 iter.) & Spectral+Racon\\
  \hline
  \multirow{8}*{\parbox{1.cm}{\textit{A. baylyi} ONT R7.3 28x}}& Relocations [\#] & 0 & 0 & 2 & 2 & 2 & 0\\
  & Translocations [\#] & 0 & 0 & 0 & 0 & 0 & 0\\
  & Inversions [\#] & 0 & 0 & 0 & 0 & 0 & 0\\
  & Missmbld. contigs [\#] & 0 & 0 & 1 & 1 & 1 & 0\\
  & Missmbld. contigs length [bp] & 0 & 0 & 3513432 & 1993457 & 1994286 & 0\\
  & Local misassemblies [\#] & 0 & 7 & 5 & 0 & 0 & 0\\
  & Mismatches [\#] & 0 & 0 & 0 & 0 & 0 & 0\\
  & Indels [\#] & 0 & 0 & 0 & 0 & 0 & 0\\
  & Indels length [bp] & 0 & 0 & 0 & 0 & 0 & 0\\\hline
  \multirow{8}*{\parbox{1.cm}{\textit{E. coli} ONT R7.3 30x}}& Relocations [\#] & 0 & 2 & 6 & 3 & 3 & 2\\
  & Translocations [\#] & 0 & 0 & 0 & 0 & 0 & 0\\
  & Inversions [\#] & 0 & 0 & 2 & 0 & 0 & 0\\
  & Missmbld. contigs [\#] & 0 & 1 & 2 & 2 & 2 & 1\\
  & Missmbld. contigs length [bp] & 0 & 2160837 & 4625543 & 3743081 & 3740186 & 2148788\\
  & Local misassemblies [\#] & 0 & 50 & 2 & 2 & 2 & 3\\
  & Mismatches [\#] & 0 & 55 & 0 & 0 & 0 & 0\\
  & Indels [\#] & 0 & 1 & 1 & 0 & 0 & 0\\
  & Indels length [bp] & 0 & 30 & 1 & 0 & 0 & 0\\\hline
  \multirow{8}*{\parbox{1.cm}{\textit{S. cerevisiae} ONT R7.3 68x}}& Relocations [\#] & 0 & 0 & 17 & 6 & 7 & 1\\
  & Translocations [\#] & 0 & 7 & 17 & 12 & 12 & 10\\
  & Inversions [\#] & 0 & 0 & 0 & 0 & 0 & 0\\
  & Missmbld. contigs [\#] & 0 & 7 & 16 & 11 & 11 & 10\\
  & Missmbld. contigs length [bp] & 0 & 1223452 & 4852688 & 4638491 & 4638515 & 909031\\
  & Local misassemblies [\#] & 0 & 57 & 17 & 9 & 10 & 12\\
  & Mismatches [\#] & 0 & 63 & 0 & 0 & 0 & 0\\
  & Indels [\#] & 0 & 3 & 2 & 3 & 2 & 1\\
  & Indels length [bp] & 0 & 90 & 124 & 167 & 132 & 54\\\hline
  \multirow{8}*{\parbox{1.cm}{\textit{S. cerevisiae} ONT R9 86x}}& Relocations [\#] & 0 & 5 & 22 & 9 & 9 & 4\\
  & Translocations [\#] & 0 & 18 & 17 & 9 & 10 & 32\\
  & Inversions [\#] & 0 & 0 & 0 & 0 & 0 & 0\\
  & Missmbld. contigs [\#] & 0 & 11 & 11 & 10 & 11 & 10\\
  & Missmbld. contigs length [bp] & 0 & 3149392 & 5957900 & 4545988 & 4563372 & 2661541\\
  & Local misassemblies [\#] & 0 & 41 & 88 & 11 & 11 & 30\\
  & Mismatches [\#] & 0 & 0 & 0 & 0 & 0 & 0\\
  & Indels [\#] & 0 & 2 & 4 & 3 & 3 & 2\\
  & Indels length [bp] & 0 & 161 & 250 & 208 & 207 & 157\\\hline
  \multirow{8}*{\parbox{1.cm}{\textit{E. coli} PacBio 161x}}& Relocations [\#] & 0 & 3 & 2 & 2 & 2 & 2\\
  & Translocations [\#] & 0 & 0 & 0 & 0 & 0 & 0\\
  & Inversions [\#] & 0 & 2 & 2 & 2 & 2 & 2\\
  & Missmbld. contigs [\#] & 0 & 1 & 1 & 1 & 1 & 1\\
  & Missmbld. contigs length [bp] & 0 & 2848876 & 4670125 & 4653228 & 4645420 & 2818134\\
  & Local misassemblies [\#] & 0 & 66 & 2 & 3 & 2 & 2\\
  & Mismatches [\#] & 0 & 0 & 0 & 0 & 0 & 0\\
  & Indels [\#] & 0 & 0 & 0 & 1 & 0 & 0\\
  & Indels length [bp] & 0 & 0 & 0 & 66 & 0 & 0\\\hline
  \multirow{8}*{\parbox{1.cm}{\textit{S. cerevisiae} PacBio 127x}}& Relocations [\#] & 0 & 17 & 31 & 21 & 20 & 18\\
  & Translocations [\#] & 0 & 40 & 44 & 39 & 38 & 50\\
  & Inversions [\#] & 0 & 0 & 1 & 1 & 1 & 0\\
  & Missmbld. contigs [\#] & 0 & 28 & 24 & 22 & 21 & 31\\
  & Missmbld. contigs length [bp] & 0 & 6470761 & 10214689 & 9569247 & 9421896 & 6683508\\
  & Local misassemblies [\#] & 0 & 157 & 26 & 42 & 30 & 33\\
  & Mismatches [\#] & 0 & 0 & 0 & 5 & 0 & 0\\
  & Indels [\#] & 0 & 3 & 8 & 9 & 6 & 2\\
  & Indels length [bp] & 0 & 132 & 260 & 416 & 245 & 78\\
  \botrule
\end{tabular}
}{
This report was obtained with QUAST \citep{GurevichQUAST} (only a subset of the report is shown).
Given the accuracy of the Miniasm assembly, it is likely that the zeros in the Miniasm column are due to the fact that the algorithm failed to correctly match the sequences, rather than the absence of misassemblies.
On all ONT datasets, the Spectral and Spectral+Racon methods are among those yielding the least global misassemblies (relocation, translocation or inversions).
}
\end{table}

\begin{table}[tbh]
\centering
\processtable{Misassemblies report of the different assemblers across the datasets corrected with Canu}
{\footnotesize\label{tab:corrMisassemblies}
\begin{tabular}{cc K{1.7cm}K{1.7cm}K{1.7cm}K{1.7cm}K{1.7cm}K{1.7cm}}
  \toprule
   & & Miniasm & Spectral & Canu & Miniasm+Racon & Miniasm+Racon (2 iter.) & Spectral+Racon\\
  \hline
  \multirow{8}*{\parbox{1.cm}{\textit{A. baylyi} ONT R7.3 28x (26x)}}& Relocations [\#] & 2 & 1 & 2 & 2 & 2 & 1\\
  & Translocations [\#] & 0 & 0 & 0 & 0 & 0 & 0\\
  & Inversions [\#] & 0 & 0 & 0 & 0 & 0 & 0\\
  & Missmbld. contigs [\#] & 1 & 1 & 1 & 1 & 1 & 1\\
  & Missmbld. contigs length [bp] & 1949981 & 3245660 & 2802152 & 1976843 & 1977319 & 3244955\\
  & Local misassemblies [\#] & 4 & 1 & 3 & 2 & 1 & 0\\
  & Mismatches [\#] & 0 & 0 & 0 & 0 & 0 & 0\\
  & Indels [\#] & 0 & 0 & 0 & 0 & 0 & 0\\
  & Indels length [bp] & 0 & 0 & 0 & 0 & 0 & 0\\\hline
  \multirow{8}*{\parbox{1.cm}{\textit{E. coli} ONT R7.3 30x (27x)}}& Relocations [\#] & 2 & 2 & 2 & 2 & 2 & 2\\
  & Translocations [\#] & 0 & 0 & 0 & 0 & 0 & 0\\
  & Inversions [\#] & 0 & 0 & 2 & 0 & 0 & 0\\
  & Missmbld. contigs [\#] & 1 & 1 & 2 & 1 & 1 & 1\\
  & Missmbld. contigs length [bp] & 3945897 & 4613973 & 4627578 & 3962753 & 3962721 & 4613521\\
  & Local misassemblies [\#] & 5 & 2 & 2 & 2 & 2 & 2\\
  & Mismatches [\#] & 58 & 0 & 0 & 77 & 77 & 77\\
  & Indels [\#] & 3 & 1 & 1 & 2 & 2 & 2\\
  & Indels length [bp] & 13 & 1 & 1 & 2 & 2 & 2\\\hline
  \multirow{8}*{\parbox{1.cm}{\textit{S. cerevisiae} ONT R7.3 68x (38x)}}& Relocations [\#] & 6 & 7 & 14 & 7 & 6 & 9\\
  & Translocations [\#] & 13 & 15 & 12 & 13 & 14 & 15\\
  & Inversions [\#] & 0 & 0 & 0 & 0 & 0 & 0\\
  & Missmbld. contigs [\#] & 11 & 15 & 14 & 11 & 11 & 15\\
  & Missmbld. contigs length [bp] & 5025689 & 2643657 & 2808407 & 5053047 & 5052895 & 2634865\\
  & Local misassemblies [\#] & 12 & 26 & 10 & 6 & 7 & 10\\
  & Mismatches [\#] & 21 & 0 & 0 & 0 & 0 & 0\\
  & Indels [\#] & 3 & 1 & 1 & 3 & 1 & 1\\
  & Indels length [bp] & 122 & 78 & 78 & 235 & 78 & 78\\\hline
  \multirow{8}*{\parbox{1.cm}{\textit{S. cerevisiae} ONT R9 86x (40x)}}& Relocations [\#] & 11 & 7 & 13 & 11 & 11 & 8\\
  & Translocations [\#] & 10 & 25 & 13 & 11 & 11 & 30\\
  & Inversions [\#] & 0 & 0 & 0 & 0 & 0 & 0\\
  & Missmbld. contigs [\#] & 10 & 12 & 12 & 9 & 9 & 13\\
  & Missmbld. contigs length [bp] & 4954988 & 3199985 & 3534917 & 4573865 & 4573600 & 3361506\\
  & Local misassemblies [\#] & 12 & 58 & 8 & 9 & 10 & 16\\
  & Mismatches [\#] & 55 & 0 & 0 & 0 & 0 & 0\\
  & Indels [\#] & 1 & 0 & 1 & 1 & 1 & 0\\
  & Indels length [bp] & 7 & 0 & 54 & 54 & 54 & 0\\\hline
  \multirow{8}*{\parbox{1.cm}{\textit{E. coli} PacBio 161x (38x)}}& Relocations [\#] & 2 & 2 & 2 & 2 & 2 & 2\\
  & Translocations [\#] & 0 & 0 & 0 & 0 & 0 & 0\\
  & Inversions [\#] & 0 & 2 & 2 & 2 & 2 & 2\\
  & Missmbld. contigs [\#] & 1 & 1 & 1 & 1 & 1 & 1\\
  & Missmbld. contigs length [bp] & 4642736 & 4663427 & 4670125 & 4642423 & 4642443 & 4662179\\
  & Local misassemblies [\#] & 13 & 5 & 2 & 2 & 2 & 3\\
  & Mismatches [\#] & 0 & 0 & 0 & 0 & 0 & 0\\
  & Indels [\#] & 0 & 0 & 0 & 0 & 0 & 0\\
  & Indels length [bp] & 0 & 0 & 0 & 0 & 0 & 0\\\hline
  \multirow{8}*{\parbox{1.cm}{\textit{S. cerevisiae} PacBio 127x (37x)}}& Relocations [\#] & 29 & 22 & 31 & 33 & 33 & 24\\
  & Translocations [\#] & 44 & 52 & 44 & 42 & 42 & 56\\
  & Inversions [\#] & 1 & 1 & 1 & 1 & 1 & 0\\
  & Missmbld. contigs [\#] & 22 & 33 & 24 & 22 & 22 & 34\\
  & Missmbld. contigs length [bp] & 10163939 & 9816851 & 10214692 & 10180811 & 10178266 & 9840033\\
  & Local misassemblies [\#] & 49 & 59 & 26 & 24 & 25 & 28\\
  & Mismatches [\#] & 28 & 0 & 0 & 0 & 0 & 0\\
  & Indels [\#] & 8 & 6 & 8 & 5 & 6 & 7\\
  & Indels length [bp] & 462 & 216 & 260 & 147 & 153 & 222\\
  \botrule
\end{tabular}
}{
This report was obtained with QUAST (only a subset of the report is shown).
The number of local misassemblies is smaller than with the uncorrected data, but the number of global ones is not.
None of the assemblers has a significantly smaller or larger number of misassemblies compared to the others.
}
\end{table}

\begin{table}[tbh]
  \centering
  \caption{Assembly of each chromosome of \textit{S. cerevisiae} (for each chromosome, we used the subset of reads from the \textit{S. cerevisiae} ONT R7.3 dataset that were mapped to it).
  The assembled contigs were evaluated with QUAST and DNAdiff for each chromosome (only a subset of the QUAST descriptive statistics is shown here).
  This experiments sheds light on how our method would behave if there were no repeats between chromosomes, or if we knew to which chromosomes some reads belong to thanks to, e.g., optical mapping.}
{\footnotesize\label{tbl:EukSepChr}
\begin{tabular}{ccccccc}
\toprule
 Chr. & Ref size [bp] & Contigs [\#] & Aln. bp ref [bp] & Aln. bp query [bp] & Misassemblies [\#] & Avg. identity [\%]\\
 \hline
  I & 230218 & 1 & 228273(99.16\%) & 225845(98.43\%) & 0 & 98.21\\
  II & 813184 & 1 & 806340(99.16\%) & 797624(98.91\%) & 0 & 98.17\\
  III & 316620 & 4 & 313707(99.08\%) & 326011(93.47\%) & 3 & 98.33\\
  IV & 1531933 & 6 & 1519577(99.19\%) & 1539642(99.04\%) & 0 & 98.24\\
  V & 576874 & 1 & 574944(99.67\%) & 575037(99.30\%) & 3 & 98.37\\
  VI & 270161 & 3 & 270161(100.00\%) & 285160(98.97\%) & 0 & 98.36\\
  VII & 1090940 & 8 & 1088278(99.76\%) & 1115166(98.37\%) & 0 & 98.09\\
  VIII & 562643 & 2 & 556839(98.97\%) & 561348(99.48\%) & 2 & 98.22\\
  IX & 439888 & 2 & 437971(99.56\%) & 443785(97.81\%) & 0 & 98.38\\
  X & 745751 & 2 & 740696(99.32\%) & 738859(99.16\%) & 0 & 98.35\\
  XI & 666816 & 2 & 665942(99.87\%) & 667003(99.46\%) & 0 & 98.35\\
  XII & 1078177 & 5 & 1067559(99.02\%) & 1084233(98.50\%) & 2 & 98.27\\
  XIII & 924431 & 4 & 922948(99.84\%) & 937417(99.58\%) & 1 & 98.12\\
  XIV & 784333 & 2 & 779066(99.33\%) & 783072(99.35\%) & 0 & 98.41\\
  XV & 1091291 & 3 & 1089941(99.88\%) & 1088832(99.49\%) & 0 & 98.34\\
  XVI & 948066 & 11 & 942078(99.37\%) & 1015108(97.50\%) & 1 & 97.83\\
  Chrmt. & 85779 & 5 & 65196(76.00\%) & 69107(80.98\%) & - & 90.32\\
\botrule
\end{tabular}
}{
}
\end{table}

\begin{figure}[htb]
  \centering
  \begin{subfigure}[b]{0.4\textwidth}
    \centering \includegraphics[width=\textwidth]{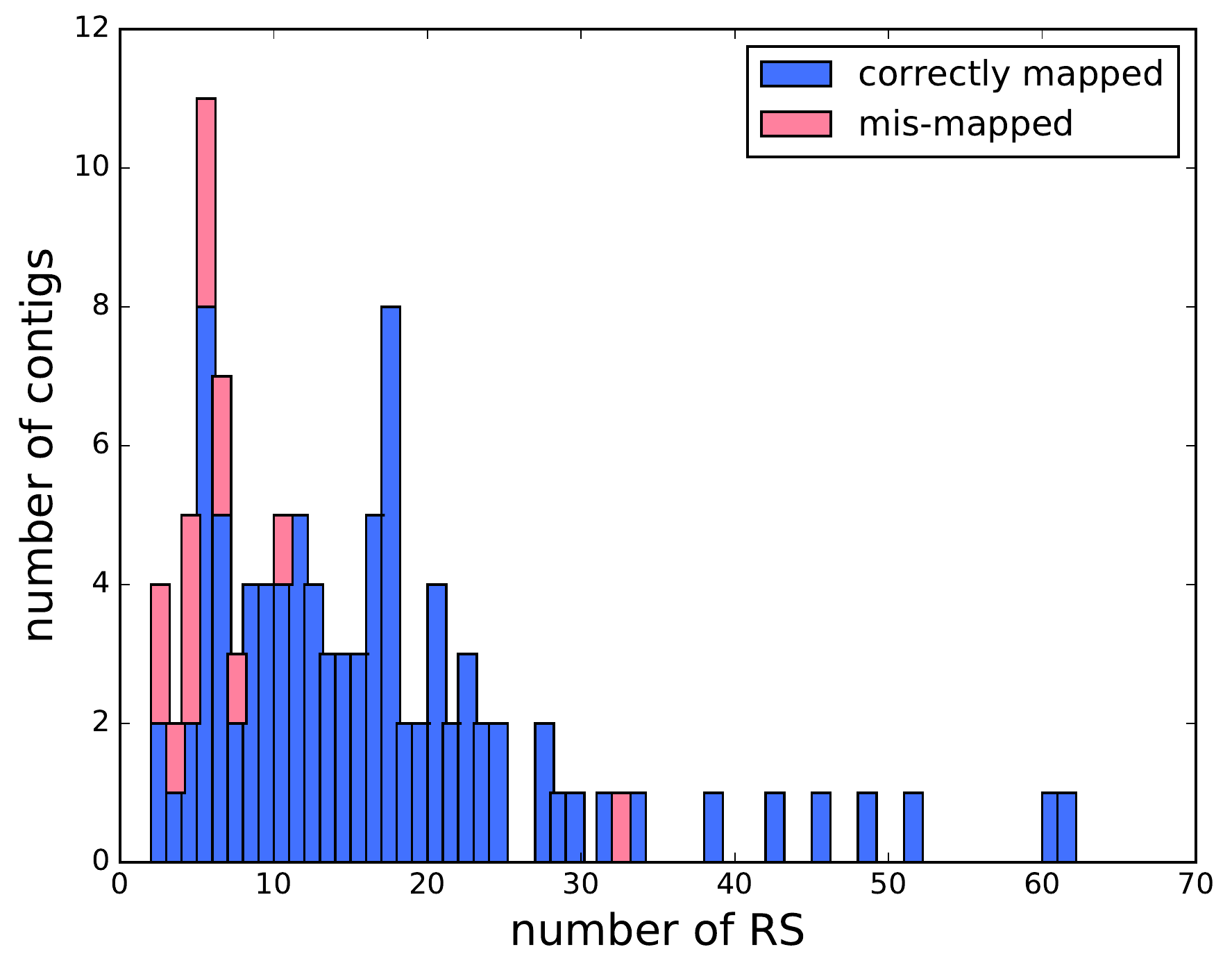}
    \caption{\textit{A. baylyi} ONT}
  \end{subfigure}
  ~
  \begin{subfigure}[b]{0.4\textwidth}
    \centering \includegraphics[width=\textwidth]{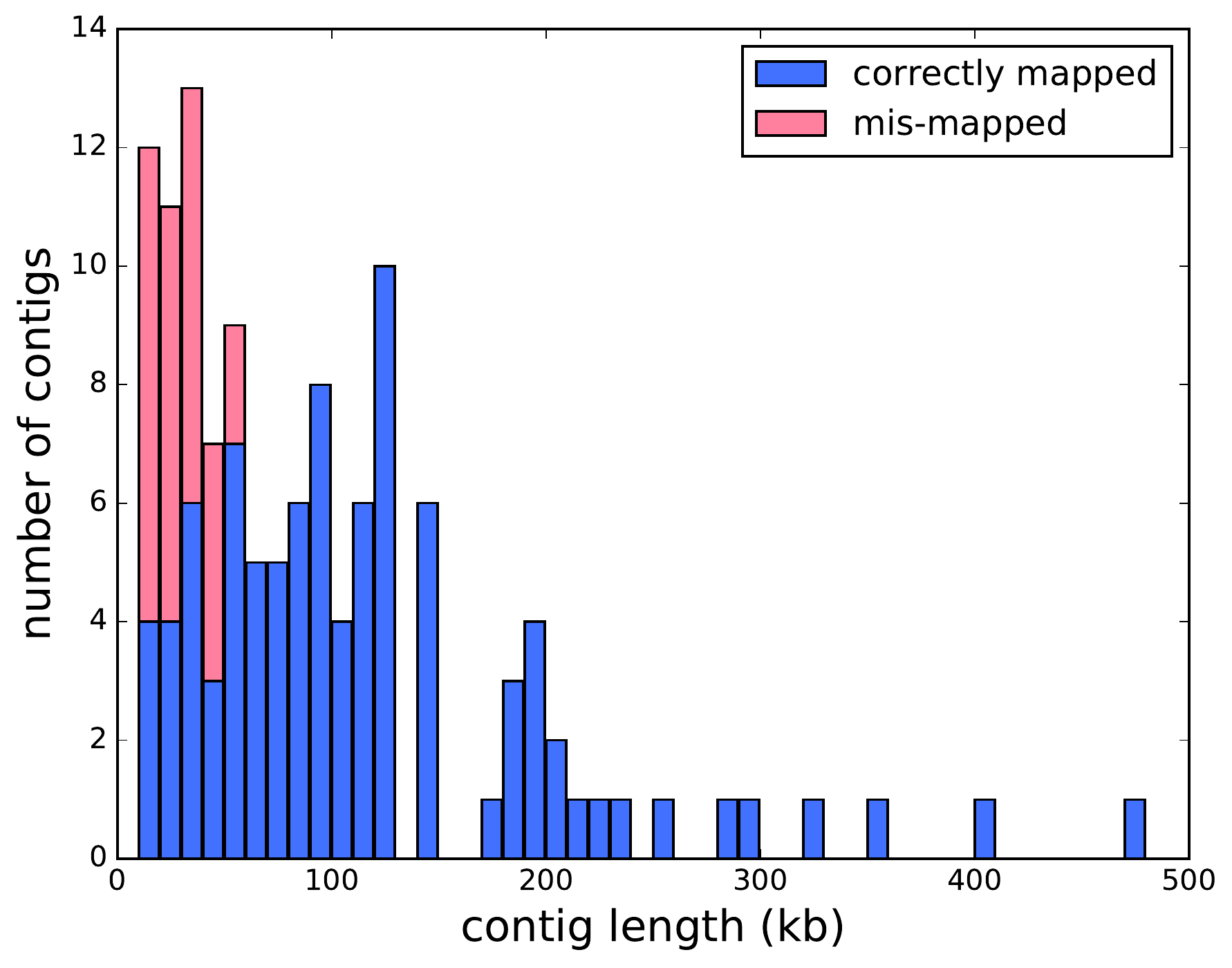}
    \caption{\textit{S. cerevisiae} ONT R7.3}
  \end{subfigure}
  \caption{
  Result of an experiment to evaluate the extent to which optical mapping could improve long-range anchoring of the 127 \textit{S. cerevisiae} ONT R7.3 contigs and provide an alternative consistency check of the assembly.
  A restriction map was generated \textit{in silico}  from the reference \textit{S. cerevisiae} genome with the BamHI restriction site (GGATCC), yielding one map per chromosome.
  This simulated optical map represents a best-case scenario since real optical measurements lack some precision and are obtained through an error-prone assembly process.
  Using the same algorithm as for the \textit{A. baylyi} genome, we obtained this bar plot
  showing the number of contigs as a function of the number of distinct restriction sites (RS) in their sequence (a) or contig length (b).
  For a given number of RS occurrences (a) or contig length (b), the blue part of the bar shows the fraction of contigs correctly aligned to the theoretical restriction map, whereas the red part corresponds to the complementary fraction of unperfectly aligned contigs.
  All contigs longer than 60kbp are correctly mapped.}
  \label{fig:opticalMap}
\end{figure}

\FloatBarrier
\subsection*{Implementation and reproducibility}\label{subsec:Implementation}
Spectrassembler is implemented in python and available on https://github.com/antrec/spectrassembler with a usage example of how to reproduce the results obtained with \textit{ E. coli} ONT data.
We used the following software :
\begin{itemize}
  \item SPOA - https://github.com/rvaser/spoa - commit b29e10ba822c2c47dfddf3865bc6a6fea2c3d69b
  \item Minimap - https://github.com/lh3/minimap - commit 1cd6ae3bc7c7a6f9e7c03c0b7a93a12647bba244
  \item Miniasm - https://github.com/lh3/miniasm - commit 17d5bd12290e0e8a48a5df5afaeaef4d171aa133
  \item Canu v1.4 - https://github.com/marbl/canu - commit r8037 4ece307bc793c3bc61628526429c224c477c2224
  \item Racon - https://github.com/isovic/racon - commit e55bb714ef534ae6d076ff657581836f324e0776
  \item MUMmer's DNAdiff version 1.2, NUCmer version 3.07 - http://mummer.sourceforge.net/
  \item QUAST - https://sourceforge.net/projects/quast/files/
  \item GraphMap - https://github.com/isovic/GraphMap - commit 84f058f92dc5be02022e944dd1d6b9414476432a
  \item errorrates.py from samscripts - https://github.com/isovic/samscripts - commit cd7440fbbffafd76f40b15973c93acbe6111265a
  \item NanoSim - https://github.com/bcgsc/NanoSim - commit 48b9a4c3fcaeff623b9207b7db6d6d88b89a5647
\end{itemize}
SPOA is used in our pipeline for performing multiple sequence alignment.
For generating the consensus in windows, it was run with the options : \texttt{-l 2 -r 0 -x -3 -o -5 -e -2} (semi-global alignment with custom gap and mismatch penalties).
minimap was run with options \texttt{-Sw5 -L100 -m0 -t12} (long reads specific values and multithreading with 12 threads).
miniasm was run with default parameters when used as a comparative method.
Canu was run with saveReadCorrections=True option and data specifications (e.g., \texttt{genomeSize=3.6m -nanopore-raw}).
Racon was run with the alignment generated with minimap (to map the draft assembly, either from miniasm or from our pipeline) with default parameters.
GraphMap \citep{sovic2016fast} was used to generate alignment between the reads and the reference genome in order to have the position of the reads and their error rate (which was computed with the script errorrates.py).
DNAdiff and QUAST were used to evaluate the assemblies.
To concatenate the contigs obtained with our method, we extracted their ends (end length used : 35kbp) and used minimap with options \texttt{-Sw5 -L500} to compute overlaps between them, and ran miniasm with options -1 -2 -e 0 -c 0 -r 1,0 (no pre-selection, no cutting small unitigs, no overlap drop).
The related script is available in the tools folder of our GitHub code.
We also publish the other scripts we used (although they may be poorly written and undocumented), including our implementation of the optical mapping algorithm of \citet{Nagarajan08}, in the tools folder.

\clearpage

\end{document}